\documentclass[]{interact}

\usepackage[numbers,sort&compress]{natbib}
\usepackage{graphicx}
\usepackage{subcaption}
\usepackage{epstopdf}
\usepackage{csquotes}
\usepackage{amsmath, amsfonts, amssymb}
\usepackage{multirow,bigstrut}
\usepackage{tabu}
\usepackage{booktabs}
\usepackage{xcolor}
\usepackage{algorithm}
\usepackage{algpseudocode}
\usepackage{ragged2e}
\usepackage{float}
\newcommand*{\skipnumber}[2][1]{%
	{\renewcommand*{\alglinenumber}[1]{}\State #2}%
	\addtocounter{ALG@line}{-#1}}

\graphicspath{{Figure/}}

\bibpunct[, ]{[}{]}{,}{n}{,}{,}
\renewcommand\bibfont{\fontsize{10}{12}\selectfont}
\makeatletter
\def\NAT@def@citea{\def\@citea{\NAT@separator}}
\makeatother

\theoremstyle{plain}

\theoremstyle{definition}

\theoremstyle{remark}

\begin{document}
	
	\articletype{Pre-print IJPEDS}
	
	\title{A robust image-based cryptology scheme based on cellular non-linear network and local image descriptors}
	
	\author{
		\name{Mohammad Mahdi Dehshibi\textsuperscript{a}\thanks{M. M. Dehshibi. dehshibi@iranprc.org, mohammad.dehshibi@yahoo.com}, Jamshid Shanbehzadeh\textsuperscript{b}\thanks{Corresponding author. Email: jamshid@khu.ac.ir}, and Mir Mohsen Pedram\textsuperscript{b}}
		\affil{
			\textsuperscript{a}Department of Computer Engineering, Faculty of Computer and Electrical Engineering, Science and Research Branch, Islamic Azad University, Tehran, Iran; \newline
			\textsuperscript{b}Department of Electrical and Computer Engineering, Faculty of Engineering, Kharazmi University, Tehran, Iran}
	}
	
	\maketitle

\begin{abstract}
Cellular nonlinear network (CNN) provides an infrastructure for Cellular Automata to have not only an initial state but an input which has a local memory in each cell with much more complexity. This property has many applications which we have investigated it in proposing a robust cryptology scheme. This scheme consists of a cryptography and steganography sub-module in which a 3D CNN is designed to produce a chaotic map as the kernel of the system to preserve confidentiality and data integrity in cryptology. Our contributions are three-fold including (1) a feature descriptor is applied to the cover image to form the secret key while conventional methods use a predefined key, (2) a 3D CNN is used to make a chaotic map for making cipher from the visual message, and (3) the proposed CNN is also used to make a dynamic $k$-LSB steganography. Conducted experiments on 25 standard images prove the effectiveness of the proposed cryptology scheme in terms of security, visual, and complexity analysis.
\end{abstract}

\section{Introduction}
Sensitive and personal data are parts of data transmission over the Internet which have always been suspicious to be intercepted. Confidentiality and data integrity are two prerequisites for users to be protected against unauthorized access during information exchange over the Internet or other open channels. To make communications secure, cryptology, i.e., cryptography and steganography, has widely been utilized \cite{cheddad2010digital}, \cite{olaniyi2013survey}, \cite{zielinska2014trends}. In cryptography, original data is transformed into an unintelligible form, known as the cipher, using mathematical operations. In the mathematical operation, data get encrypted using a key, which is only known by the sender and receiver \cite{nissenbaum2009privacy}. Generally, the cryptographic algorithms are divided into public and private key categories based on the key generation method. Public key uses asymmetric algorithms in which any person can encrypt a message intended for a specific receiver using the public key while the encrypted message can only be decrypted with the receiver's private key. Due to the computational complexity of asymmetric encryption, public key cryptography is usually used for small blocks of data. In private Key cryptography, the secret key is just known to sender and receiver; therefore, if the key is lost, the system becomes void. To resolve this defect, cryptography systems have been forced to use a public key along with the private key, i.e., public key infrastructure (PKI). Steganography, on the other hand, makes information hidden within another information \cite{subhedar2014current}. While both security scheme can provide confidentiality and identification during a secure communication, cryptography has two extra security features which are data integrity and non-repudiation. 

Neither Cryptography nor Steganography is the swift yet comprehensive solution to observe the privacy issues of open systems. While cryptography is very powerful for securing data, the algorithm can be broken with an exhaustive brute-force-attack \cite{hoffman2009survey}. There are similar conditions for steganography, which could compromise the entire system's security through a chance-based attack. Therefore, the combination of cryptography and steganography forms a cross-platform that can efficiently hide an encrypted message within a digital media file and takes cryptography a step farther by hiding the fact that communication is happening \cite{zielinska2014trends}, \cite{jung2016survey}. In this study, we take the similar approach with a three-fold contribution in which (i) a chaotic cellular nonlinear network-based (CNN) architecture is proposed to be used as the kernel of encryption and steganography sub-systems, (ii) the cryptography key formation algorithm is known to both sender and receiver while in the standard approaches, they know the key itself. (iii) Besides standard comparison approaches, we also conduct a series of complexity-driven experiments which could reveal that the proposed Crypto+Stegano approach satisfies privacy criteria while it is complex enough.

It was mentioned that using asymmetric algorithms in the public key encryption could result in increasing the computational complexity concerning the size of data. Therefore, in this study, we focus on the private key encryption in a way that the chance of losing key decreases. In the proposed symmetric key generation algorithm, instead of using a pre-defined phrase as the raw key, we use a local image descriptor, SIFT, which extracts the raw key from the cover image that is used in the steganography sub-module. While the cipher is made, it gets hidden in a cover image and propagates through the communication channel. The receiver just knows which feature descriptor was applied and how the candidate descriptor was selected. Therefore, the problem of ``key distribution'' would be resolved in decryption phase. Furthermore, utilizing different image descriptors as the key of the encryption and decryption modules will influence on the complexity of CNN and, in turn, proposed security schema. Summary of the proposed system is as follows:

\begin{enumerate}
	\item SIFT \cite{lowe1999object} descriptor is applied to the cover image. The most promising descriptors from each feature vector are then selected to make a raw key.
	\item The key in conjunction with plain data is passed into a chaotic system of 3-D CNN to make a chaotic cipher.
	\item The encrypted data is hidden within the cover carrier by a dynamic LSB steganography method in which the proposed CNN defines which bits could be manipulated.
	\item The first step of the de-cryptology is done by splitting the message from the cover. Then the descriptors on which both sender and receiver were dealt is extracted from the cover image and is passed to CNN to decrypt the cipher.
\end{enumerate}

This work also puts emphasis on analyzing the complexity of the whole cryptology system, as a complex system, by using temporal and Kolmogorov complexity measures. The performance of the proposed method is also tested on 25 benchmark images of different sizes through investigating visual and security analysis. The secret key-space, embedding ratio, correlation analysis, occlusion attack are those criteria used in security analysis. In terms of the visual test, we quantify the image quality using full-reference and non-reference metrics.

\section{Background}

Historical concerns for preserving privacy could have brought about different information encryption and hiding techniques which currently can be applied to different file formats. Meanwhile, there is the fact that a data with a higher degree of redundancy is more suited to cryptography and steganography objectives. The redundant bits of a file are those bits that their alterations cannot be detected easily. The image file is the most popular file on the Internet which can also mainly satisfy the redundancy requirement \cite{chen2004symmetric}. Accordingly, our focus in this work revolves around steganography in digital images, and our literature review sticks around image-based crypto+steganography methods.

Steganography methods can be broadly partitioned based on the embedding techniques into spatial and transform domains \cite{al2016steganography}. In spatial domain steganography methods, the message is directly embedded in the pixels of the cover image. The spatial domain consists various approaches such as Least Significant Bit (LSB) \cite{morkel2005overview}, \cite{amirtharajan2013steganography}, gray-level modification \cite{muhammad2015secure}, pixel value differencing, quantization index modulation \cite{chen2001quantization}, multiple based notational system, and predictive coding \cite{soleymani2012survey}. Transform domain steganography methods, on the other hand, transform the Cover image into another domain by performing one or more transforms and then embed Message by modifying the transformed coefficient values. Discrete Transform Domain (DCT), Discrete Wavelet Transform (DWT), and Singular Value Decomposition (SVD) are samples of such transformations \cite{kanan2014novel}. 

A chaotic system is applicable to securing information which facilitates secure communication and cryptography. Mathews \cite{matthews1989derivation} presented the earliest chaotic stream cipher. Pecora and Corel \cite{pecora1991driving}, then, presented two identical chaotic systems that synchronized a secure communication. Three components can be found between chaotic maps and cryptographic algorithms that are proved in \cite{fridrich1997image}, \cite{kocarev2001chaos}, \cite{alvarez2006some}. The first one is Ergodicity in chaos compared with confusion in cryptography. The second component depends on the first state of the algorithm and parameters of chaotic map compared to the diffusion property of a cryptography system. The last one is generating random sequences as a key in cryptography due to the behaviour of chaotic systems. These chaotic properties have been used in designing cryptographic system \cite{yang1997cryptography}, \cite{kocarev2001logistic}. 

Scharinger \cite{scharinger1998fast} includes key-permutations on large data blocks by specific Kolmogorov flows to encrypt image and videos. Yen and Guo \cite{guo2000new} utilized a chaotic system to make a binary sequence as a key in which the pixels of an image is rearranged based on the key, and the scale of grey values are XORed or XNORed bit by bit to one of the two predetermined keys. Pareek \cite{pareek2006image} an external secret key of 80-bit and two chaotic logistic maps are employed to build an image encryption method. In this algorithm, the external key and eight different types of operations were applied to the initial conditions of both logistic maps for encrypting the pixels of an image. To make the cipher more robust against any attack, the secret key is modified after encrypting each block of sixteen pixels of the image. Pisarchik \cite{pisarchik2006encryption} presented a chaotic map lattice in which all pixel colours were converted one by one to lattices of chaotic maps. To handle small key space and weak security problems, Gao \cite{gao2006new} presented an algorithm in which power and tangent functions were used instead of a linear function. Then, Gao and Chen \cite{gao2008new}, improved this algorithm by shuffling the position of pixels in an image whereas a hyper-chaotic Chen system was used to disarrange the relationship between the original image and encrypted image. The chaotic CNN is one of the approaches can be used to construct a cryptography system. Peng et al. \cite{peng2009digital} can prove that the security performance is better compared to other chaos-based image encryption algorithm where the correlation of adjacent pixels is rearranged, and the grey scale value of pixels is changed.

Even though both cryptography and steganography provide security, using Cryptography and Steganography together is beneficial where they can add multiple layers to security. Concerning this hypothesis, some researchers concentrate their objective on proving its validity. Bloisi and Iocchi \cite{bloisi2007image} presented a crypto-steganography system by using images as cover objects for steganography and as keys for cryptography. The strength of the proposed system was in the new concept of key image in which both cover and the key were considered as the cover image to change the cover coefficients randomly. This modification opposes for a steganalytic tool to search for a predictable set of modifications. Bansod et al. \cite{bansod2012modified} suggested a hybrid crypto-steganography scheme in which Block Plane Coding Technique (BPCS) was used to hide more secret data with PSNR values of 43.2, 34.8, and 43.9 dB for Red, Green, and Blue channels, respectively. Bhattacharyya et al. \cite{bhattacharyya2011pixel} improved the robustness of secret data in the BPCS algorithm by investigating the role of Pixel Mapping. In \cite{ranjbar2016evaluation} compressive sensing (CS) theory was used to propose a secure and high capacity crypto-steganography system which preserves the cover image imperceptibility. The discrete Cosine transform (DCT) and random sensing matrix were utilized as the sparse domain and measurement domain, respectively.

\section{Proposed Method}

The proposed method is a two-level cryptology scheme in which a 3D CNN-based architecture is used to encrypt the plain image and, then, to hide the cipher into a cover image. In the proposed symmetrical cryptography algorithm, an image descriptor is applied to the Cover image and the Key is then selected from the descriptors set. Before describing the algorithm, an explanation about symbols which are used in the algorithm is provided. Suppose $I$ is the original image with the size of $ih \times iw$, where $ih$ and $iw$ are the height and the width of the given image, respectively. We can, then, rewrite $I$ as a set of $\{I_{i,j}, \quad i=1,2, \cdots,ih\}$ and $j=1, 2, \cdots, iw$. $T$ denotes the 128 bit key, and can be written as $T_{1}T_{2}\cdots T_{16}$. Other symbols are described in the context of each sub-algorithm.

\subsection{Key Formation}
The first pillar of a cryptographic system is the choice of Key. The complexity of the key can significantly reduce the likelihood of its resolution. In a standard cryptography scheme, the key is a predefined word agreed between the receiver and the sender. As mentioned earlier, the key is not a predefined word, but an algorithm that both sender and receiver have agreed upon applying it. We used Scale-invariant feature transform (SIFT) descriptor which is invariant to partial occlusion, uniform scaling, orientation, illumination changes, and affine distortion. Therefore, if the cover image damaged a bit during transformation, the receiver will have less concern \cite{mikolajczyk2005performance}.

\begin{algorithm}[H]
	\caption{Scale-invariant feature transform algorithm.} \label{alg:sift}
	\begin{flushleft}
		\textbf{Input:}
		\begin{itemize}
			\item[] $I \gets$ input image.
			\item[] $(x,y) \gets$ coordinate locations of $I$
		\end{itemize} 
			
		\textbf{Output:}
		\begin{itemize}
			\item[] $m \gets$ set of gradient magnitude of keypoints.
			\item[] $\theta \gets$ set orientation of keypoints.
		\end{itemize}
		
		\vspace{1em}
		\textbf{Procedure:}
	\end{flushleft}

	\begin{algorithmic}[1]
		\State Convolving the image with variable scale Gaussian kernel function. 
			$$L(x,y,\sigma) = G(x,y,\sigma)\ast I (x,y),$$
			$$G(x,y,\sigma) = \frac{1}{2\pi \sigma}e^{-(x^2+y^2)/2\sigma^2}$$
		\State Calculate the difference between of DoG function, $D(x,y,\sigma)$, of one scale and $k$ times scale to detect stable keypoint locations.
			$$D(x,y,\sigma) = L(x,y,k\sigma)-L(x,y,\sigma).$$
		\State Selecting keypoint candidates.
			\begin{itemize}
				\justifying
				\item[] $\mathbf{min_{D}} = \min\{D(x-1,y-1,\sigma), D(x-1,y,\sigma), D(x-1,y+1,\sigma), D(x,y-1,\sigma), D(x,y+1,\sigma), D(x+1,y-1,\sigma), D(x+1,y,\sigma), D(x+1,y+1,\sigma)\},$
				\item[] $\mathbf{max_{D}} = \max\{D(x-1,y-1,\sigma), D(x-1,y,\sigma), D(x-1,y+1,\sigma), D(x,y-1,\sigma), D(x,y+1,\sigma), D(x+1,y-1,\sigma), D(x+1,y,\sigma), D(x+1,y+1,\sigma)\},$
				\item[] $\mathbf{min_{D_{k}}} = \min\{D_{k}(x-1,y-1,k\sigma), D_{k}(x-1,y,k\sigma), D_{k}(x-1,y+1,k\sigma), D_{k}(x,y-1,k\sigma), D_{k}(x,y,k\sigma), D_{k}(x,y+1,k\sigma), D_{k}(x+1,y-1,k\sigma), D_{k}(x+1,y,k\sigma), D_{k}(x+1,y+1,k\sigma)\},$
				\item[] $\mathbf{max_{D_{k}}} = \max\{D_{k}(x-1,y-1,k\sigma), D_{k}(x-1,y,k\sigma), D_{k}(x-1,y+1,k\sigma), D_{k}(x,y-1,k\sigma), D_{k}(x,y,k\sigma), D_{k}(x,y+1,k\sigma), D_{k}(x+1,y-1,k\sigma), D_{k}(x+1,y,k\sigma), D_{k}(x+1,y+1,k\sigma)\},$
				\item[] \textbf{keypoint} $\gets \{D(x,y,\sigma) \mid (D(x,y,\sigma) \neq \mathbf{min_{D}}) \vee (D(x,y,\sigma) \neq \mathbf{max_{D}}) \vee (D(x,y,\sigma) \neq \mathbf{min_{D_{k}}}) \vee (D(x,y,\sigma) \neq \mathbf{max_{D_{k}}})\},$
			\end{itemize}
		\State Eliminating points that have low contrast or poorly localized on edges.
		
		$$D(\mathbf{x}) = D + \frac{\partial D^T}{\partial{ \mathbf{x}}}\mathbf{x} + \frac{1}{2}\mathbf{x}^T\frac{\partial^2 D}{\partial\mathbf{x}^2}\mathbf{x},$$ 
		\Comment{\textcolor{gray}{$\%\mathbf{x} = (x,y,\sigma)^T$ is the offset from the particular point.}}
		\algstore{bkbreak}
	\end{algorithmic}
\end{algorithm}

\setcounter{algorithm}{0}
\begin{algorithm}[H]
	\caption{(continue) Scale-invariant feature transform algorithm.}
	\begin{algorithmic}[1]
		\algrestore{bkbreak}
		\State continued from 4 
			
					$$\mathbf{\hat{x}} = \frac{\partial^2 D^{-1}}{\partial\mathbf{x}^2}\frac{\partial D}{\partial{\mathbf{x}}}$$
					\Comment{\textcolor{gray}{$\% \mathbf{\hat{x}}$ is the the location of the extreme.}}
					
			\skipnumber[4]{\If {$(\mathbf{\hat{x}} > 0.5) \wedge (|D(\mathbf{\hat{x}})|) > \zeta)$} \newline \Comment{\textcolor{gray}{$\%$ Here, $\zeta = 0.03$.}} \newline \Comment{\textcolor{gray}{$\%$ When $\mathbf{\hat{x}} > 0.5$ the extreme lies closer to a different keypoint.}}
				\State $D(\mathbf{\hat{x}}) = D + \frac{1}{2}\frac{\partial D^T}{\partial\mathbf{x}}\mathbf{\hat{x}}$
			\EndIf}
		
		\State Finding the gradient magnitude, $ m(x,y) $, and orientation, $ \theta(x,y) $, for each \textbf{keypoint} at a particular scale.
		\begin{itemize}
			\justifying
			\item[6-1] Calculating Hessian matrix: $\mathbf{H} = \begin{bmatrix} D_{xx} & D_{xy} \\ D_{xy} & D_{yy} \end{bmatrix},$
			\item[6-2] Calculating eigenvalues of $\mathbf{H}$ which are the principal curvatures of $D$. Here, $\alpha$ and $\beta$ are the largest and smallest magnitude of eigenvalues, respectively; where $\alpha + \beta = \mathrm{Tr(\mathbf{H})}= D_{xx}+D_{yy},$ and $\alpha\beta = \mathrm{Det(\mathbf{H})}= D_{xx}D_{yy} - (D_{xy})^2.$
			\item[6-3] If $\alpha\beta < 0$, $\mathbf{keypoint} \gets \mathbf{keypoint}-\{D(\mathbf{x})\}$
			\item[6-4] Let $\alpha = r\beta$, then $\frac{\mathrm{Tr(\mathbf{H})^2}}{\mathrm{Det(\mathbf{H})}} = \frac{(\alpha + \beta)^2}{\alpha \beta} = \frac{(r+1)^2}{r}$. The value of $r$ has to be below the threshold, which is in this study is 10.
			\item[6-5] Suppose $L_x = L(x+1,y)-L(x-1,y)$ and $L_y = L(x,y+1)-L(x,y-1)$. \\ \Return{$m(x,y) = \sqrt{L_x^2 + L_y^2 }$} \\ \Return{$\theta(x,y) = \tan^{-1}(L_y/L_x)$}.
		\end{itemize}
	\end{algorithmic}
\end{algorithm}

An orientation histogram is formed from the gradient orientations of sample points within a region around the keypoint. Another histogram is then formed by weighting each sample by a Gaussian-weighted circular window with a $\sigma$ that is 1.5 times that of the scale of the keypoint. The highest peak in the histogram is detected, and then any other local peak that is within 80\% of the highest peak is used to also create another keypoint orientation histogram. Finally, a parabola is fit to the 3 histogram values closest to each peak to interpolate the peak position to improve the accuracy. While the image location, scale, and orientation have been assigned to each keypoint, a 2D coordinate system can be imposed to describe the local image region invariant to these parameters. The descriptor is made by assigning a weight to each sample point magnitude by applying a Gaussian weighting function with $\sigma$ equal to one half the width of the descriptor window.

\begin{figure}[H]
	\centering
	\begin{subfigure}[b]{0.49\textwidth}
		\includegraphics[width=\textwidth]{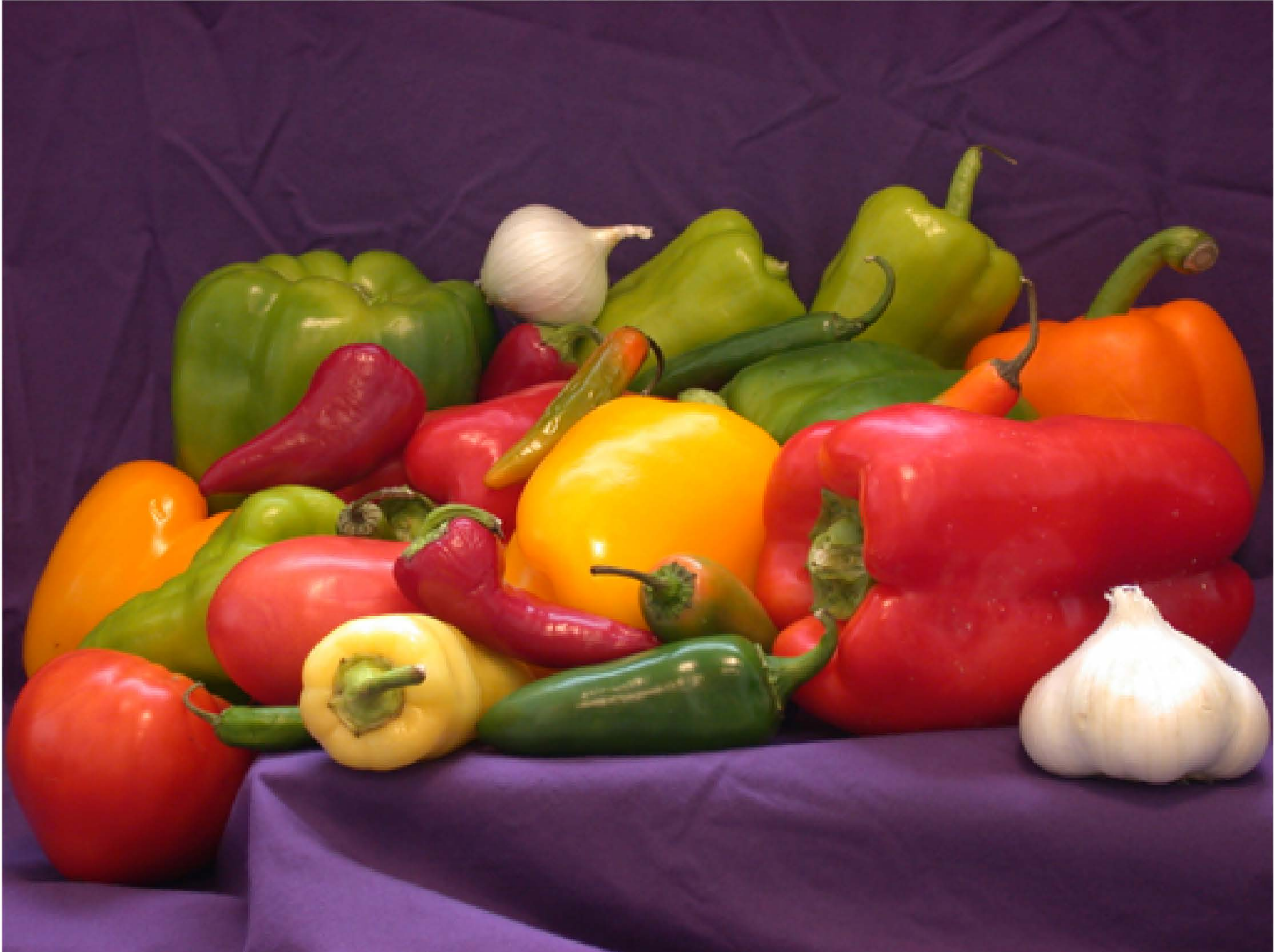}
		\caption{}
	\end{subfigure}
	\hfill
	\begin{subfigure}[b]{0.49\textwidth}
		\includegraphics[width=\textwidth]{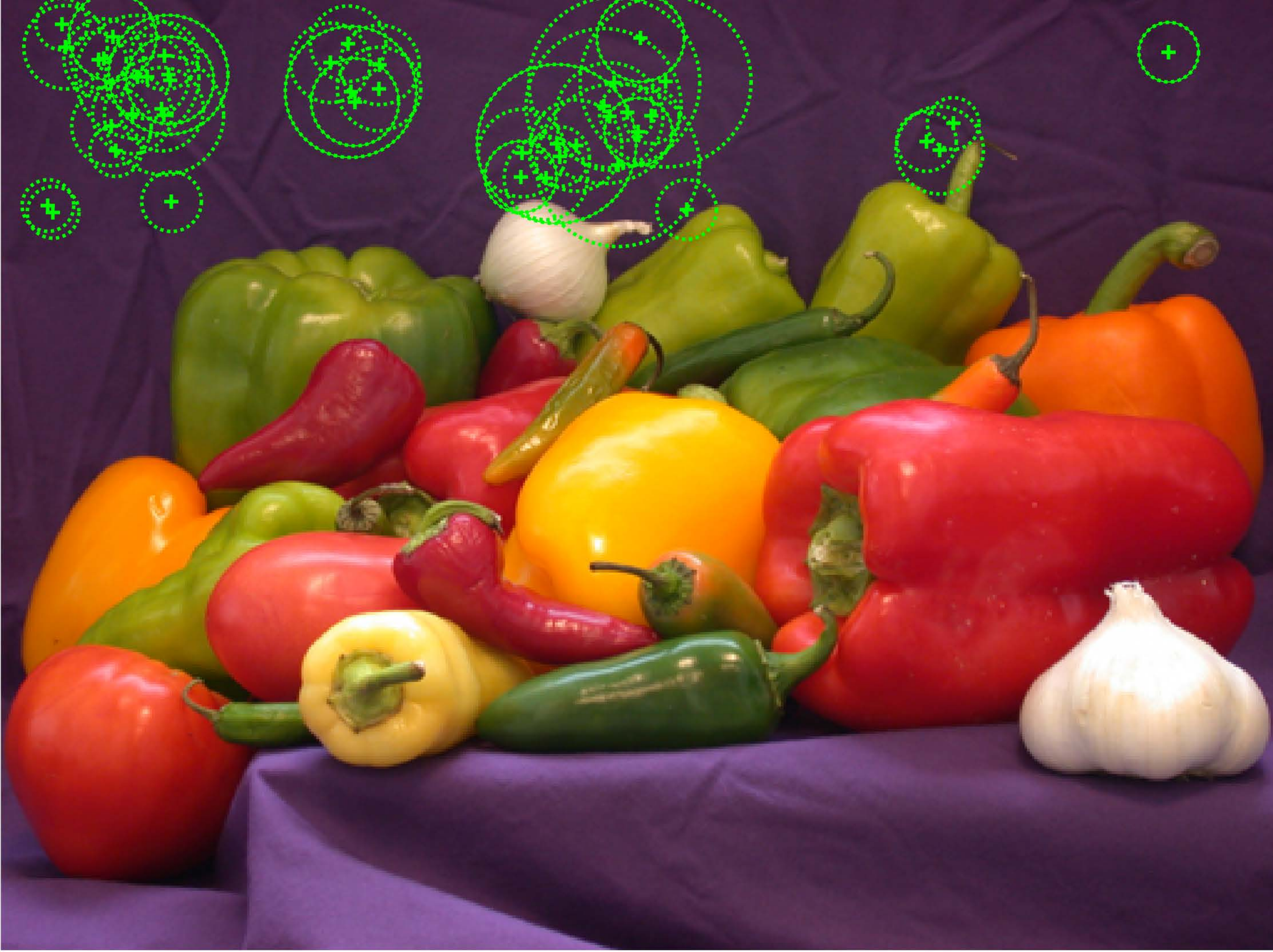}
		\caption{}
	\end{subfigure}
	\hfill
	\begin{subfigure}[b]{0.49\textwidth}
		\includegraphics[width=\textwidth]{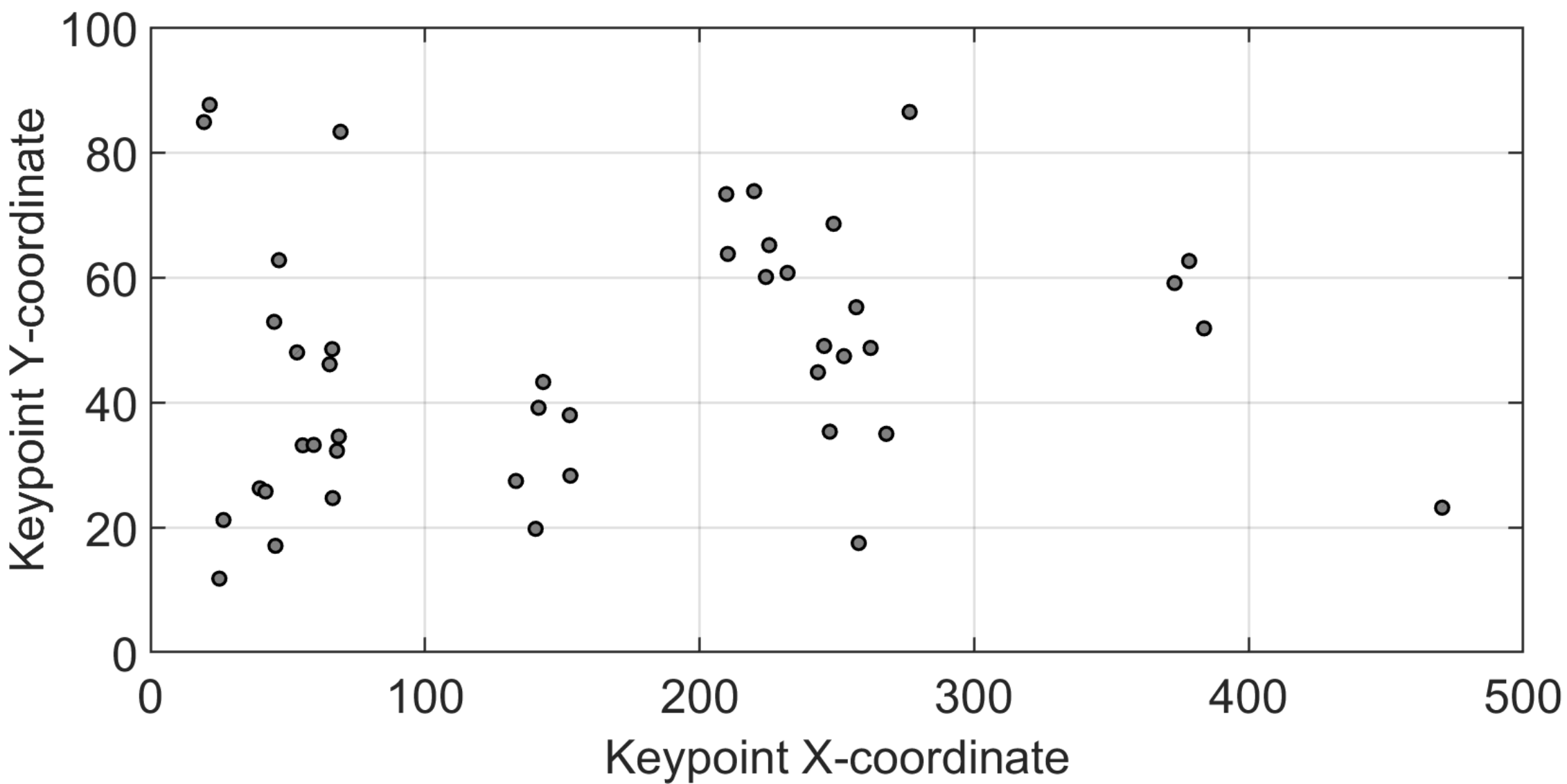}
		\caption{}
	\end{subfigure}
	\hfill
	\begin{subfigure}[b]{0.49\textwidth}
		\includegraphics[width=\textwidth]{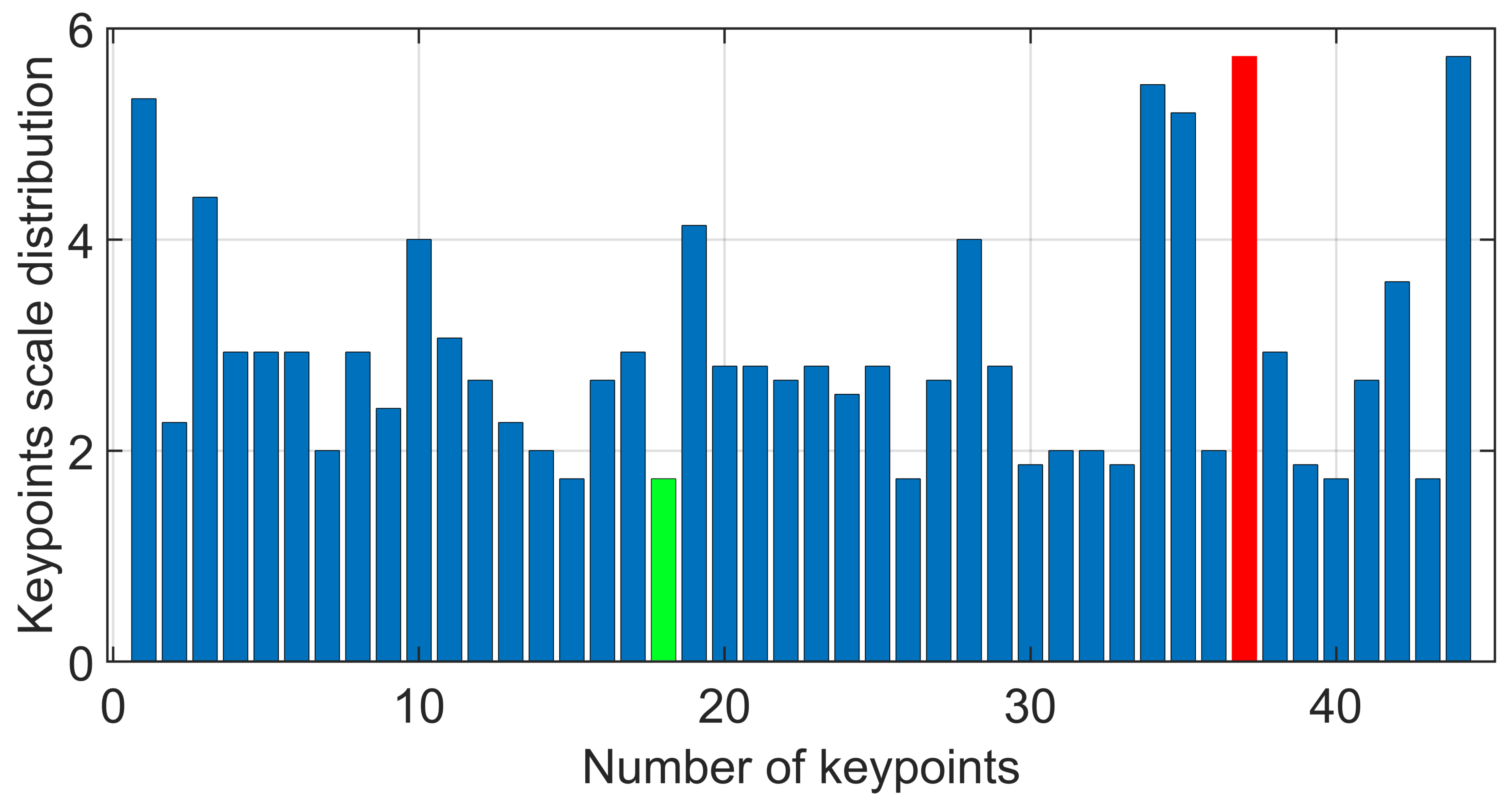}
		\caption{}
	\end{subfigure}
	\caption{(a) Original image, which is used as the Cover, where PSNR = 2.8455 dB (reference image is a same-sized white one) and entropy = 7.3785 (b) Result of applying SIFT to the original image. (c) Distribution of keypoint over $[x,y]$ coordinates. (d) Distribution of keypoint scales. The minimum scale is 1.73 which is shown in a green bin and the maximum scale, which is highlighted with red, is 5.73. Those scales which are below 1.5 were eliminated to ensure that keypoint set is free from low contrast and poorly localized edges.}\label{fig:sift}
\end{figure}

To avoid boundary affects where the descriptor changes from one histogram or orientation to another, tri-linear interpolation is applied to distribute the value of each gradient sample into adjacent histogram bins. Each value in a bin is multiplied by a weight of $1-d$ for each dimension, $d$ being the distance of the sample from the central value of the bin based on the histogram bin spacing. Following Lowe \cite{lowe1999object}, a $4 \times 4$ array of histograms with eight orientations in each bin is used to form a keypoint with the length of $4\times 4 \times 8 = 128$. To limit the effect of illumination changes, the descriptor vector gets normalized to unit length. Consequently, large gradient magnitudes are thresholded by a factor of experimentally determined, 0.2, and the entire feature vector is re-normalized. Figure \ref{fig:sift} shows a sample of applying SIFT to an image.

\subsection{Cryptology Algorithm}

The main module of the proposed cryptology algorithm is a cellular non-linear network (CNN) which is a dynamic model with full connection and exhibits chaotic phenomenon. The use of CNN in image processing was introduced in \cite{lee1996color}. In the proposed method, a three-order CNN dynamic model is defined Each cell of a CNN is characterized by a set of nonlinear differential equations which the mathematical model of a cell state can be formulated with Eq. \ref{equ:1}.

\begin{equation}
    \label{equ:1}
    \frac{dx_{j}}{dt}=-x_{j}+a_{j}y_{j}+\sum_{k=1, k\neq j}^3 a_{jk}y_{k}+\sum_{k=1}^3 b_{jk}x_{k}+z_{j}, \quad j=1,2,3
\end{equation}
where $x_{j}$ and $y_{j}$ are the state variables and cell outputs of CNN, respectively. The relationship between $x_{j}$ and $y_{j}$ is defined by Eq. \ref{equ:2}:

\begin{equation}
    \label{equ:2}
    y_{j} = \frac{|x_{j}+1|-|x_{j}-1|}{2}, \quad j=1,2,3
\end{equation}
where $a_{jk}$ and $b_{jk}$ are the elements of space-invariant template feedback matrix, $\mathbf{A}$, and the input synaptic operators, $\mathbf{B}$, respectively. Here, $z$ is the bias term that is constant for all the CNN cells. These templates must cause the CNN converging to a stable equilibrium state while it must have chaotic attractors or limit cycles. Therefore, the CNN must belong to the uncoupled class in which $a_{jk} = 0$ except for $j = k$. If we set the bias of each layer to zero, $z_{j}=0$ and the following elements of $\mathbf{B}$ are set as $b_{13} = b_{31} = b_{22} = 0; b_{33} = b_{21} = b_{23} = 1$, then, Eq. \ref{equ:1} can be rewritten as a system in Eq. \ref{equ:3}.

\begin{equation}
    \label{equ:3}
    \frac{dx_{j}}{dt} =
    \begin{cases}
        \frac{dx_{1}}{dt} = −x_{1} + a_{1}y_{1} + b_{11}x_{1} + b_{12}x_{2}, & (a)\\
        \frac{dx_{2}}{dt} = −x_{2} +x_{1} + x_{3}, & (b)\\
        \frac{dx_{3}}{dt} = −x_{3} + b_{32}x_{2} + x_{3}.  & (c)
    \end{cases} 
\end{equation}

The coefficient matrix of the linear sub-system of model in Eq. \ref{equ:3} is:
\begin{equation}
	\label{equ:4}
	\mathbf{W} = \begin{bmatrix} b_{11}-1 & b_{12} & 0 \\ 1 & -1 & 1 \\ 0 & b_{32} & b_{33}-1 \end{bmatrix}
\end{equation}

Calculating eigenvalues of matrix $\mathbf{W}$ will result in finding $b_{11}, b_{12}, b_{32},$ and $b_{33}$. As our aim is to provide a chaotic map, if we set $b_{33}$ to 1 while keeping the identical value of $b_{11}, b_{12},$ and $b_{32}$, calculated eigenvalues are -2.5245 , 0.0787+2.5506$i$, and 0.0787-2.5506$i$, respectively . Therefore, we can see at least one eigenvalue has a positive real part which causes the trajectory evolution of model becomes chaotic. This CNN is used in both cryptography and steganography sub-modules. Cryptography scheme has five main steps which are:

\begin{itemize}
	\item [\textbf{Step 1:}] Once the descriptor is selected from the set of features, its Modula-2 is calculated to from a 128-bit key. This key is then re-written as a set of $T=\{T_{1}, T_{2}, \cdots, T_{16}\}$, where each $T_{i}$ contains 8 sequentially selected bits of the key.
	\item[\textbf{Step 2:}]  CNN parameters calculate by Eq. \ref{equ:5}.
		\begin{equation}
			\label{equ:5}
			\begin{split}
				&H_{1} = \oplus_{i=1}^{ih} \oplus_{j=1}^{iw} I_{i,j}, \\
				&H_{2} = (\sum_{i=1}^{ih} \sum_{j=1}^{iw} I_{i,j}) \bmod 256, \\
				&I = (T_{1} + T_{2} + \cdots + T_{16}) \bmod 256, \\
				&P = (T_{1} \oplus T_{2} \oplus \cdots \oplus T_{16}), \\
				&\lambda = (T_{10} + T_{11} + T_{12} + H_{1} \times H_{2}) \bmod 256, \\
				&h = (T_{13} + T_{14} + T_{15} + T_{16} + H_{1} \times H_{2}) / 256.
			\end{split}
		\end{equation}
	\item[\textbf{Step 3:}] Determine the initial condition $(x_{1,0}, x_{2,0}, x_{3,0})$ and iteration number, $N_{0}$, of the CNN using Eq. \ref{equ:6}.
	\begin{equation}
		\label{equ:6}
		\begin{split}
			&x_{1,0} = (T_{1} \times T_{4} \times T_{7} \times I \times P)/256^{5}, \\
			&x_{2,0} = (T_{2} \times T_{5} \times T_{8} \times I \times P)/256^{5}, \\
			&x_{3,0} = (T_{3} \times T_{6} \times T_{9} \times I \times P)/256^{5}, \\
			&N_{0} = (I^{2} + P^{2}) \bmod 256
		\end{split}
	\end{equation}
	CNN iterates for $iw \times ih$ times from the initial condition $(x_{1,0} , x_{2,0} , x_{3,0})$ by using fourth-order Runge-Kutta Method (the time step size is 0.005) to solve the CNN differential equation as in Eq. \ref{equ:3}. Here, to avoid the transient effect, the first $N_{0}$ iteration is considered. \newline
	\item[\textbf{Step 4:}] Let $\left\{(x_{1,i},x_{2,i},x_{3,i})\right\}_{i=1}^{iw\times ih}$ is the CNN solutions set. This set is of size of $iw\times ih$, where the $ith$ element has three points $(x_{1,i}, x_{2,i}, x_{3,i})$, \quad $i = 1, 2, \cdots, iw \times ih$. The $(x_{1,i}, x_{2,i}, x_{3,i})$ point is then mapped to a character that will be used to encrypt image pixel in the future. Mapping procedure is described as follows:
	Assume $v_{i} = {h(x_{1,i}^2 + x_{2,i}^2 + x_{3,i}^2)}^{1/2} + \lambda$, and $u_{i}$ denotes the decimal fraction of $v_{i}$, then $u_{i}$ can be represented as a binary sequence $u_{i} = 0.b_{i,1}b_{i,2}...b_{i,p}$ where $p$ is a certain precision (here, $p=32$). Let $ w_{i} = (w_{i,0}w_{i,1}...w_{i,7})_{2}$, where $w_{i,j}$ denotes the $jth$ bit of $w_{i}$, and $w_{i,j} = b_{i, 4j+1}\oplus b_{i, 4j+2}\oplus b_{i, 4j+3}\oplus b_{i, 4j+4}\oplus, \quad j = 0, 1, \cdots, 7$. Hence, we obtain a character $w_{i}$ from $(x_{1,i},x_{2,i},x_{3,i})$. Finally, when the variable $i$ changes from 1 to $iw \times ih$, we will obtain a pseudo-random key-stream $\left\{w_{i}\right\}_{i=1}^{iw \times ih}$ to encrypt the original image. \newline
	\item[\textbf{Step 5:}] Suppose $e_{i}$, $w_{i}$, and $m_{i}$ denote the encrypted image, pseudo-random key-stream, and original image, respectively, where $i= 1,2, \cdots, iw \times ih$. Then the encryption process is done by Eq. \ref{equ:7}. \newline
	\begin{equation}
		\label{equ:7}
		e_{i}= \mathrm{xor}(w_{i},m_{i}) \bmod 256 
	\end{equation}
\end{itemize}

Results of applying the proposed cryptography algorithm is shown in Figure \ref{fig:cryotography}. As is evident both visually and in the histogram, the cipher is a chaotic map. 

\begin{figure}[H]
	\centering
	\hspace*{2em}\begin{subfigure}[b]{0.33\textwidth}
		\includegraphics[width=\textwidth]{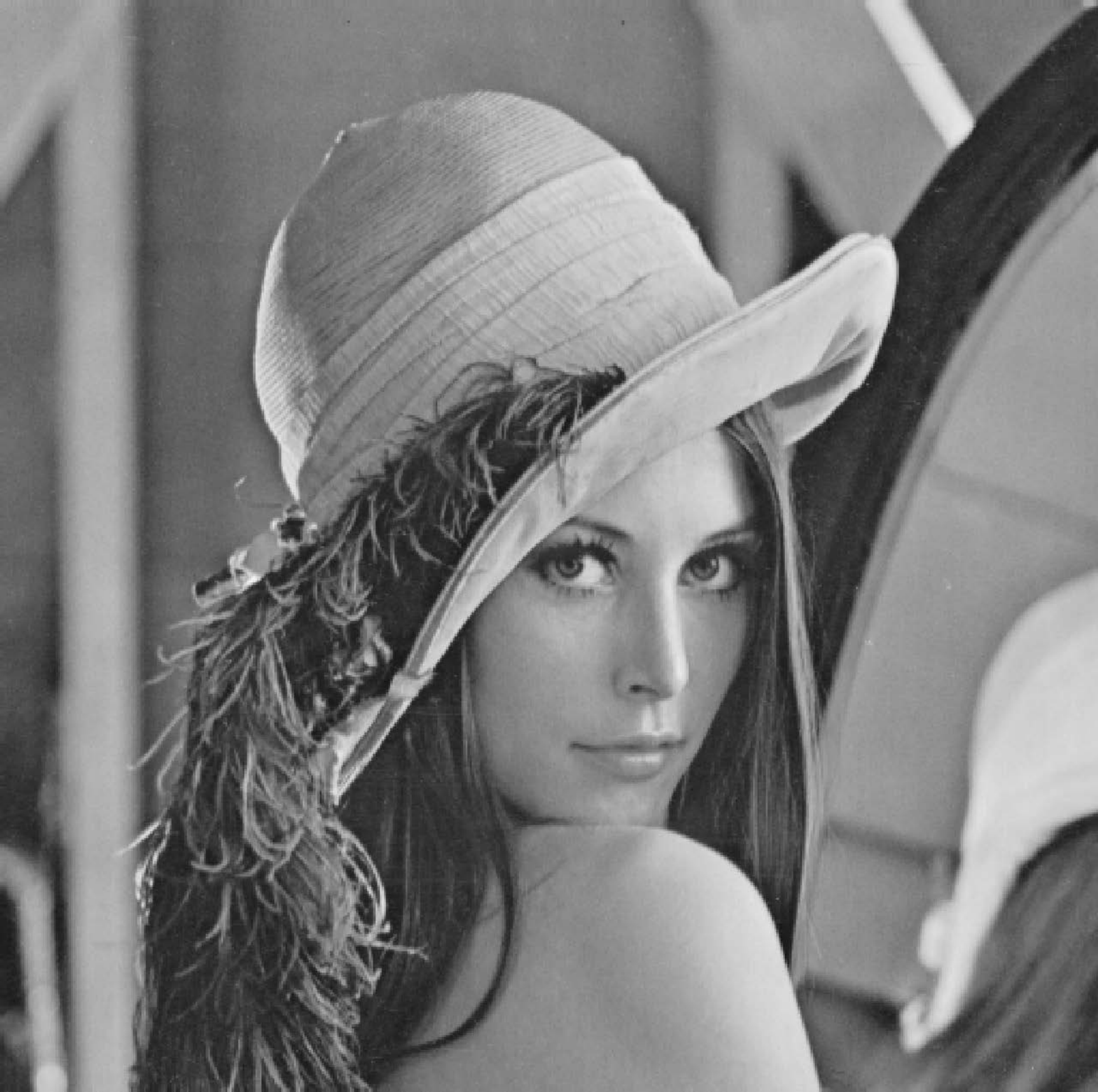}
		\caption{}
	\end{subfigure}
	\hfill
	\begin{subfigure}[b]{0.45\textwidth}
		\includegraphics[width=\textwidth]{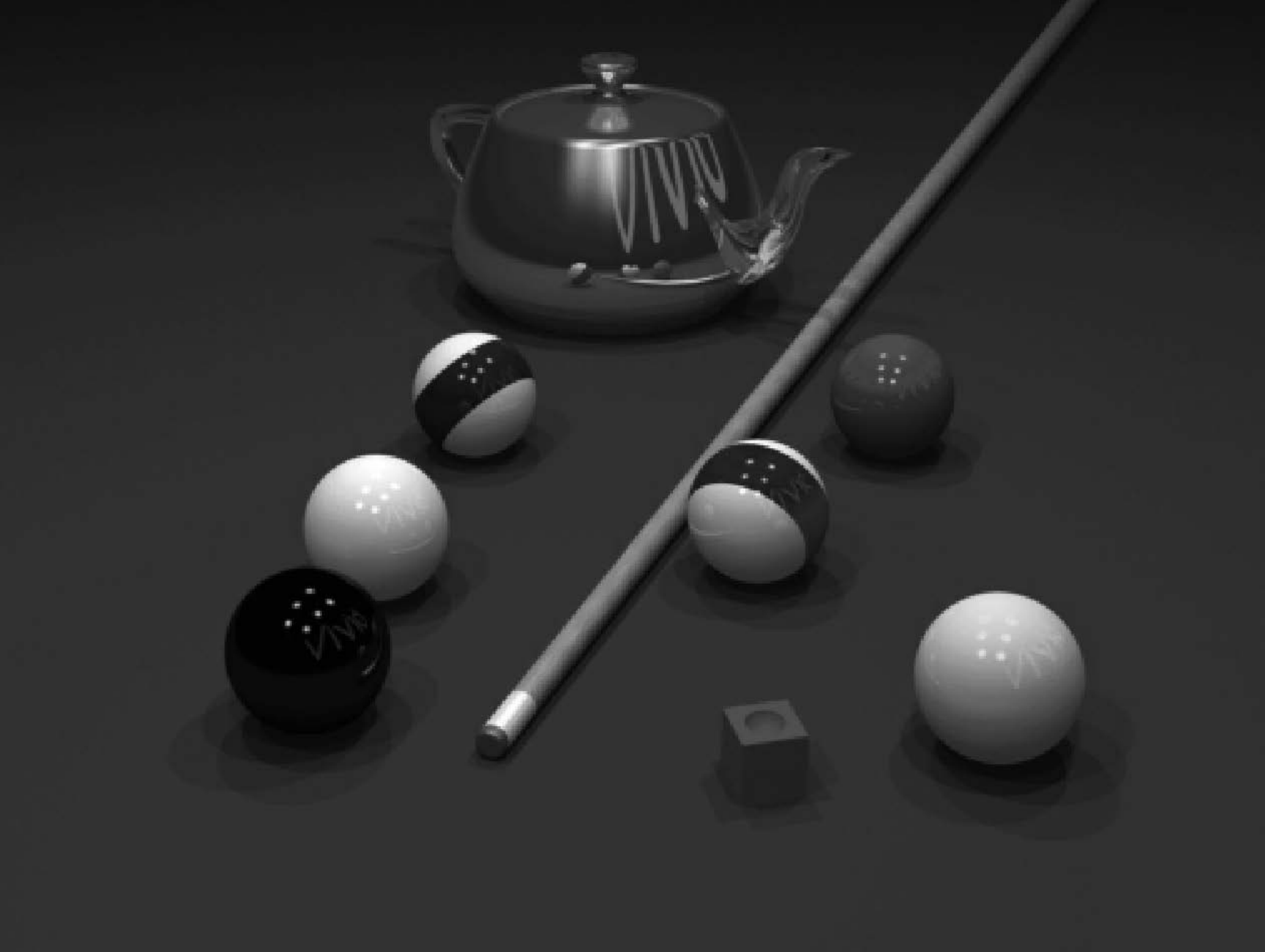}
		\caption{}
	\end{subfigure}
\end{figure}
\begin{figure}[H]\ContinuedFloat
	\begin{subfigure}[b]{0.49\textwidth}
		\includegraphics[width=\textwidth]{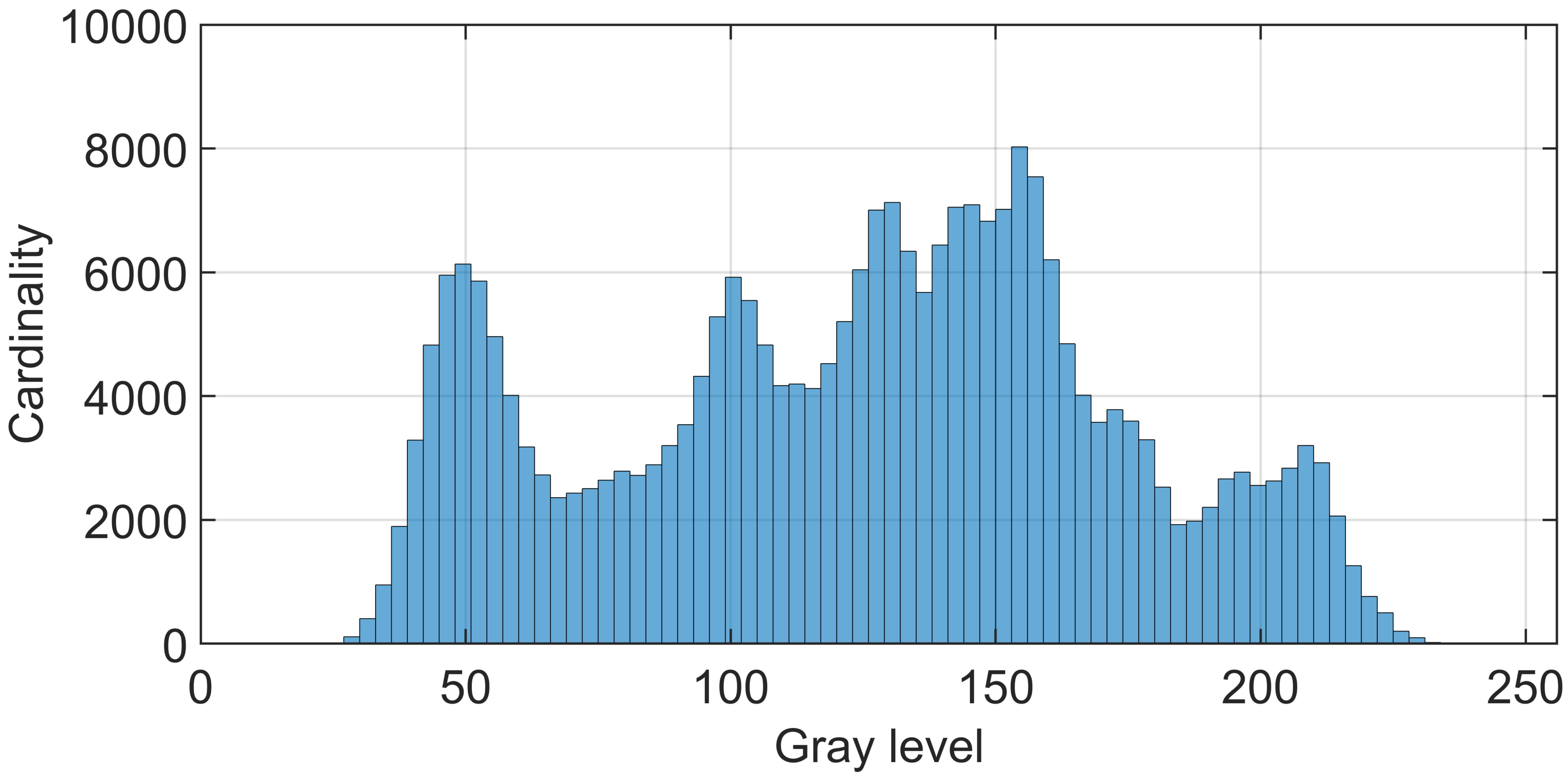}
		\caption{}
	\end{subfigure}
	\hfill
	\begin{subfigure}[b]{0.49\textwidth}
		\includegraphics[width=\textwidth]{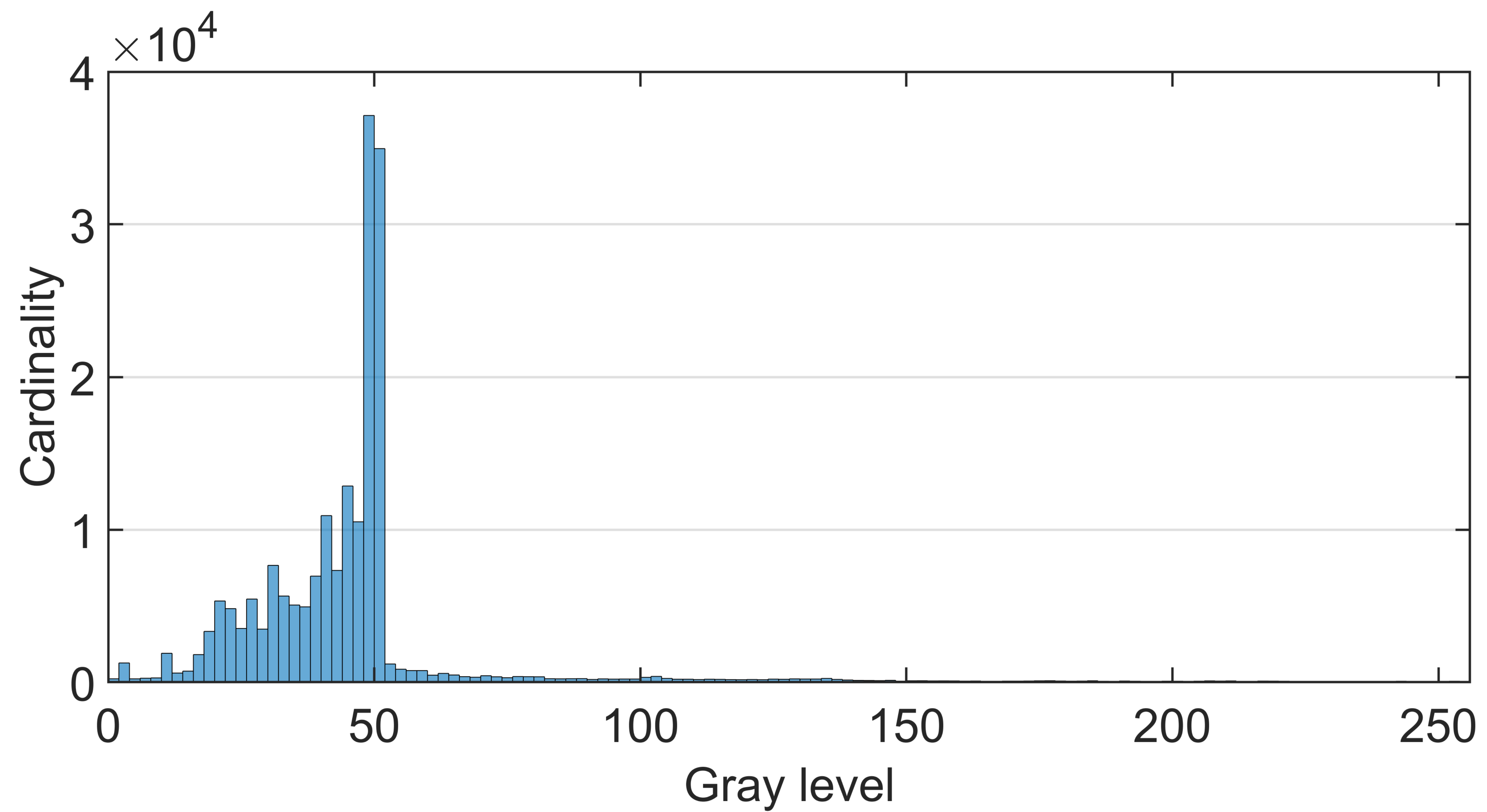}
		\caption{}
	\end{subfigure}
	\hfill
	\hspace*{2em}\begin{subfigure}[b]{0.33\textwidth}
		\includegraphics[width=\textwidth]{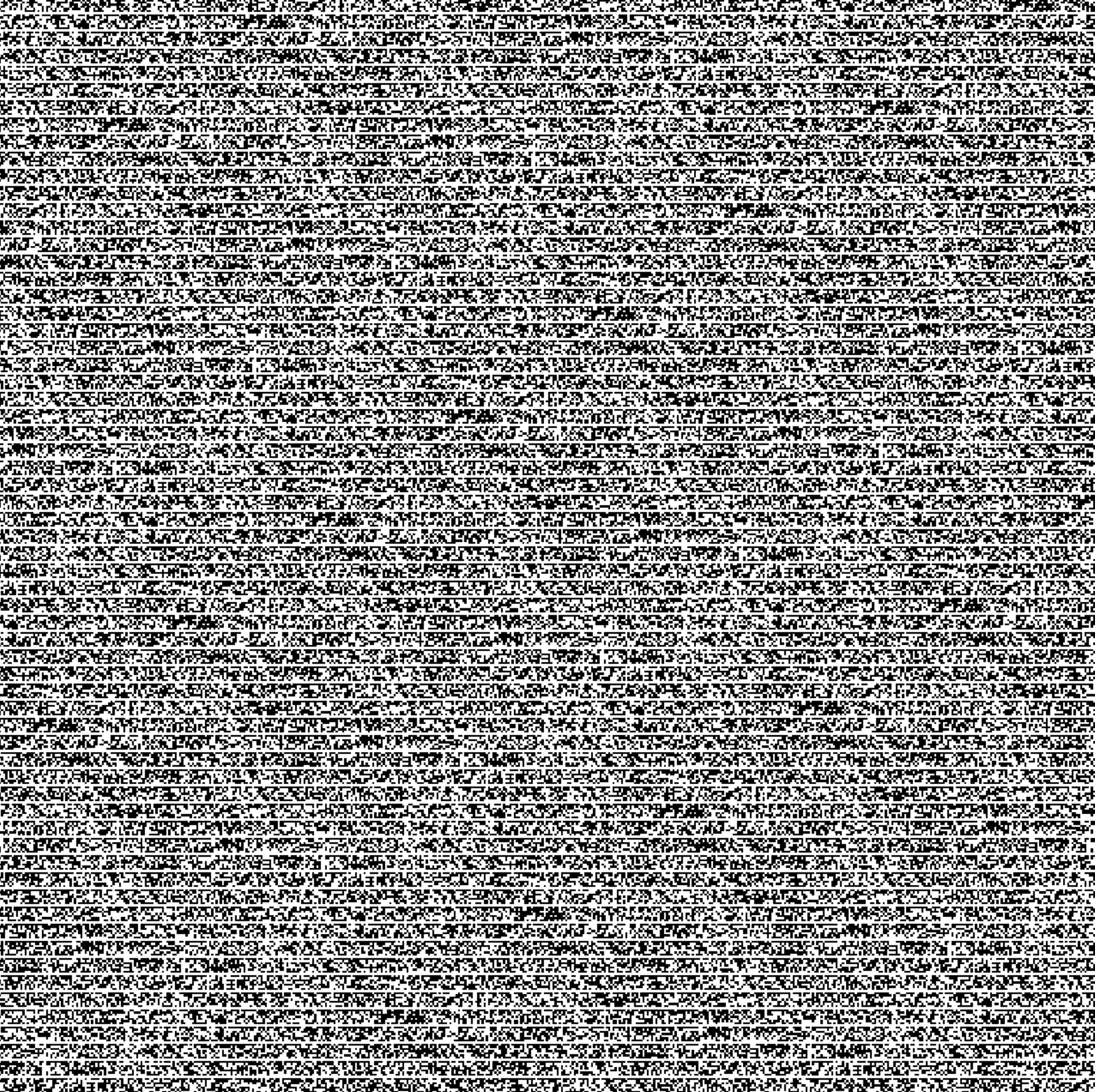}
		\caption{}
	\end{subfigure}
	\hfill
	\begin{subfigure}[b]{0.45\textwidth}
		\includegraphics[width=\textwidth]{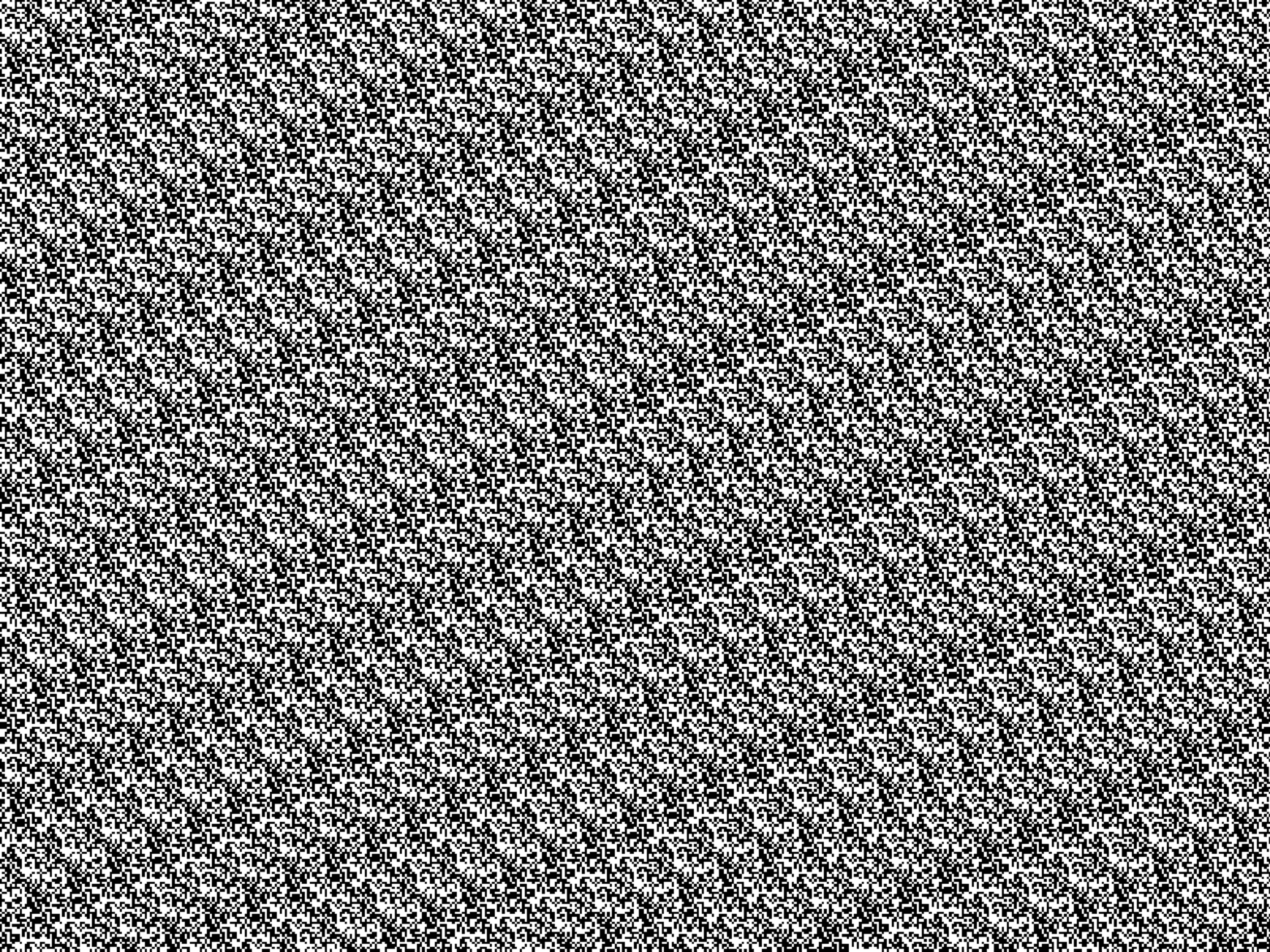}
		\caption{}
	\end{subfigure}
	\hfill
	\begin{subfigure}[b]{0.49\textwidth}
		\includegraphics[width=\textwidth]{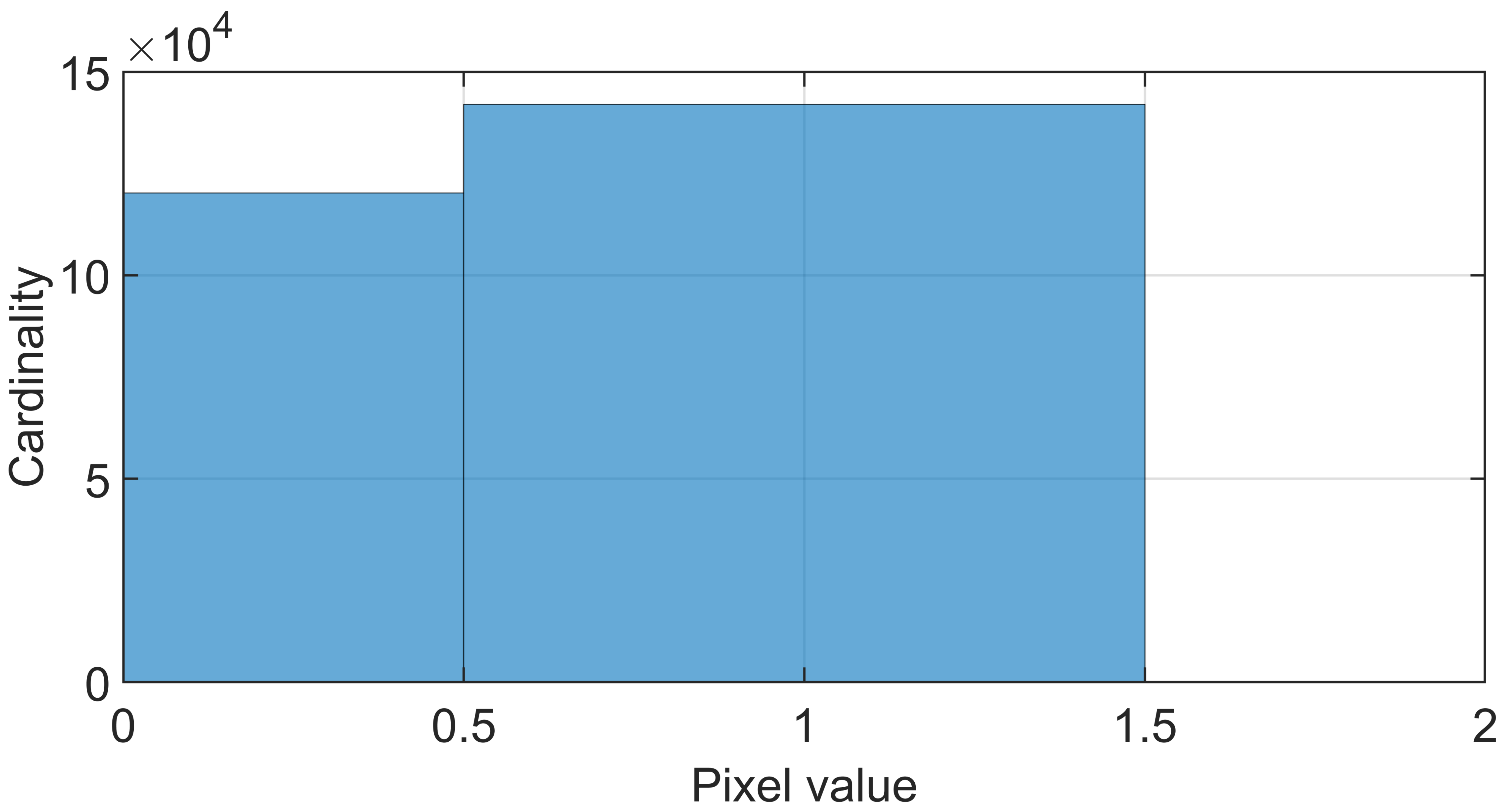}
		\caption{}
	\end{subfigure}
	\hfill
	\begin{subfigure}[b]{0.49\textwidth}
		\includegraphics[width=\textwidth]{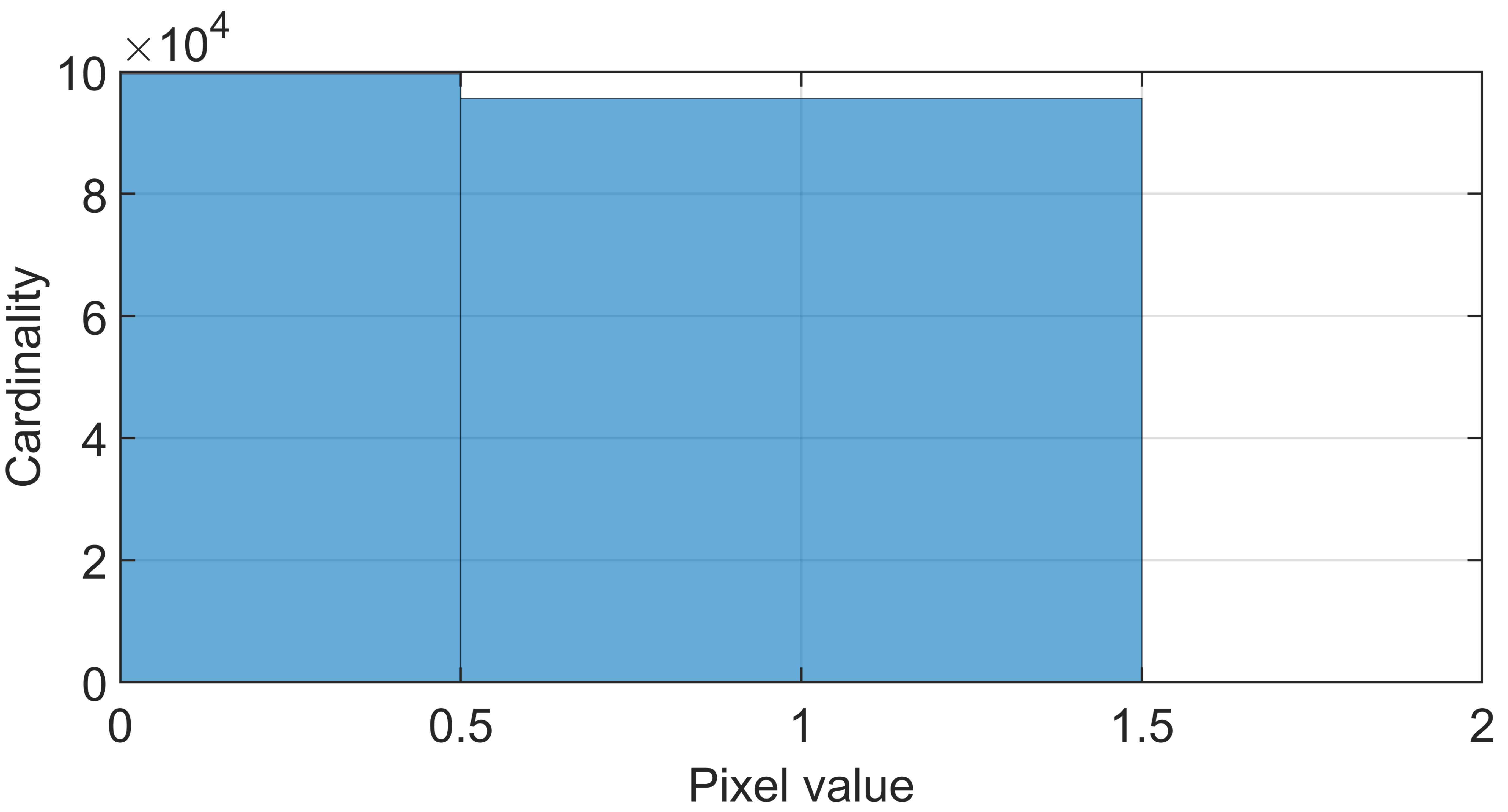}
		\caption{}
	\end{subfigure}
	\caption{(ab) Original images, where entropies are 7.4455 and 5.2720 respectively. (cd) Histogram of the original images. (ef) Results of applying cryptography algorithm to images, where PSNR (reference image is the original image) values are 5.6896 dB, 13.6566 dB, respectively, and entropies are 0.9950 and 0.9997, respectively. (gh) Histogram of the encrypted images.}\label{fig:cryotography}
\end{figure}

The Least Significant Bit (LSB) hiding is the most common methodology to implement steganography in which the LSBs of some or all pixels of the cover image is manipulated to embed the message. In this study, we propose a dynamic $k$-LSB approach in which $k$ least significant bits of Cover image are substituted with the cipher. Selecting $k$ for each pixel is done using the chaotic map constructed by CNN. This approach has two advantages over state-of-the-art ones. First, the entropy of stego-image shows the least change. Second, dynamic manipulating LSB of an image, itself, produces an image with the least amount of visual changes. Algorithm \ref{alg:stego} provides a detailed description of the dynamic $k$-LSB steganography method.

\begin{algorithm}[H]
	\caption{Dynamic $k$-LSB steganography algorithm.} \label{alg:stego}

	\hspace*{\algorithmicindent} \textbf{Input:} \\
	\hspace*{3em} $I_{c} \gets$ cover image. \\
	\hspace*{3em} $I_{m} \gets$ encrypted image, i.e., message. \\
	
	\hspace*{\algorithmicindent} \textbf{Output:} \\
	\hspace*{3em} $I_{stego} \gets$ results of steganography.\\
	
	\begin{algorithmic}[1]
		\State 	Set coefficients of Eq. \ref{equ:3} to $S_{11} = -1.65, S_{12} = 8.78, S_{32} = -13.25,$ and $ S_{33} = 1$.
		\For{$i = 1$ to $row_{I_{c}}$}
			\For{$j = 1$ to $col_{I_{c}}$}
				\State pixel = $I_{c}(i, j)$
				\State $b_{1}b_{2} \cdots b_{8} \gets$ \textbf{dec2bin}(pixel) \Comment{\textcolor{gray}{\% \textbf{dec2bin}() returns binary value of a decimal.}}
				\State $x_{1,0} = b_{6}, \quad x_{2,0} = b_{7}, \quad x_{3,0} = b_{8}$
				\State Let $\left\{(x_{1,i},x_{2,i},x_{3,i})\right\}_{i=1}^{N_{0}}$ is the CNN solution set.
				\State $\gamma = \mathbf{abs}(\lceil x_{1,N_{0}} + x_{2,N_{0}} + x_{3,N_{0}} \rceil) \bmod 255$ \Comment{\textcolor{gray}{\% \textbf{abs}() returns the absolute value.}}
				\State $d_{1}d_{2} \cdots d_{8} \gets \mathbf{dec2bin}(\gamma)$
				\State count = $(b_{5}\otimes d_{5}) + (b_{6}\otimes d_{6}) + (b_{7}\otimes d_{7}) + (b_{8}\otimes d_{8}$)
				\State pixel $\gets \mathbf{shift_{R}}$(pixel , count) \Comment{\textcolor{gray}{\% $\mathbf{shift_{R}}$() returns a bit-wise right shift of input.}}
				\State $I_{c}(i,j) \gets \mathbf{bitset}(I_{c}(i,j) , \mathrm{count}, 0)$ \Comment{\textcolor{gray}{\% $\mathbf{bitset}$() sets \textit{count}-LSB bits to 0.}}
				\State $I_{c}(i, j)$ = pixel
			\EndFor
		\EndFor
	\State Making $I_c$ and $I_m$ the same size.
	\State $I_{stego} \gets \mathbf{add}(I_{c}, I{m}) $ \Comment{\textcolor{gray}{\% $\mathbf{add}$() Add two images by preserving underlying data.}} \newline
	\Return{$I_{stego}$}
	\end{algorithmic}
\end{algorithm}

Result of applying the proposed dynamic $k$-LSB manipulation method is depicted in Figure \ref{fig:klsb}(b). It is evident from Figure \ref{fig:klsb}(c-f) that PSNR and entropy of an ordinary LSB method where $k =1, 2, 3, 4,$ differ from the proposed one. Indeed, this method, besides its visual quality, is more resident against stego-attacks.

\begin{figure}[H]
	\centering
	\begin{subfigure}[b]{0.3\textwidth}
		\includegraphics[width=\textwidth]{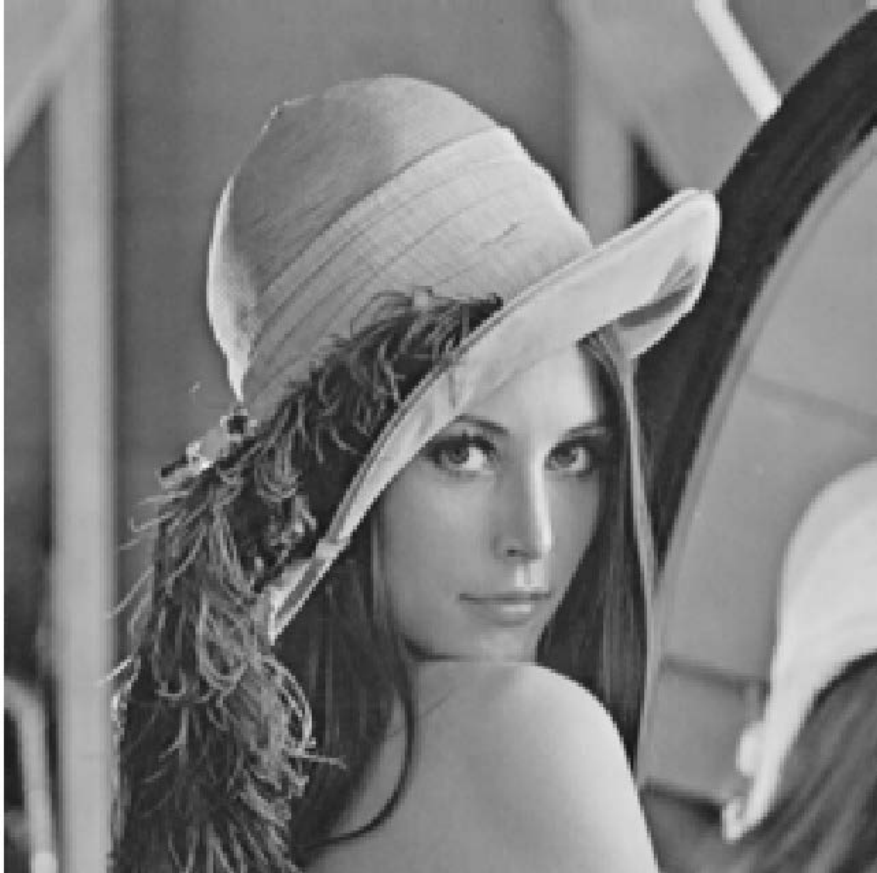}
		\caption{}
	\end{subfigure}
	\hfill
	\begin{subfigure}[b]{0.3\textwidth}
		\includegraphics[width=\textwidth]{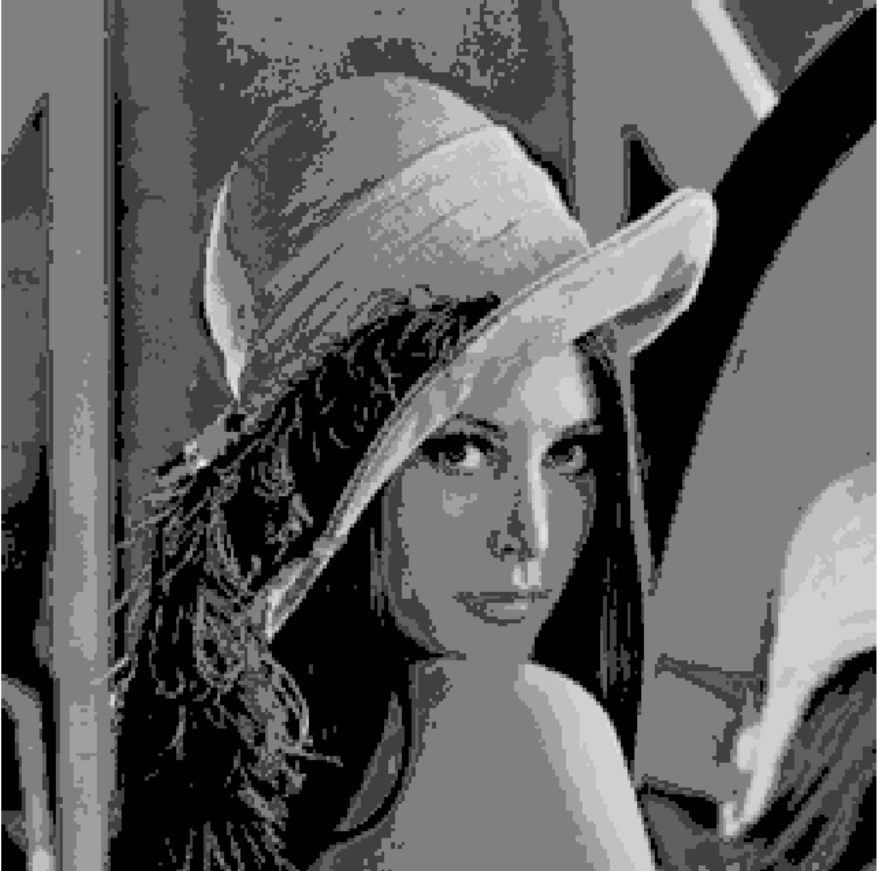}
		\caption{PSNR = 26.8997 dB}
	\end{subfigure}
	\hfill
	\begin{subfigure}[b]{0.3\textwidth}
		\includegraphics[width=\textwidth]{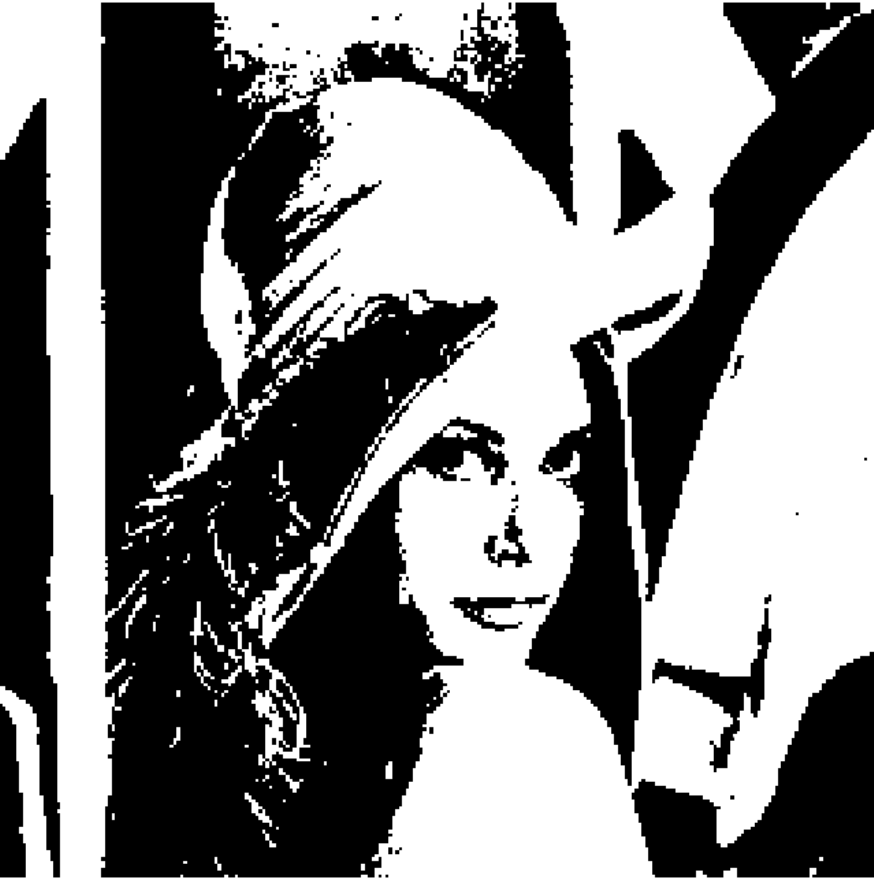}
		\caption{PSNR = 11.3441 dB}
	\end{subfigure}
\end{figure}
\begin{figure}[H]\ContinuedFloat
	\begin{subfigure}[b]{0.3\textwidth}
		\includegraphics[width=\textwidth]{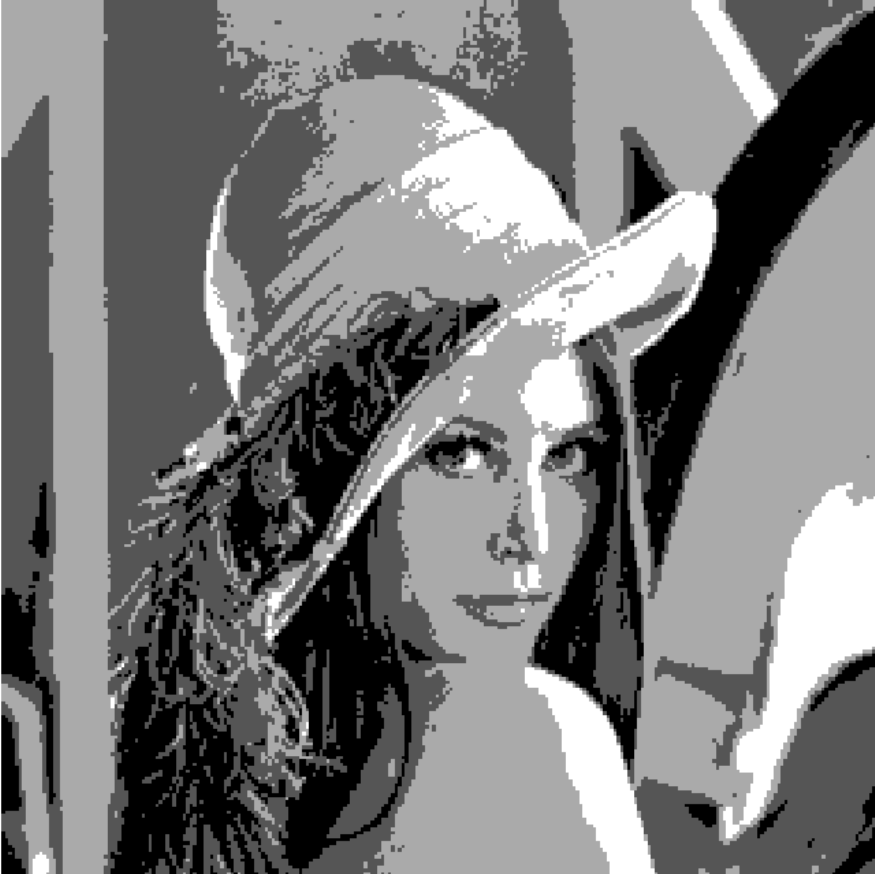}
		\caption{PSNR = 16.8538 dB}
	\end{subfigure}
	\hfill
	\begin{subfigure}[b]{0.3\textwidth}
		\includegraphics[width=\textwidth]{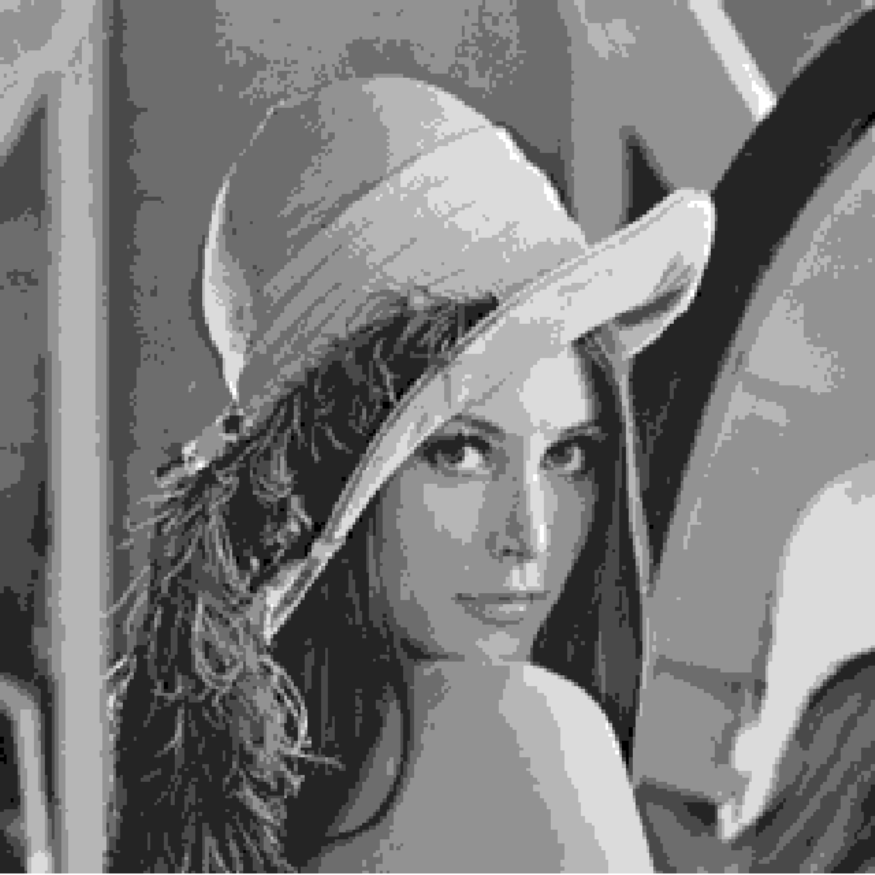}
		\caption{PSNR = 23.0104 dB}
	\end{subfigure}
	\hfill
	\begin{subfigure}[b]{0.3\textwidth}
		\includegraphics[width=\textwidth]{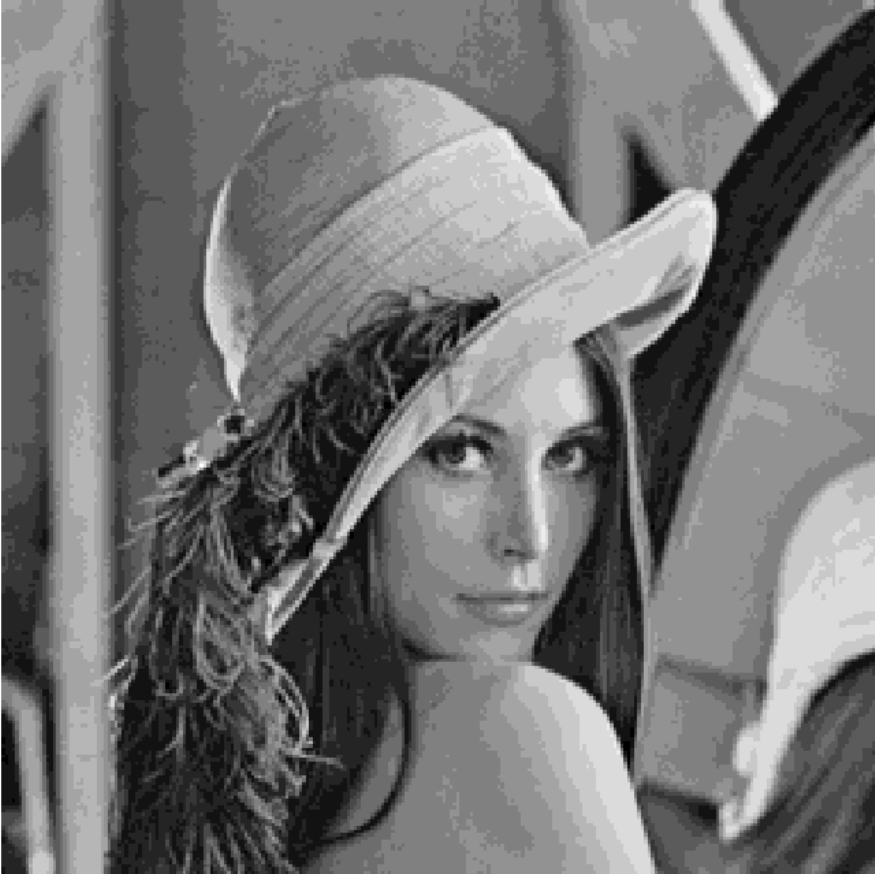}
		\caption{PSNR = 29.2365 dB}
	\end{subfigure}
	\caption{Results of applying LSB bit manipulation method on (a) original image (entropy = 7.4318) with (b) dynamic $k$ (entropy = 2.4170), (c) $k=1$ (entropy = 0.9997), (d) $k=2$ (entropy = 1.7899), (e) $k=3$ (entropy = 2.5079), and (f) $k=4$ (entropy = 3.4868).} \label{fig:klsb}
\end{figure}

\subsection{De-cryptology Algorithm}
The de-cryptology algorithm consists of two step. First, the message is extracted from the cover. Then, decryption phase is applied to appear the original message. The extraction phase consists of the following steps.
\begin{enumerate}
	\item Extract SIFT features and select the descriptor from feature sets that both sender and receiver agreed upon.
	\item Solve Eq. \ref{equ:3} of CNN to find out the solution set which also could reveal the $k$ for each pixel of stego-image.
	\item Utilize the obtained $k$ to extract $k$-LSBs in each pixel.
\end{enumerate}

Since the proposed cryptography algorithm is a symmetrical one, the decryption process is similar to the encryption process except for the extend key $T_{E}$ used to generate $w_{i}$. The original image can be regenerated by using Eq. \ref{equ:8}:

\begin{equation}
\label{equ:8}
m_{i}=\mathrm{xor}(e_{i}, w_{i}) \bmod 256
\end{equation}

In Table \ref{tbl:stego}, results of two criteria, i.e., PSNR and entropy, are tabulated where the proposed cryptology method is applied to a gray-scale image (see Figure \ref{fig:klsb}(a)) as the message and a $384 \times 512$, 8-bit color image (entropy = 7.3785) as the cover. In general, the distortion of an image is hard to be
detected by human visual system, if its PSNR is more than 30 dB \cite{chan2004hiding}. Indeed, the smaller the difference between the stego-image and cover image is, the greater the PSNR value is. PSNR of the proposed method is 54.94 dB which proves that dynamic $k$-LSB method is more robust than that of static ones are. Meanwhile, the entropy of the proposed method is 7.3928 which reveals that the obtained disorder of the proposed method is greater than that of static LSB can achieve. Indeed, we could reach to a trade-off between visual quality (see Figure \ref{fig:klsb}) and hiding capacity.

\begin{table}[H]
	\centering
	\renewcommand{\arraystretch}{1.3}
	\caption{Comparison of PSNR and Entropy values for different $k$ where the proposed cryptology method was applied to Figure \ref{fig:klsb}(a).}
	\label{tbl:stego}
	\resizebox{\textwidth}{!}{%
		\begin{tabu}{clccccc}
			\toprule
			\textbf{Image$^{\divideontimes}$} & \multicolumn{1}{l}{\textbf{Criteria}} & \textbf{Dynamic \textit{k}} & \textbf{\textit{k} = 1} & \textbf{\textit{k} = 2} & \textbf{\textit{k} = 3} & \textbf{\textit{k} = 4} \\ 
			\midrule
			\multirow{2}{*}[3.5mm]{\includegraphics[width=3cm]{Fig1a}} & \textbf{Entropy} & 7.3928 & 7.3914 & 7.3927 & 7.3926 & 7.3927 \bigstrut \\
			& \textbf{PSNR (dB)} & 54.94 & 53.81 & 54.68 & 54.92 & 54.84 \bigstrut \\
			\bottomrule
			\multicolumn{7}{l}{$^{\divideontimes}$ This image is used as the cover for the rest of experiments.}
		\end{tabu}%
	}
\end{table}

\section{Experiments}
We demonstrate the efficiency of the proposed cryptology scheme by conducting experiments in terms of (1) security analysis, (2) visual quality metrics, and (3) complexity analysis. In our tests, 25 benchmark images were used which are shown in Figure \ref{fig:data}.

\begin{figure}[H]
	\centering
	\begin{subfigure}[b]{0.19\textwidth}
		\includegraphics[width=\textwidth]{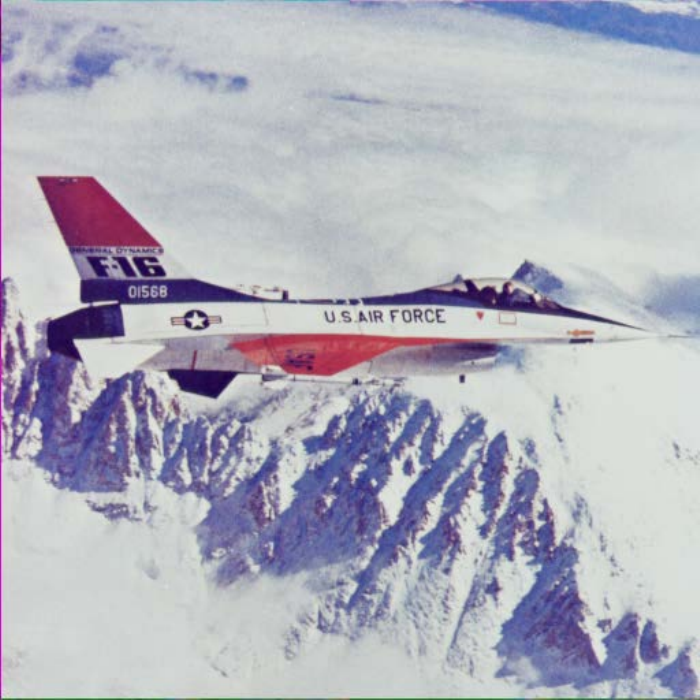}
		\caption*{Airplane}
	\end{subfigure}
	\hfill
	\begin{subfigure}[b]{0.19\textwidth}
		\includegraphics[width=\textwidth]{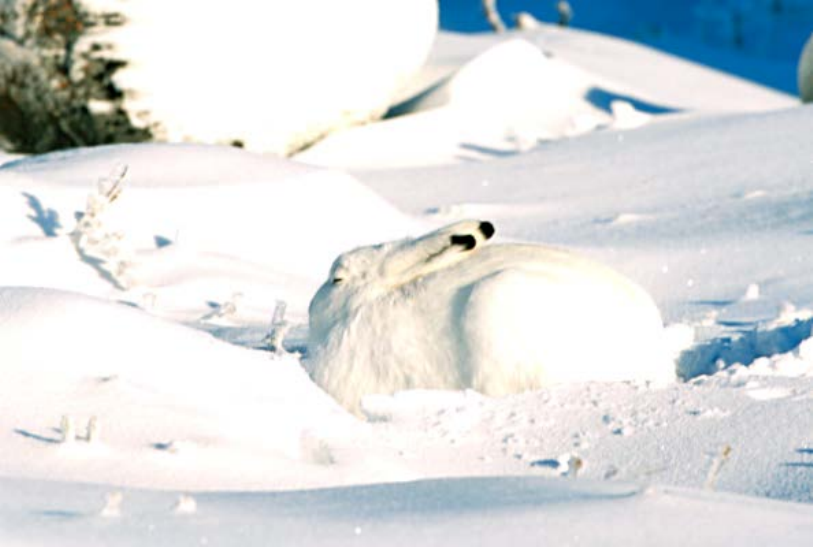}
		\caption*{Architecture}
	\end{subfigure}
	\hfill
	\begin{subfigure}[b]{0.19\textwidth}
		\includegraphics[width=\textwidth]{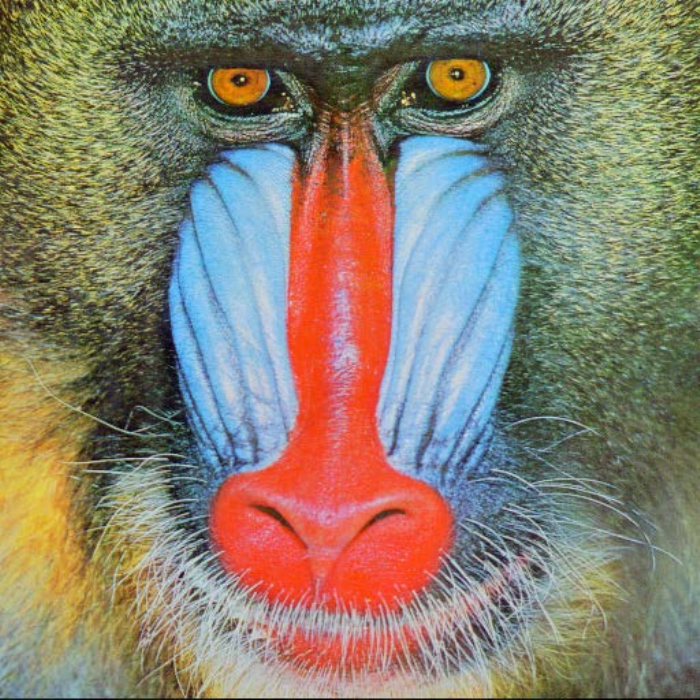}
		\caption*{Baboon}
	\end{subfigure}
	\hfill
	\begin{subfigure}[b]{0.19\textwidth}
		\includegraphics[width=\textwidth]{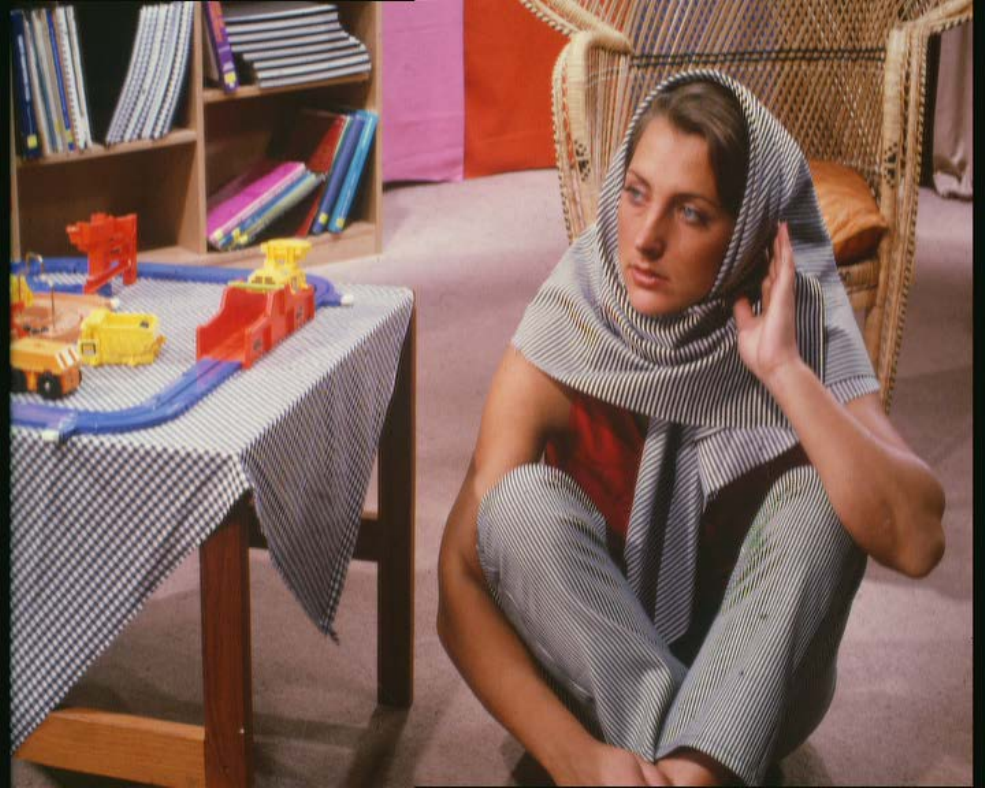}
		\caption*{Barbara}
	\end{subfigure}
	\hfill
	\begin{subfigure}[b]{0.19\textwidth}
		\includegraphics[width=\textwidth]{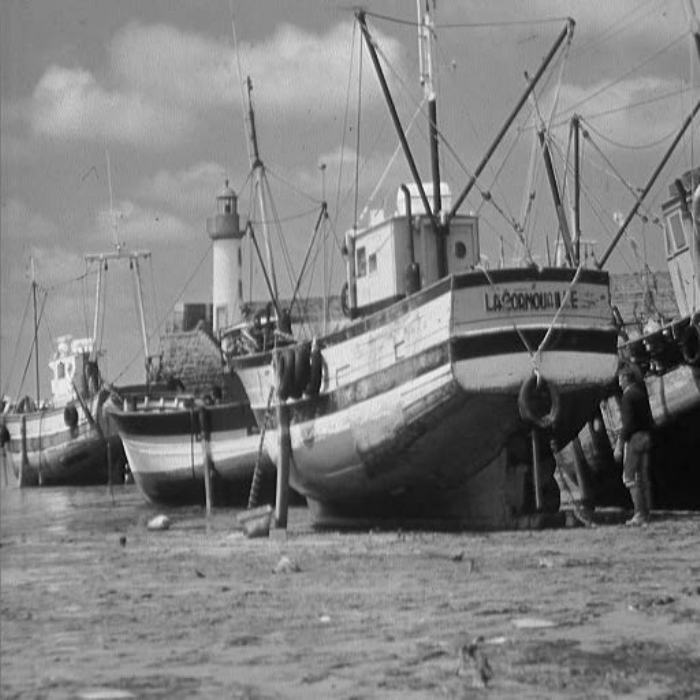}
		\caption*{Boat}
	\end{subfigure}
	\hfill
	\begin{subfigure}[b]{0.19\textwidth}
		\includegraphics[width=\textwidth]{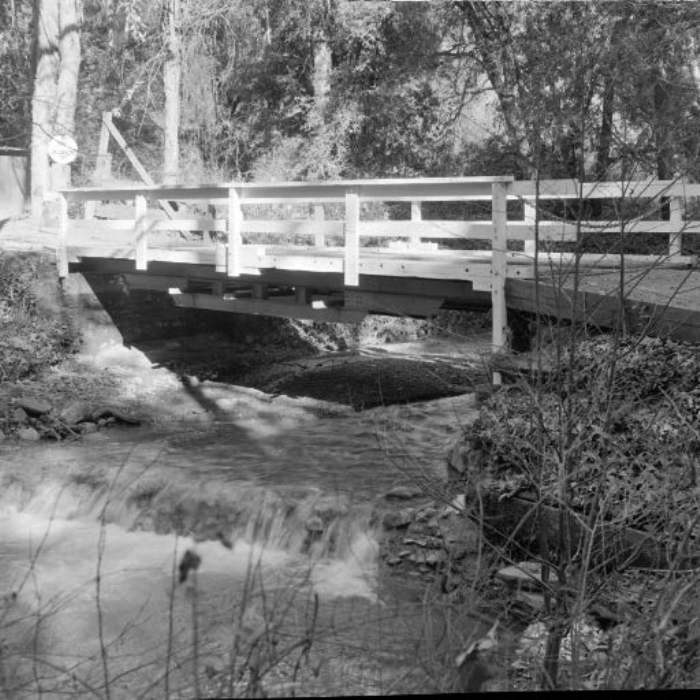}
		\caption*{Bridge}
	\end{subfigure}
	\hfill
	\begin{subfigure}[b]{0.19\textwidth}
		\includegraphics[width=\textwidth]{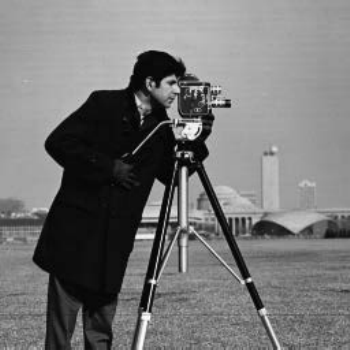}
		\caption*{Cameraman}
	\end{subfigure}
	\hfill
	\begin{subfigure}[b]{0.19\textwidth}
		\includegraphics[width=\textwidth]{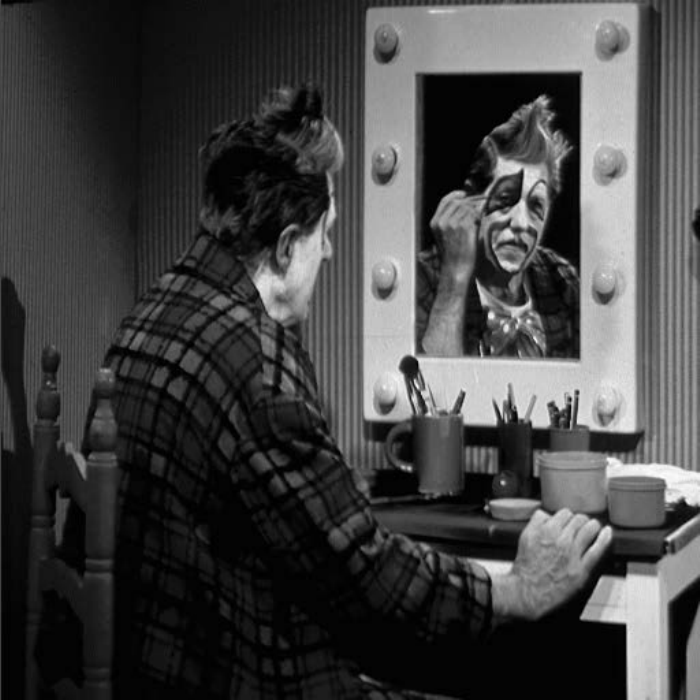}
		\caption*{Clown}
	\end{subfigure}
	\hfill
	\begin{subfigure}[b]{0.19\textwidth}
		\includegraphics[width=\textwidth]{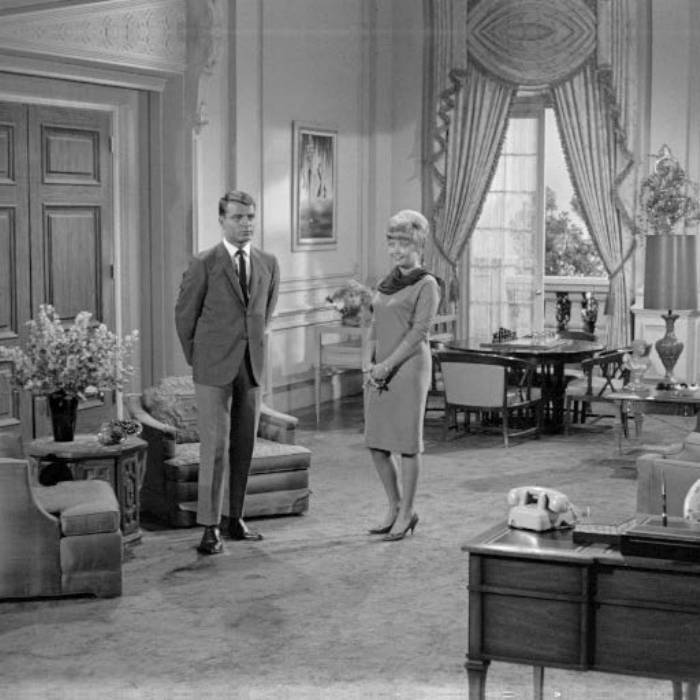}
		\caption*{Couple}
	\end{subfigure}
	\hfill
	\begin{subfigure}[b]{0.19\textwidth}
		\includegraphics[width=\textwidth]{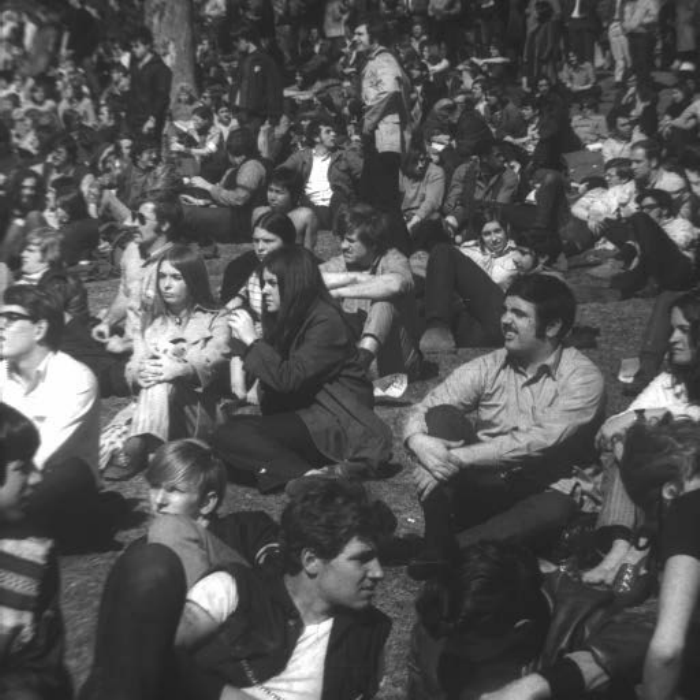}
		\caption*{Crowd}
	\end{subfigure}
	\hfill
	\begin{subfigure}[b]{0.19\textwidth}
		\includegraphics[width=\textwidth]{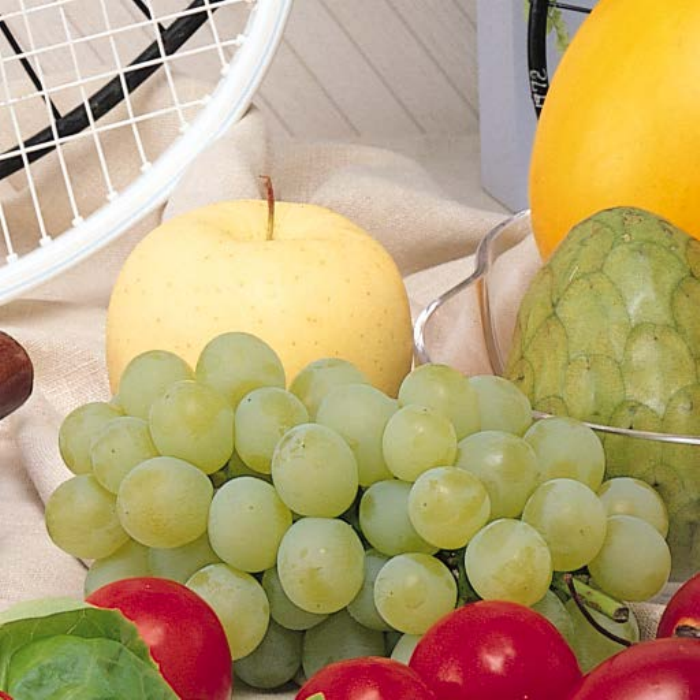}
		\caption*{Fruits}
	\end{subfigure}
	\hfill
	\begin{subfigure}[b]{0.19\textwidth}
		\includegraphics[width=\textwidth]{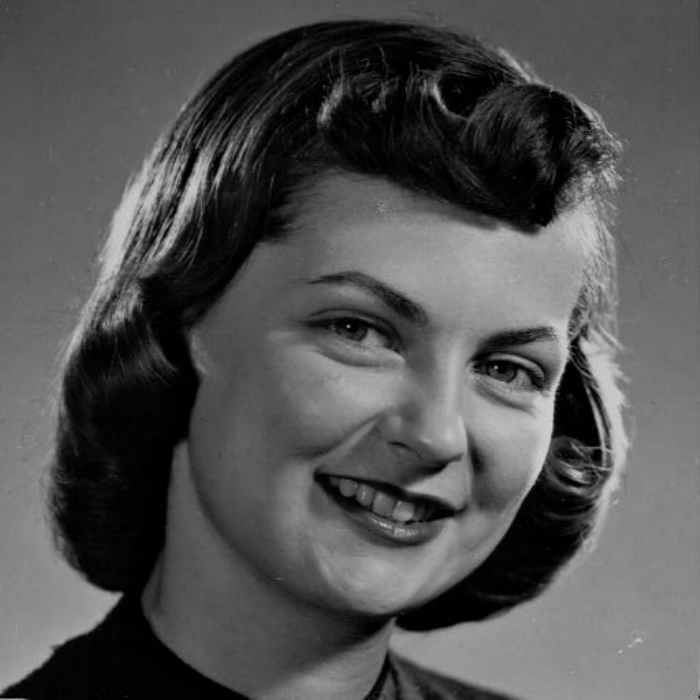}
		\caption*{Girl face}
	\end{subfigure}
	\hfill
	\begin{subfigure}[b]{0.19\textwidth}
		\includegraphics[width=\textwidth]{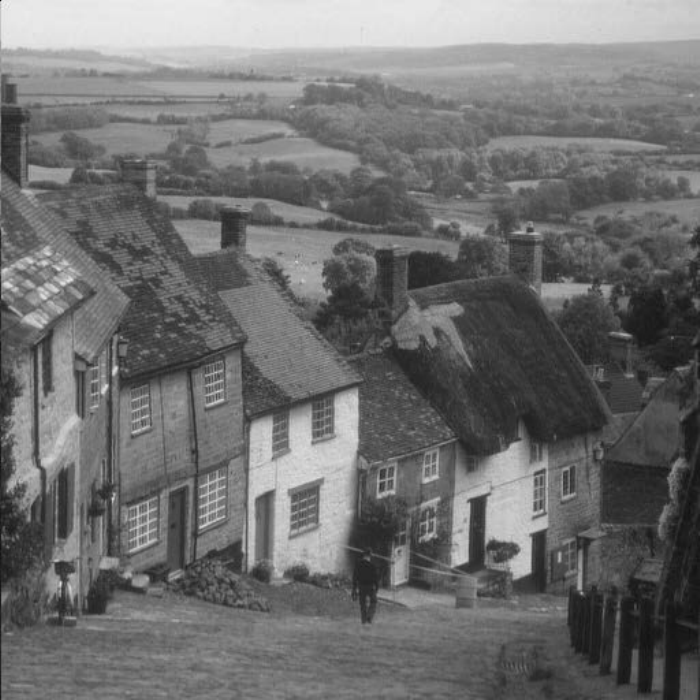}
		\caption*{Gold hill}
	\end{subfigure}
	\hfill
	\begin{subfigure}[b]{0.19\textwidth}
		\includegraphics[width=\textwidth]{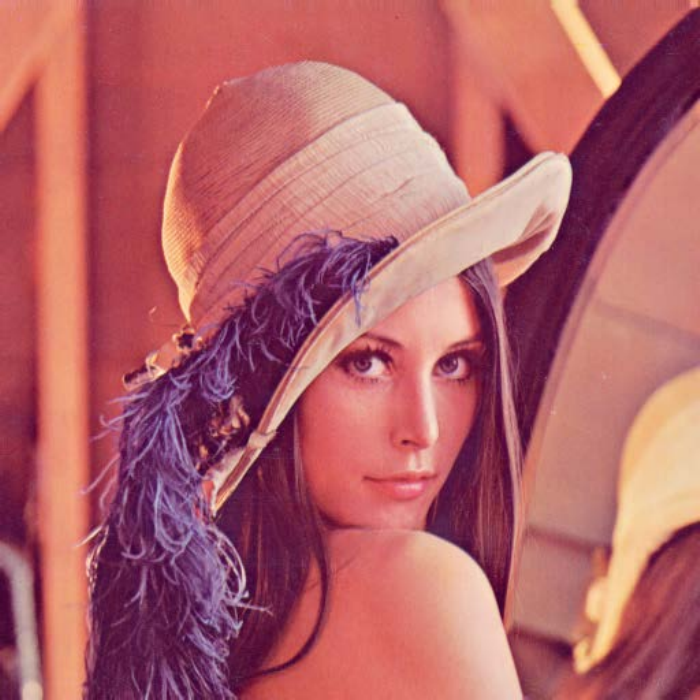}
		\caption*{Lena}
	\end{subfigure}
	\hfill
	\begin{subfigure}[b]{0.19\textwidth}
		\includegraphics[width=\textwidth]{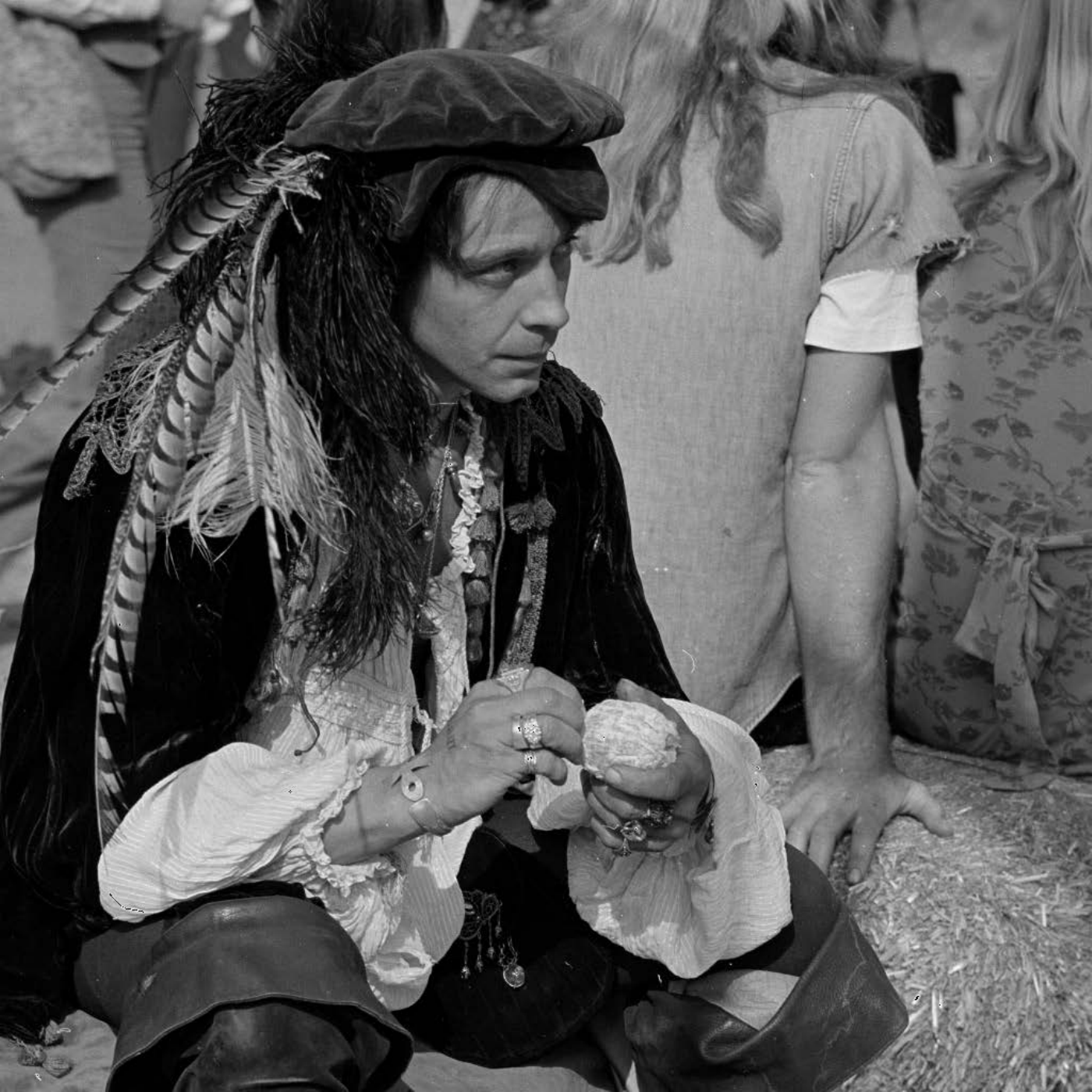}
		\caption*{Man}
	\end{subfigure}
\end{figure}
\begin{figure}[H]\ContinuedFloat
	\begin{subfigure}[b]{0.19\textwidth}
		\includegraphics[width=\textwidth]{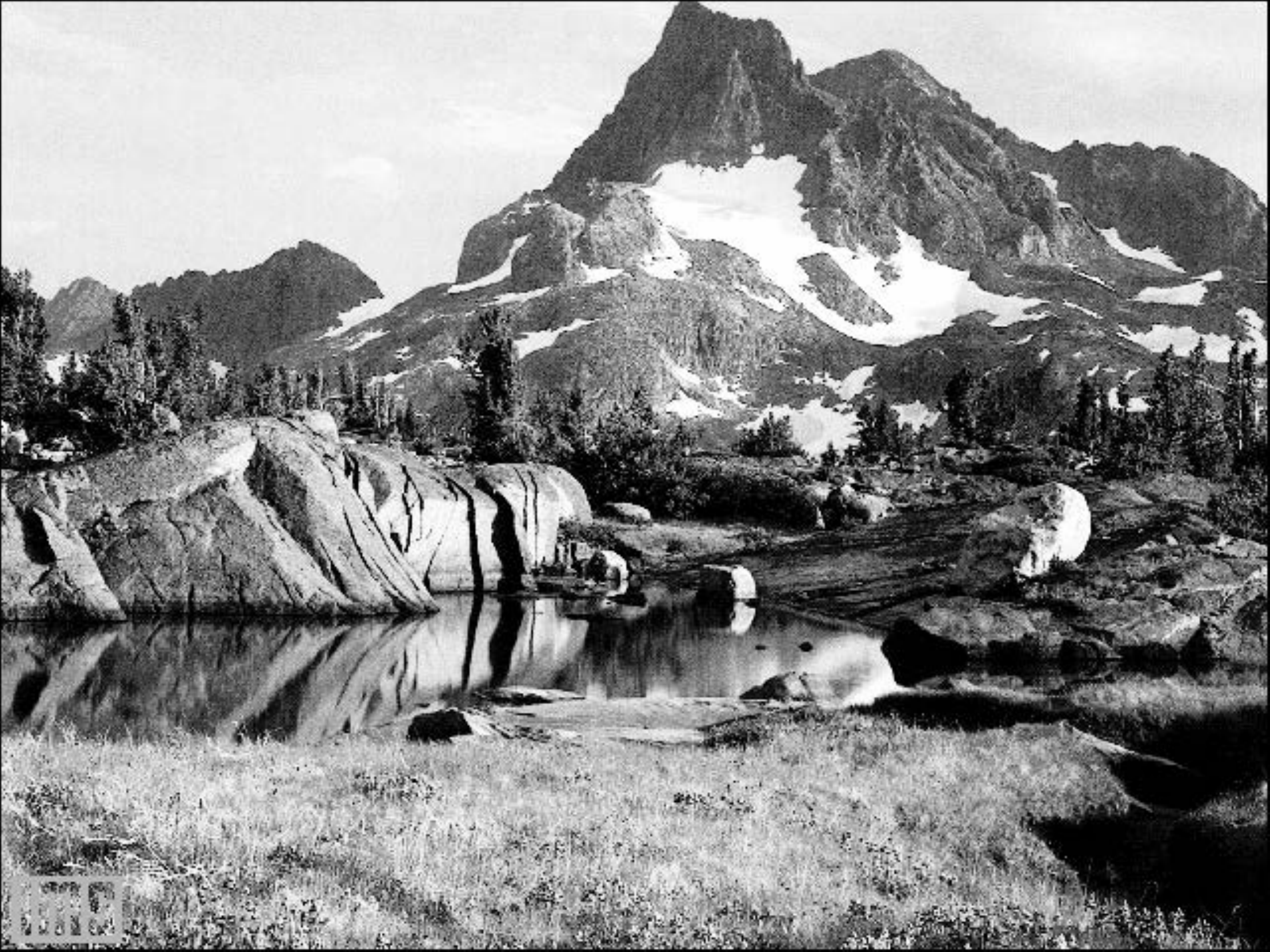}
		\caption*{Mountain}
	\end{subfigure}
	\hfill
	\begin{subfigure}[b]{0.19\textwidth}
		\includegraphics[width=\textwidth]{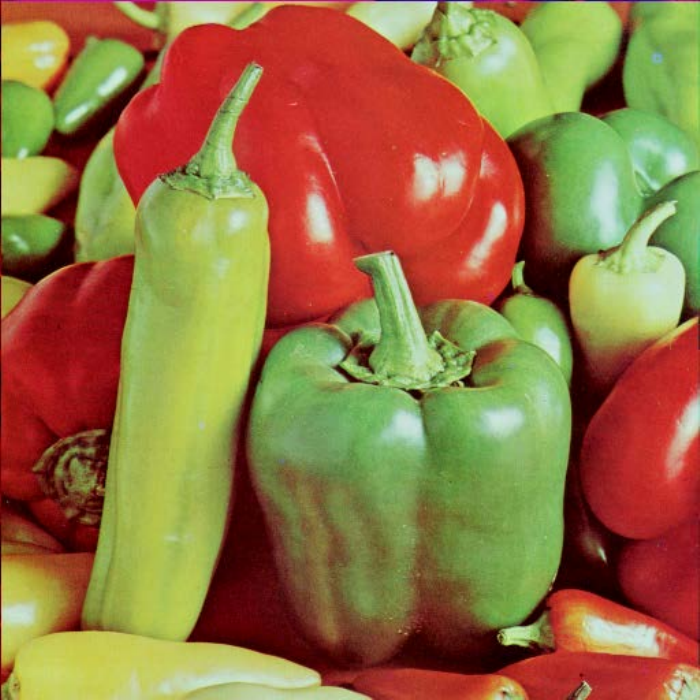}
		\caption*{Peppers}
	\end{subfigure}
	\hfill
	\begin{subfigure}[b]{0.19\textwidth}
		\includegraphics[width=\textwidth]{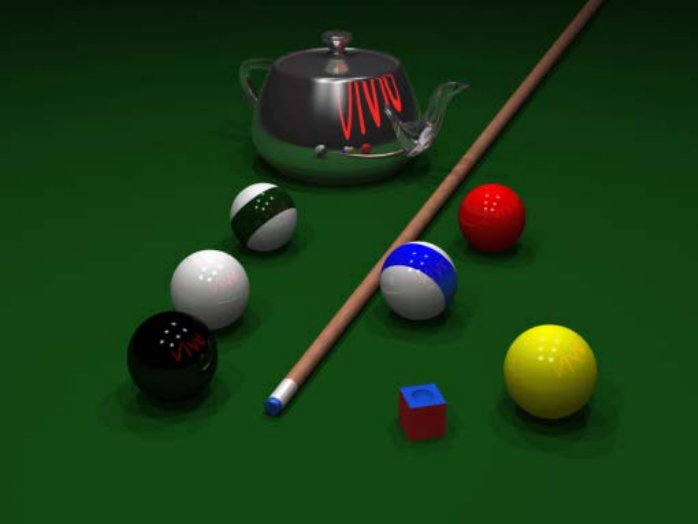}
		\caption*{Pool}
	\end{subfigure}
	\hfill
	\begin{subfigure}[b]{0.19\textwidth}
		\includegraphics[width=\textwidth]{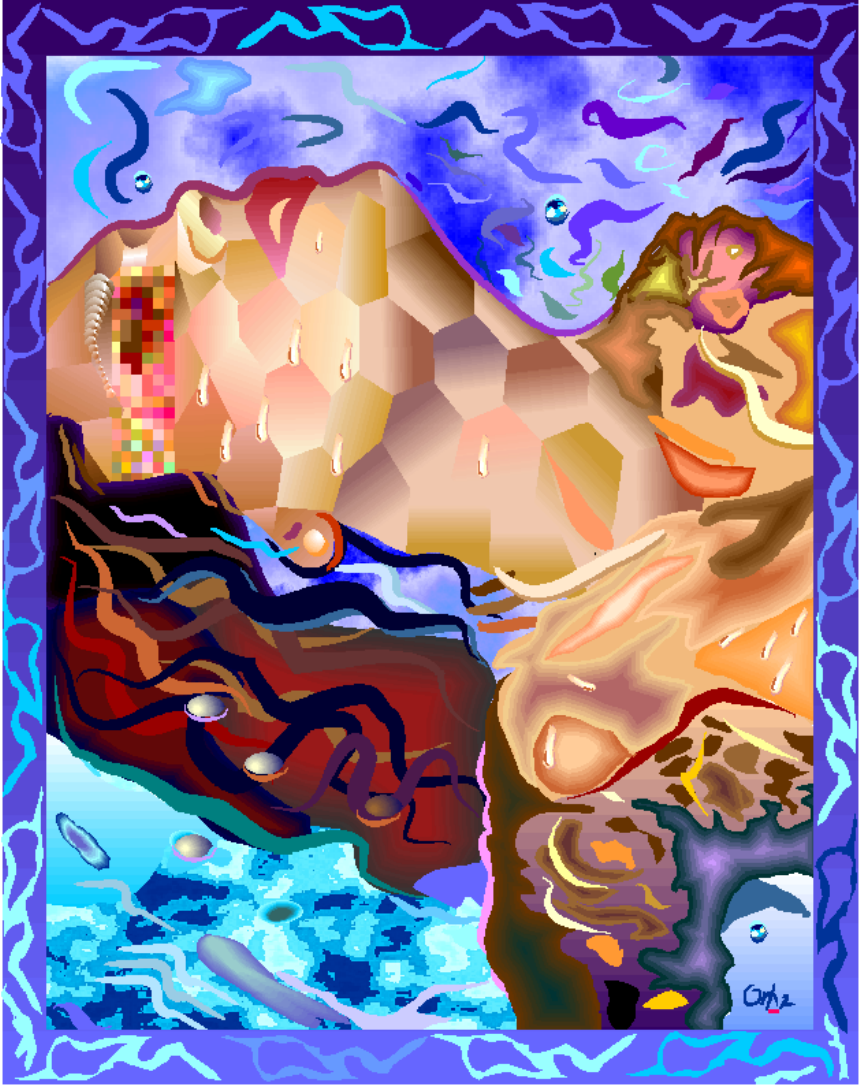}
		\caption*{Serrano}
	\end{subfigure}
	\hfill
	\begin{subfigure}[b]{0.19\textwidth}
		\includegraphics[width=\textwidth]{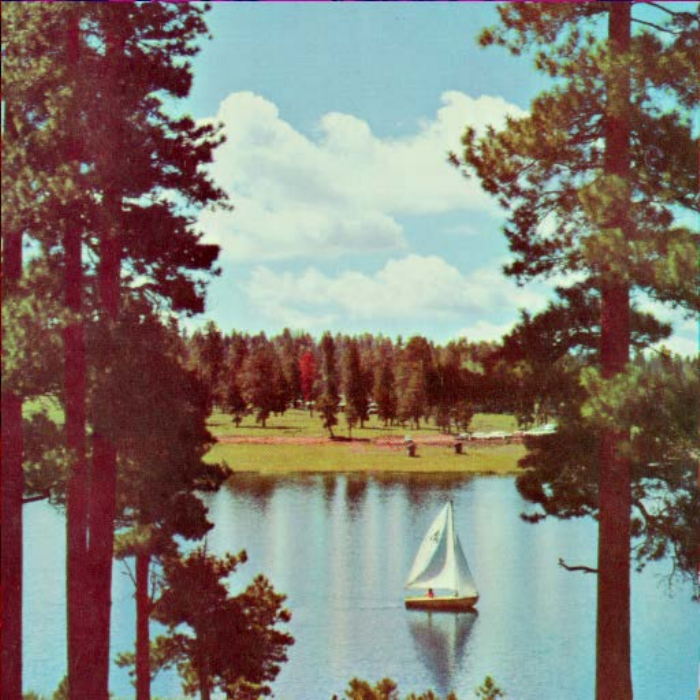}
		\caption*{Sailboat}
	\end{subfigure}
	\hfill
	\begin{subfigure}[b]{0.19\textwidth}
		\includegraphics[width=\textwidth]{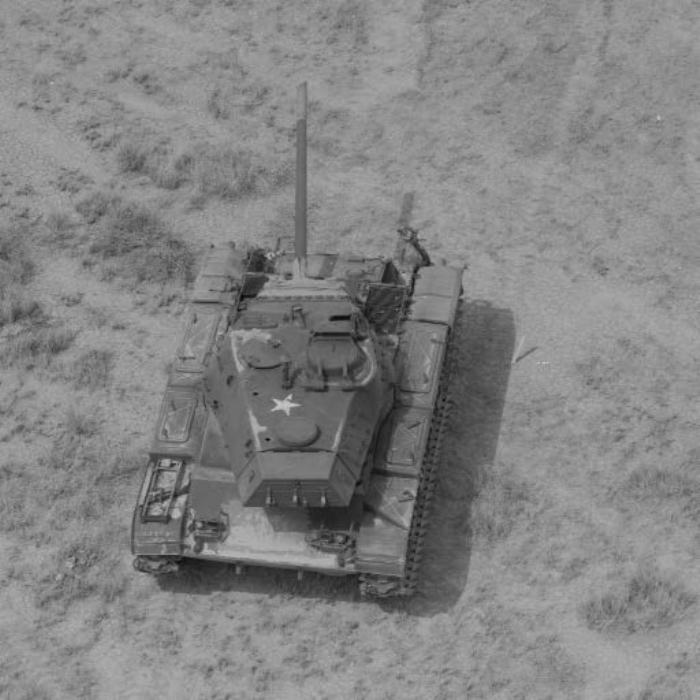}
		\caption*{Tank 1}
	\end{subfigure}
	\hfill
	\begin{subfigure}[b]{0.19\textwidth}
		\includegraphics[width=\textwidth]{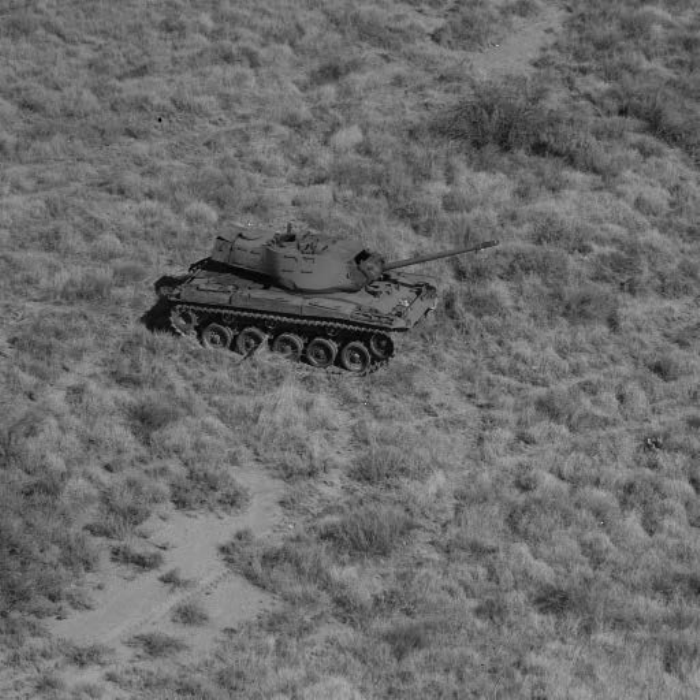}
		\caption*{Tank 2}
	\end{subfigure}
	\hfill
	\begin{subfigure}[b]{0.19\textwidth}
		\includegraphics[width=\textwidth]{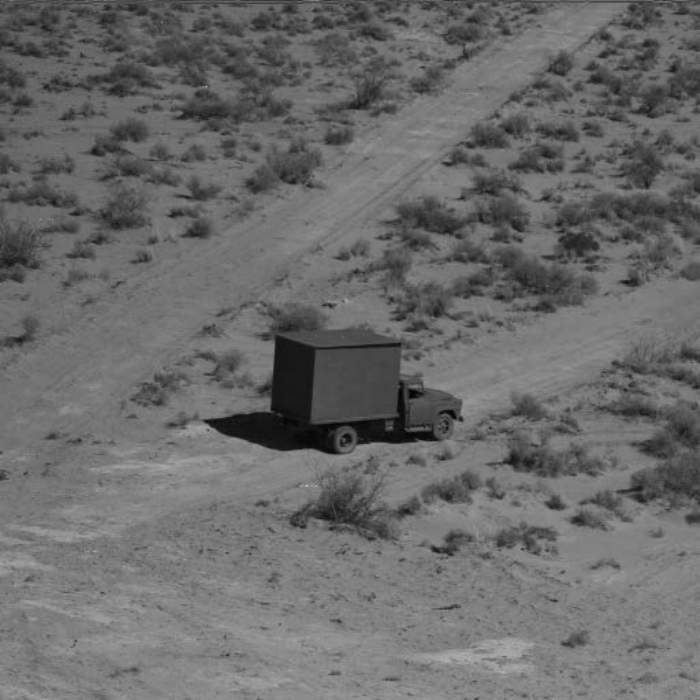}
		\caption*{Truck}
	\end{subfigure}
	\hfill
	\begin{subfigure}[b]{0.19\textwidth}
		\includegraphics[width=\textwidth]{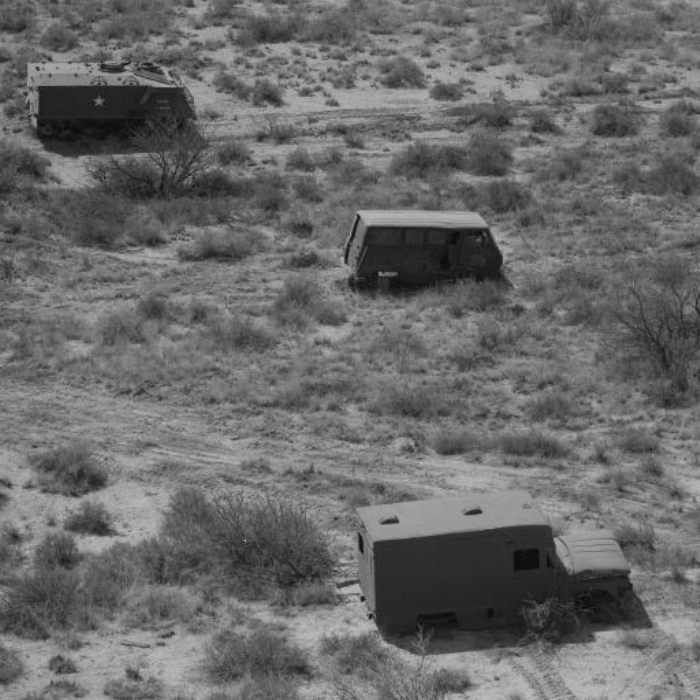}
		\caption*{Trucks}
	\end{subfigure}
	\hfill
	\begin{subfigure}[b]{0.19\textwidth}
		\includegraphics[width=\textwidth]{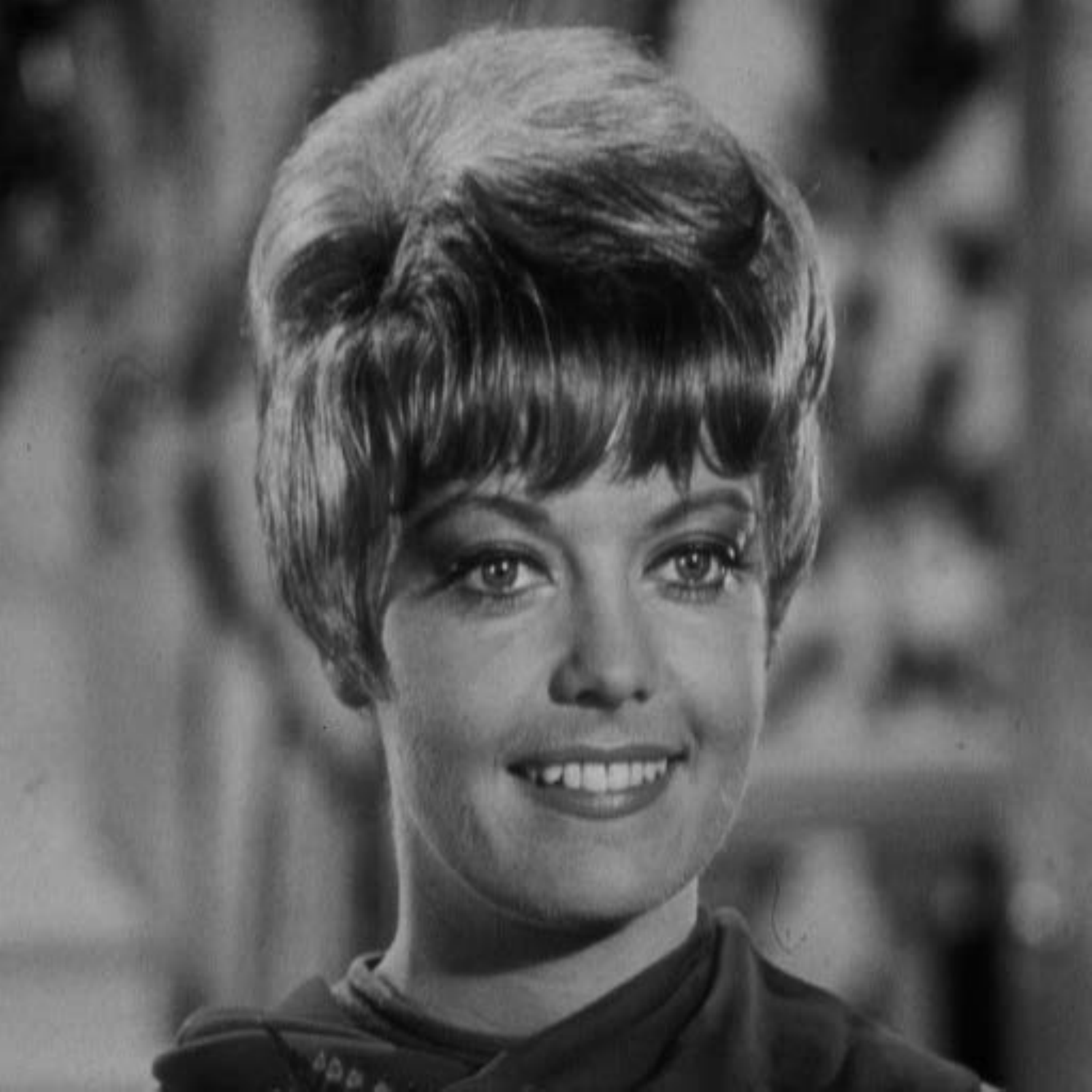}
		\caption*{Zelda}
	\end{subfigure}
	\caption{Image dataset.} \label{fig:data}
\end{figure}

\subsection{Security Analysis}
For the security analysis, we investigated four parameters which are secret key space, embedding ratio, correlation analysis, and occlusion attack.
It is obvious that a large key space can repel the brute-force attack. Since the algorithm has a 128-bit key, the key space size is $2^{128}\simeq 3.4 \times 10^{38}$. Considering control parameters of CNN as a part of the key, i.e., $(H_{1}, H_{2}, I, P, \lambda, h)$, $(x_{1,0}, x_{2,0}, x_{3,0})$, and $N_{0}$ (refer to Eq. \ref{equ:5} - \ref{equ:6}), the key space size will be even larger. If the computed precision is $10^{-14}$, the total secret key space is $2^{128} \times (10^{14})^{10} = \approx 2^{593} $ which is sufficient to make exhaustive attacks infeasible \cite{barker2012recommendation}. In Table \ref{tbl:secretkey}, we compared the proposed method and other algorithms in terms of secret key space which reveals that the proposed one is better than that of other reach. Even if $H_{1}, H_{2})$ excludes from the the calculation of the key space, as they are related to the plain Image and can be considered as constants, the key-space is approximately $2^{500}$.

\begin{table}[H]
	\centering
	\caption{Comparison of secret key space.}
	\label{tbl:secretkey}
	\begin{tabular}{lcccc}
		\hline
		\textbf{Encryption Algorithm} & \textbf{Proposed} & \textbf{Ref. \cite{bhatnagar2012chaos}} & \textbf{Ref. \cite{liu2012chaos}} & \textbf{Ref. \cite{zhou2013image}} \\ \hline
		Secret key space & $10^{150} \approx 2^{500}$ & $10^{84} \approx 2^{279}$ & $10^{72} \approx 2^{259}$ & $10^{68} \approx 2^{226}$ \\ \hline
	\end{tabular}
\end{table}

An essential criterion for understanding the fitness capability of steganography approach is embedding ratio, also known as payload, which can be calculated using Eq. \ref{equ:9}.

\begin{equation}
	\label{equ:9}
	ER = \frac{\mathbf{tb}(stego)-\mathbf{tb}(cover)}{\mathbf{tb}(cover)+\mathbf{tb}(message)}
\end{equation}
where $\mathbf{tb}(\bullet)$ counts the total bits of $\bullet$. Table \ref{tbl:er} tabulates the calculated ER values for dynamic $k$-LSB and static $k$-LSB where $k = 1, 2, 3, 4$.

\begin{table}[H]
	\centering
	\caption{Comparison of embedding ratio.}
	\label{tbl:er}
	\begin{tabular}{lccccc}
		\hline
		\textbf{Image} & \textbf{Dynamic $k$-LSB} & \textbf{$k=1$} & \textbf{$k=2$} & \textbf{$k=3$} & \textbf{$k=4$} \\ \hline
		Airplane & 50.46 & 16.66 & 33.33 & 50 & 66.66 \\
		Architecture & 52.67 & 15.10 & 30.21 & 45.31 & 60.42 \\
		Baboon & 34.12 & 16.66 & 33.33 & 50 & 66.66 \\
		Barbara & 50.87 & 26.36 & 52.73 & 79.10 & 105.46 \\
		Boat & 30.06 & 16.66 & 33.33 & 50 & 66.66 \\
		Bridge & 33.34 & 16.66 & 33.33 & 50 & 66.66 \\ 
		Cameraman & 7.06 & 4.16 & 8.33 & 12.50 & 16.66 \\
		Clown & 21.50 & 16.66 & 33.33 & 50 & 66.66 \\
		Couple & 31.36 & 16.66 & 33.33 & 50 & 66.66 \\
		Crowd & 29.93 & 16.66 & 33.33 & 50 & 66.66 \\
		Fruits & 41.21 & 16.66 & 33.33 & 50 & 66.66 \\
		Girl face & 24.56 & 16.66 & 33.33 & 50 & 66.66 \\ 
		Gold hill & 33.50 & 16.66 & 33.33 & 50 & 66.66 \\
		Lena & 33.85 & 16.66 & 33.33 & 50 & 66.66 \\
		Man & 101.48 & 66.66 & 133.33 & 200 & 266.66 \\
		Mountain & 42.15 & 16.66 & 33.33 & 50 & 66.66 \\
		Peppers & 34.00 & 12.41 & 24.83 & 37.25 & 49.67 \\
		Pool & 17.85 & 16.66 & 33.33 & 50 & 66.66 \\
		Serrano & 37.19 & 31.75 & 63.50 & 95.25 & 127.01 \\
		Sailboat & 68.49 & 16.66 & 33.33 & 50 & 66.66 \\
		Tank 1 & 29.21 & 16.66 & 33.33 & 50 & 66.66 \\
		Tank 2 & 33.67 & 16.66 & 33.33 & 50 & 66.66 \\
		Truck & 30.14 & 16.66 & 33.33 & 50 & 66.66 \\
		Trucks & 31.47 & 16.66 & 33.33 & 50 & 66.66 \\
		Zelda & 28.14 & 16.66 & 33.33 & 50 & 66.66 \\
		\hline
	\end{tabular}
\end{table}

It is obvious from tabulated results in Table \ref{tbl:er} that ER, where the static LSB approach is utilized, just depends on the size of message. Therefore, regardless of image content, an attacker can easily calculate the size of hidden message and the number of manipulated bits in the message and cover image. However, the proposed approach depends on the size and content of image which make it difficult for attacker to find out how many bits might alter. Besides, it is obvious from Figure \ref{fig:klsb} and Table \ref{tbl:er} that the proposed method establishes a balance between visual quality and resistance to possible cryptography attacks.

\begin{figure}[H]
	\centering
	\begin{subfigure}[c]{0.215\textwidth}
		\includegraphics[width=\textwidth]{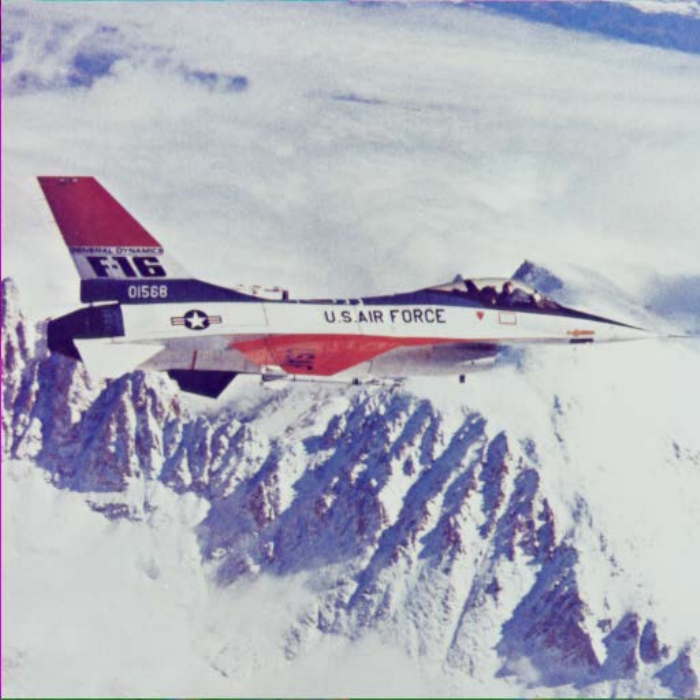}
		\caption{}
	\end{subfigure}
	\hfill
	\begin{subfigure}[c]{0.215\textwidth}
		\includegraphics[width=\textwidth]{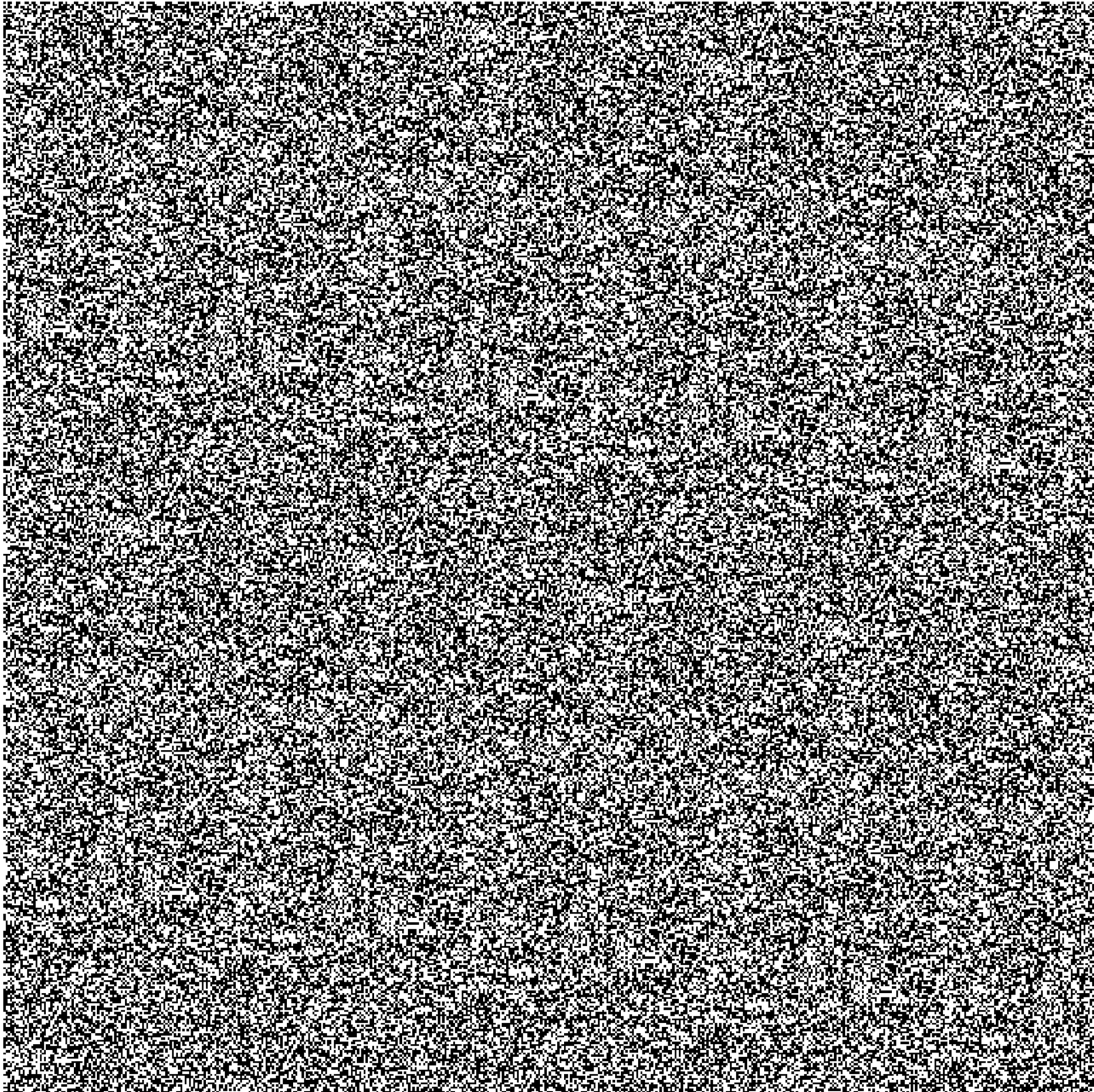}
		\caption{}
	\end{subfigure}
	\hfill
	\begin{subfigure}[c]{0.55\textwidth}
		\includegraphics[width=\textwidth]{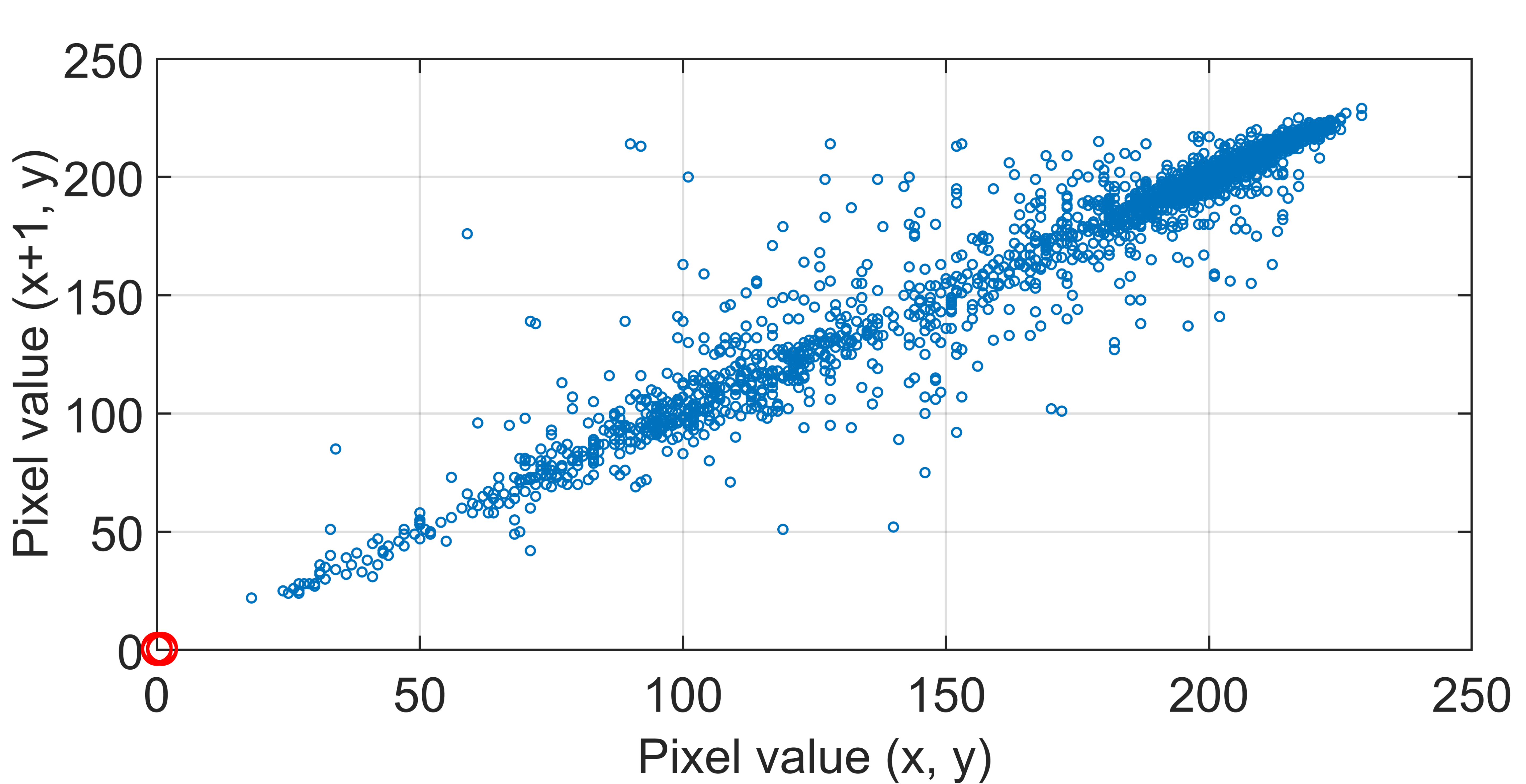}
		\caption{}
	\end{subfigure}
	\hfill
	\begin{subfigure}[c]{0.215\textwidth}
		\includegraphics[width=\textwidth]{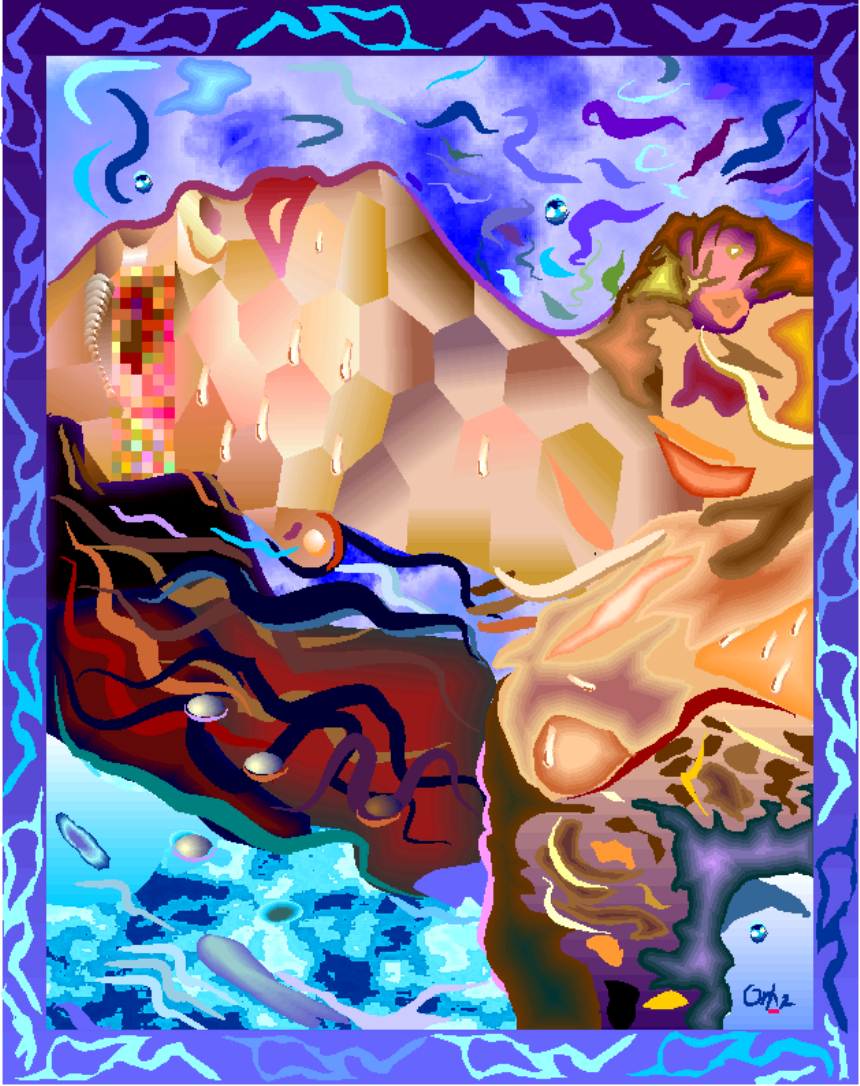}
		\caption{}
	\end{subfigure}
	\hfill
	\begin{subfigure}[c]{0.215\textwidth}
		\includegraphics[width=\textwidth]{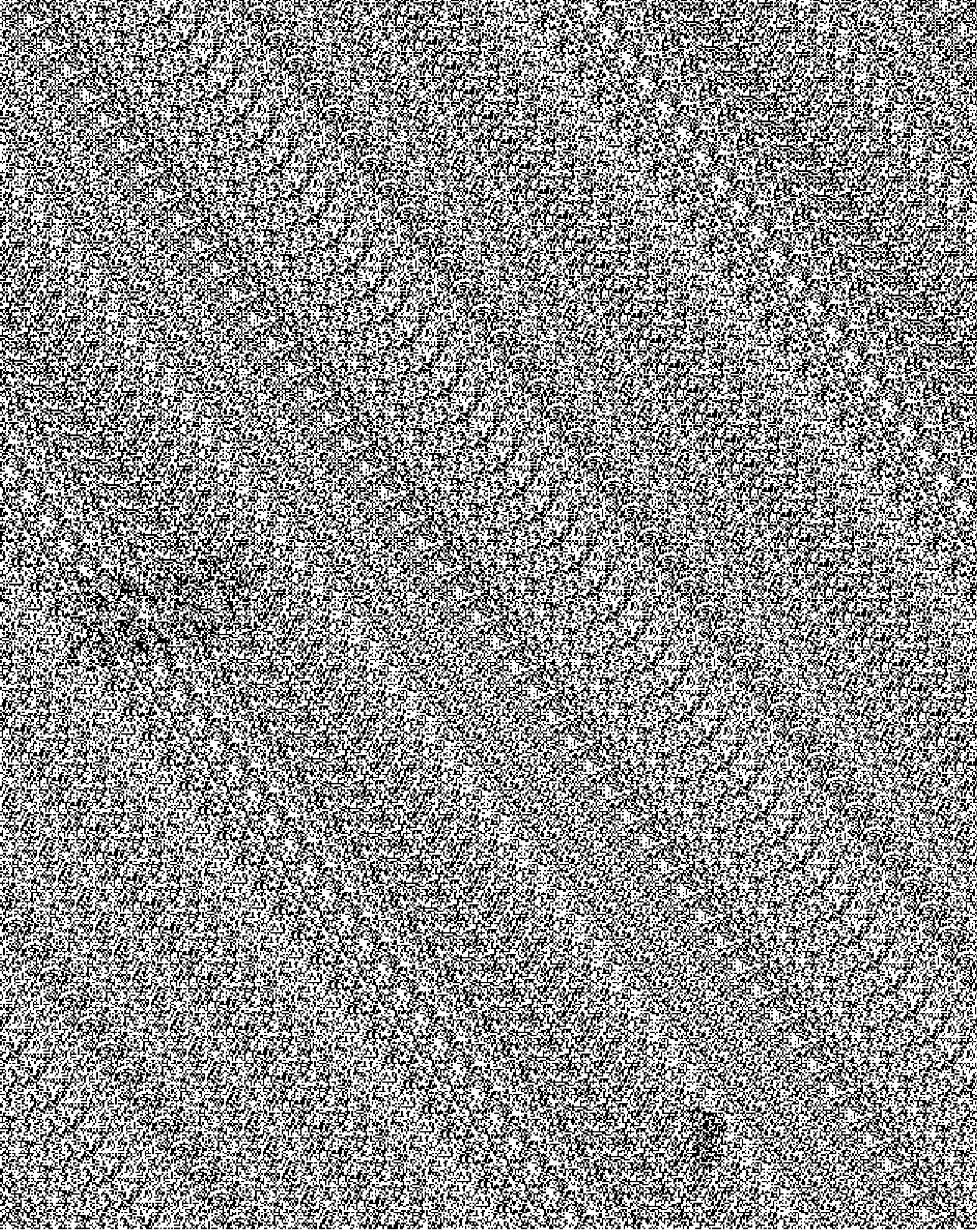}
		\caption{}
	\end{subfigure}
	\hfill
	\begin{subfigure}[c]{0.55\textwidth}
		\includegraphics[width=\textwidth]{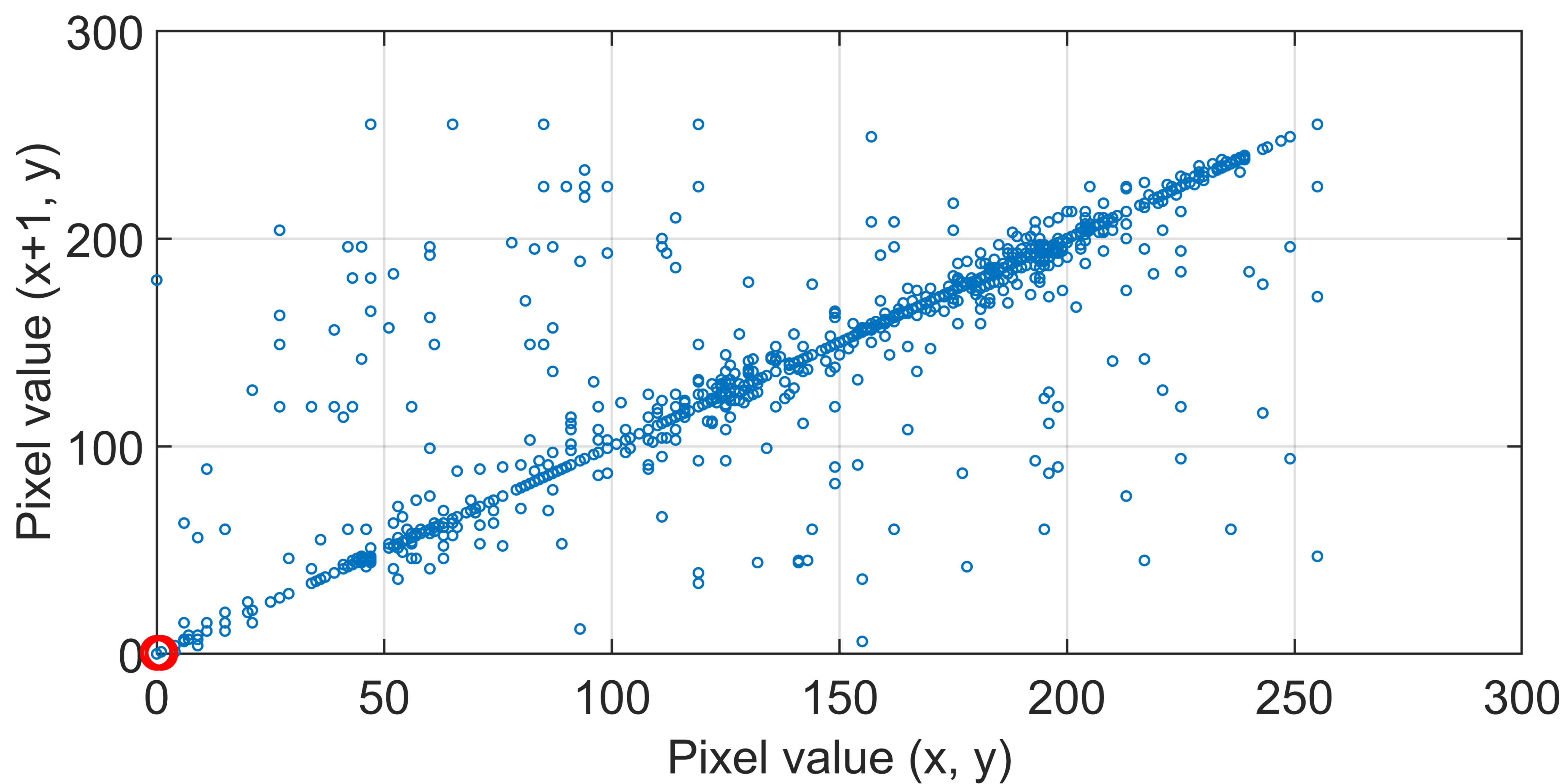}
		\caption{}
	\end{subfigure}
	\caption{Correlation analysis of two adjacent pixels for, (a)-(c) Airplane (d)-(f) Serrano. \textcolor{blue}{$\circ$} and \textcolor{red}{$\circ$} relate to the original and encrypted images, respectively.} \label{fig:correlation}
\end{figure}

It is proved that a cryptography algorithm must satisfy the 0-correlation condition \cite{seyedzadeh2012fast} while it must fulfill 1-correlation in terms of steganography. Correlation analysis is done by calculating coefficients using Eq.\ref{equ:10}.

\begin{equation}
	\label{equ:10}
	\rho(A,B)=\frac{\mathrm{cov}(A,B)}{\sigma_A\sigma_B}
\end{equation}
where $\mathrm{cov}(\cdot,\cdot)$ calculates the covariance of images $A$ and $B$. We analyze two adjacent pixels in the horizontal positions for original images (Figure \ref{fig:correlation}(a), (d)) and their encrypted images (Figure \ref{fig:correlation}(b),(e)), in which we can see that in encrypted images, the means of the correlation coefficients of two adjacent pixels are very close to 0.

Table \ref{tbl:correlation} shows the comparison of correlation coefficients on the encrypted and stegano images with the proposed method and other algorithms.

Occlusion attack tests the robustness of a cryptology scheme to evaluate the ability of recovering original image when the encrypted or stegano image lost data. This situation might happen when data is transmitted over a noisy network or when the attacker tries to interfere with the disruption. The PSNR is used to measure the quality of decrypted image in this attack and is calculated using Eq. \ref{equ:11}.

\begin{equation}
\label{equ:11}
\mathrm{PSNR}=10 \log_{10}\frac{255^{2}}{\frac{1}{iw \times ih} \sum_{i=1}^{iw} \sum_{j=1}^{ih}(o(i,j)-\tilde{o}(i,j))^{2}}
\end{equation}
where $o(i, j)$ and $\tilde{o}(i, j)$ represent the original image and distorted image, respectively. Figs. \ref{fig:stegocl}(a)-(c) show partially occluded stegano images with 1/36, 1/18, and 1/12 occlusion, Figs. \ref{fig:stegocl}(d)-(f) show the corresponding extracted message (encrypted image). Figs. \ref{fig:stegocl}(g)-(i) show the result of applying decryption method to the extracted images. PSNR values of occluded images are 38.23, 32.25, and 28.41 dB, respectively, which prove while the image experienced an altering, it remains in a range that the whole image could be seen by human. 

\begin{table}[H]
	\centering
	\caption{Comparison of correlation coefficients.}
	\label{tbl:correlation}
	\resizebox{\textwidth}{!}{
	\begin{tabular}{lcccccc}
		\hline
		\textbf{Image} & \textbf{\begin{tabular}[c]{@{}c@{}}Proposed \\ Encryption\end{tabular}} & \textbf{\begin{tabular}[c]{@{}c@{}}Proposed \\ Steganography\end{tabular}} & Ref. \cite{zhang2014symmetric} & Ref. \cite{hsiao2015fingerprint} & \begin{tabular}[c]{@{}c@{}}Ref. \cite{zhang2014symmetric} \\ + 2-LSB \end{tabular} & \begin{tabular}[c]{@{}c@{}}Ref. \cite{zhang2014symmetric} \\ + 3-LSB \end{tabular}\\ \hline
		Airplane & -0.0041 & 0.9987 & 0.0239 & 0.0042 & 0.9332 & 0.8291 \\
		Architecture & 0.0046 & 0.9999 & 0.0127 & -0.0215 & 0.9868 & 0.8822 \\
		Baboon & 0.0135 & 0.9984 & 0.0187 & 0.0249 & 0.9199 & 0.9064 \\
		Barbara & -0.0577 & 0.9994 & 0.0233 & -0.0210 & 0.9570 & 0.8993 \\
		Boat & 0.0136 & 0.9994 & 0.0219 & 0.0533 & 0.9137 & 0.9001 \\
		Bridge & 0.0127 & 0.9992 & 0.0129 & 0.0286 & 0.9170 & 0.9043 \\ 
		Cameraman & 0.0086 & 0.9957 & 0.0244 & -0.0206 & 0.9473 & 0.8387 \\
		Clown & 0.1008 & 0.9984 & 0.0252 & 0.0157 & 0.9555 & 0.8547 \\
		Couple & -0.0009 & 0.9988 &  0.0129 & -0.0223 & 0.9499 & 0.8490 \\
		Crowd & -0.0346 & 0.9989 & 0.0222 & 0.0368 & 0.9227 & 0.9101 \\
		Fruits & -0.0173 & 0.9994 & 0.0150 & 0.0197 & 0.9932 & 0.9759 \\
		Girl face & 0.1114 & 0.9847 & 0.0309 & -0.0587 & 0.9760 & 0.9646 \\ 
		Gold hill & -0.0146 & 0.9972 & 0.0112 & 0.0283 & 0.9671 & 0.8525 \\
		Lena & 0.1757 & 0.9991 & 0.0213 & 0.1818 & 0.9174 & 0.9317 \\
		Man & 0.0347 & 0.9997 & 0.0126 & 0.0951 & 0.8837 & 0.7490 \\
		Mountain & 0.0091 & 0.9987 & 0.0144 & -0.0345 & 0.9826 & 0.9735 \\
		Peppers & 0.0853 & 0.9968 & 0.0205 & 0.1003 & 0.9613 & 0.9760 \\
		Pool & -0.0363 & 0.9997 & 0.0162 & 0.0489 & 0.9055 & 0.9692 \\
		Serrano & 0.0305 & 0.9996 & 0.0135 & -0.0365 & 0.9053 & 0.8348 \\
		Sailboat & -0.0189 & 0.9932 & 0.0173 & 0.0196 & 0.9604 & 0.8415 \\
		Tank 1 & -0.0146 & 0.9994 & 0.0156 & 0.0156 & 0.9573 & 0.8827 \\
		Tank 2 & -0.0135 & 0.9994 & 0.0234 & -0.0173 & 0.9913 & 0.9778 \\
		Truck & -0.0371 & 0.9987 & 0.0331 & 0.0440 & 0.9125 & 0.9654 \\
		Trucks & 0.0145 & 0.9964 & 0.0250 & 0.0228 & 0.9744 & 0.8599 \\
		Zelda & 0.1515 & 0.9970 & 0.1652 & 0.0998 & 0.9380 & 0.8865 \\
		\hline
		\textbf{Mean} & \textbf{0.0206} & \textbf{0.9978} & 0.0253 & 0.0242 & 0.9451 & 0.8965 \\
		\hline 
	\end{tabular}
	}
\end{table}

\begin{figure}[H]
	\centering
	\begin{subfigure}[c]{0.31\textwidth}
		\includegraphics[width=\textwidth]{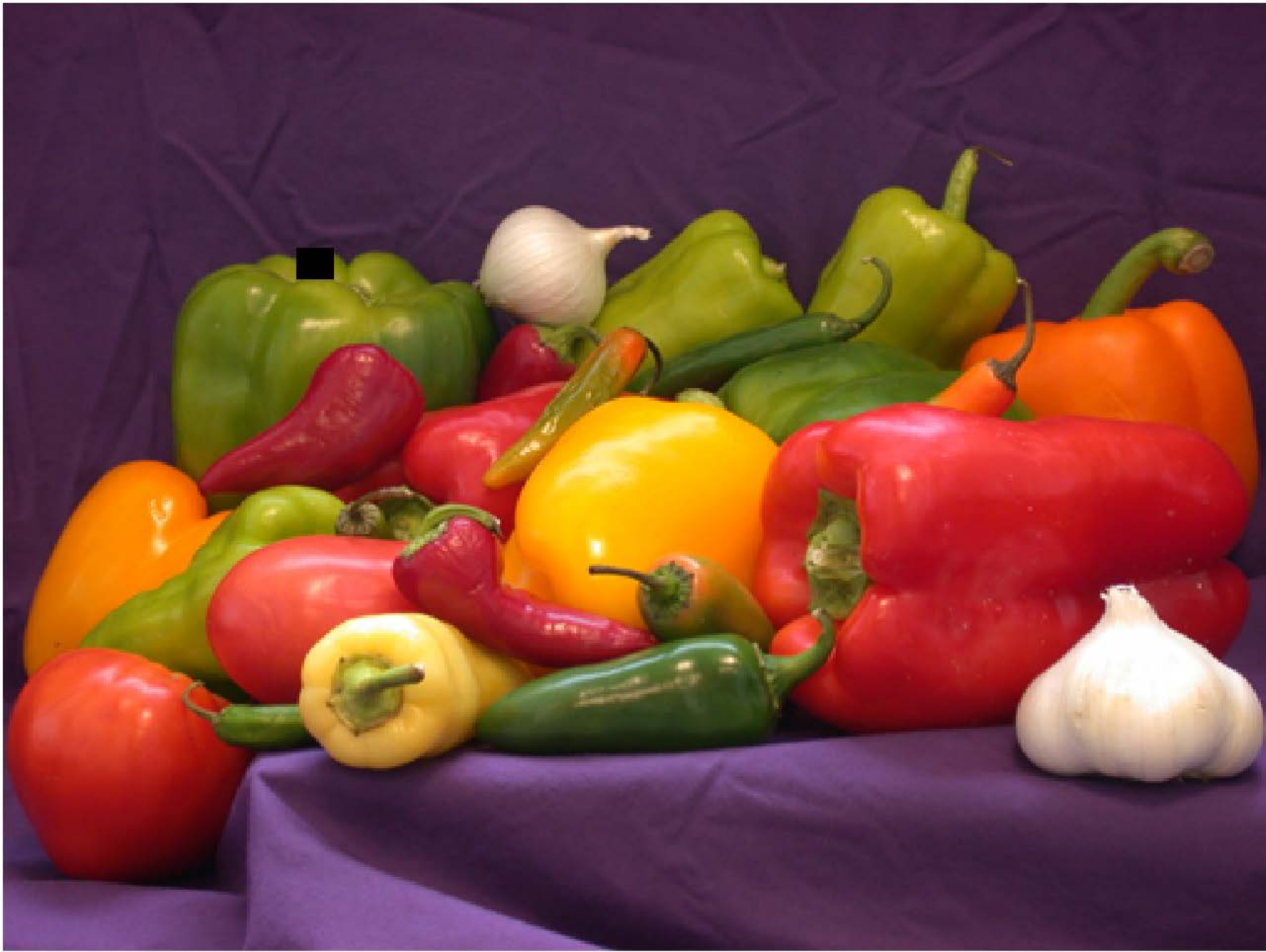}
		\caption{PSNR = 38.23 dB}
	\end{subfigure}
	\hfill
	\begin{subfigure}[c]{0.31\textwidth}
		\includegraphics[width=\textwidth]{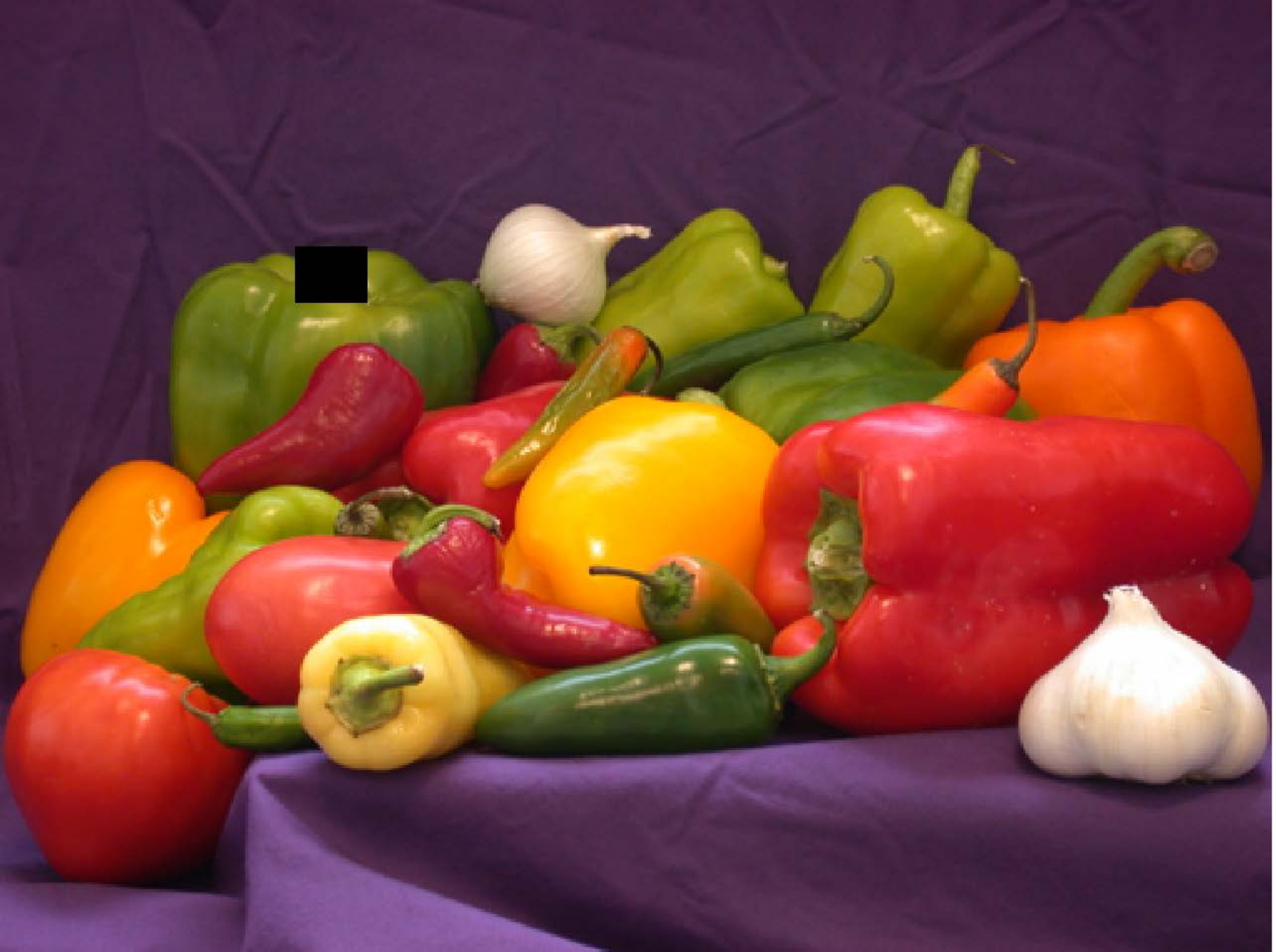}
		\caption{PSNR = 32.25 dB}
	\end{subfigure}
	\hfill
	\begin{subfigure}[c]{0.31\textwidth}
		\includegraphics[width=\textwidth]{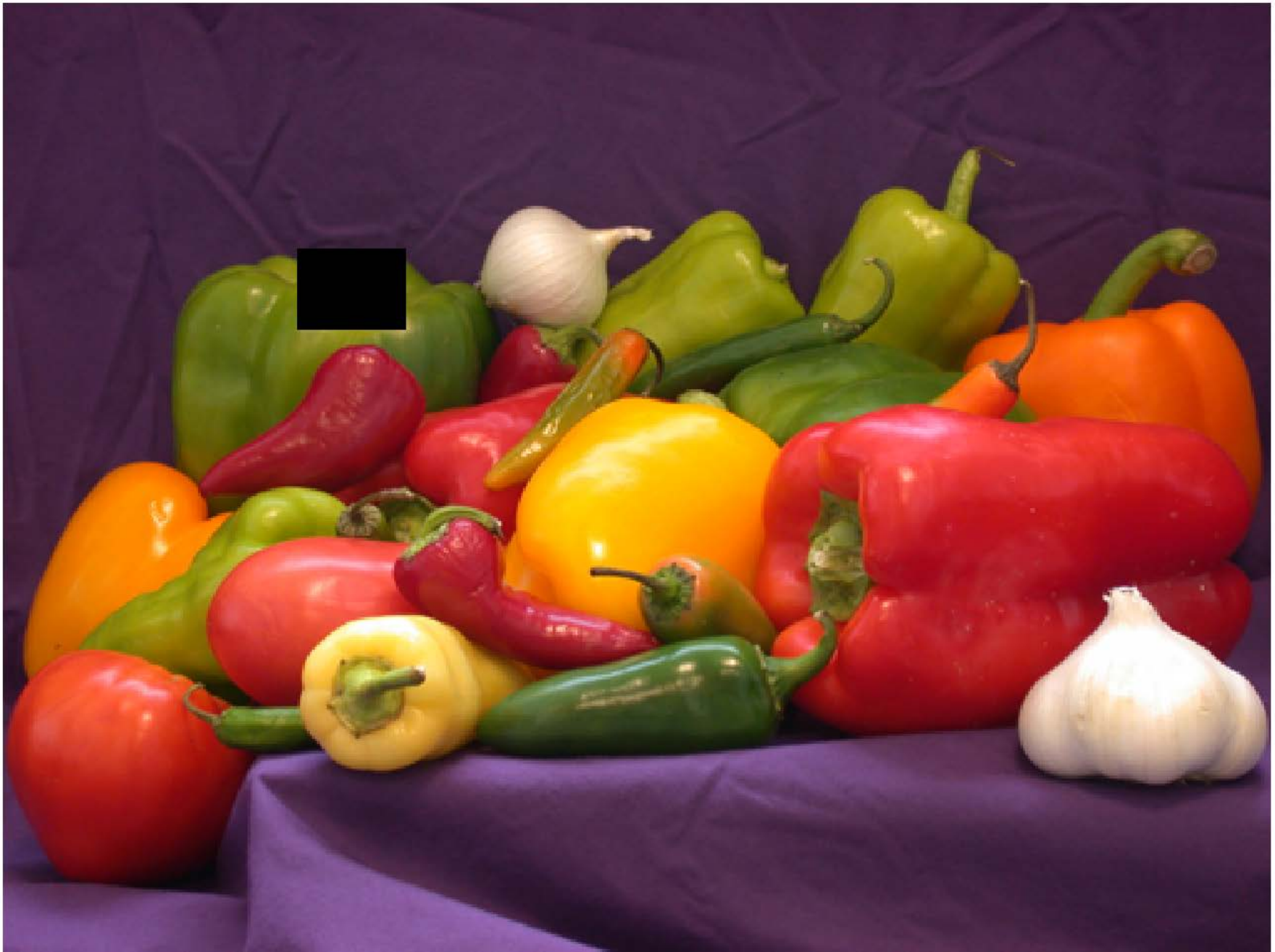}
		\caption{PSNR = 28.41 dB}
	\end{subfigure}
	\hfill
	\begin{subfigure}[c]{0.31\textwidth}
		\includegraphics[width=\textwidth]{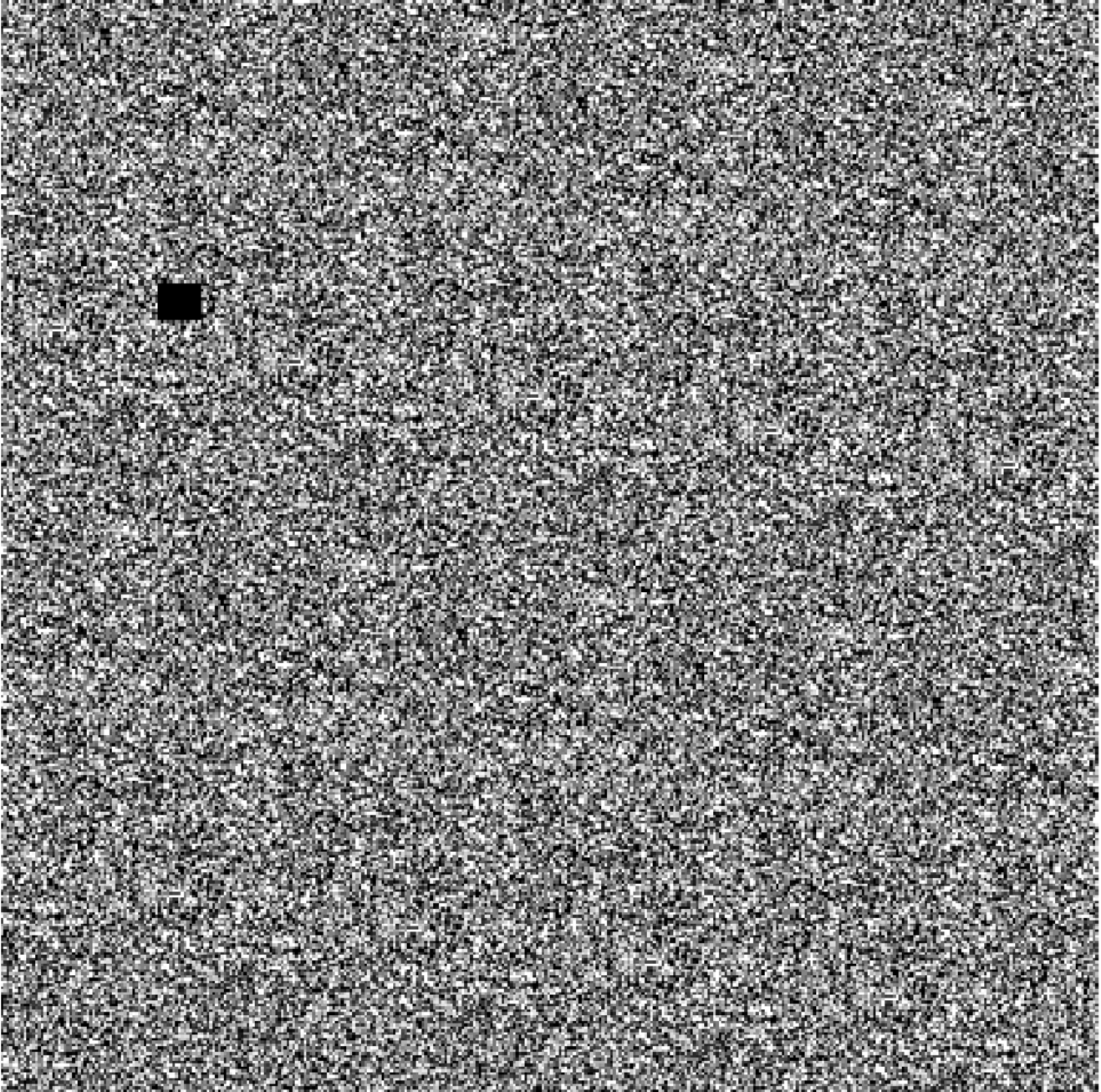}
		\caption{}
	\end{subfigure}
	\hfill
	\begin{subfigure}[c]{0.31\textwidth}
		\includegraphics[width=\textwidth]{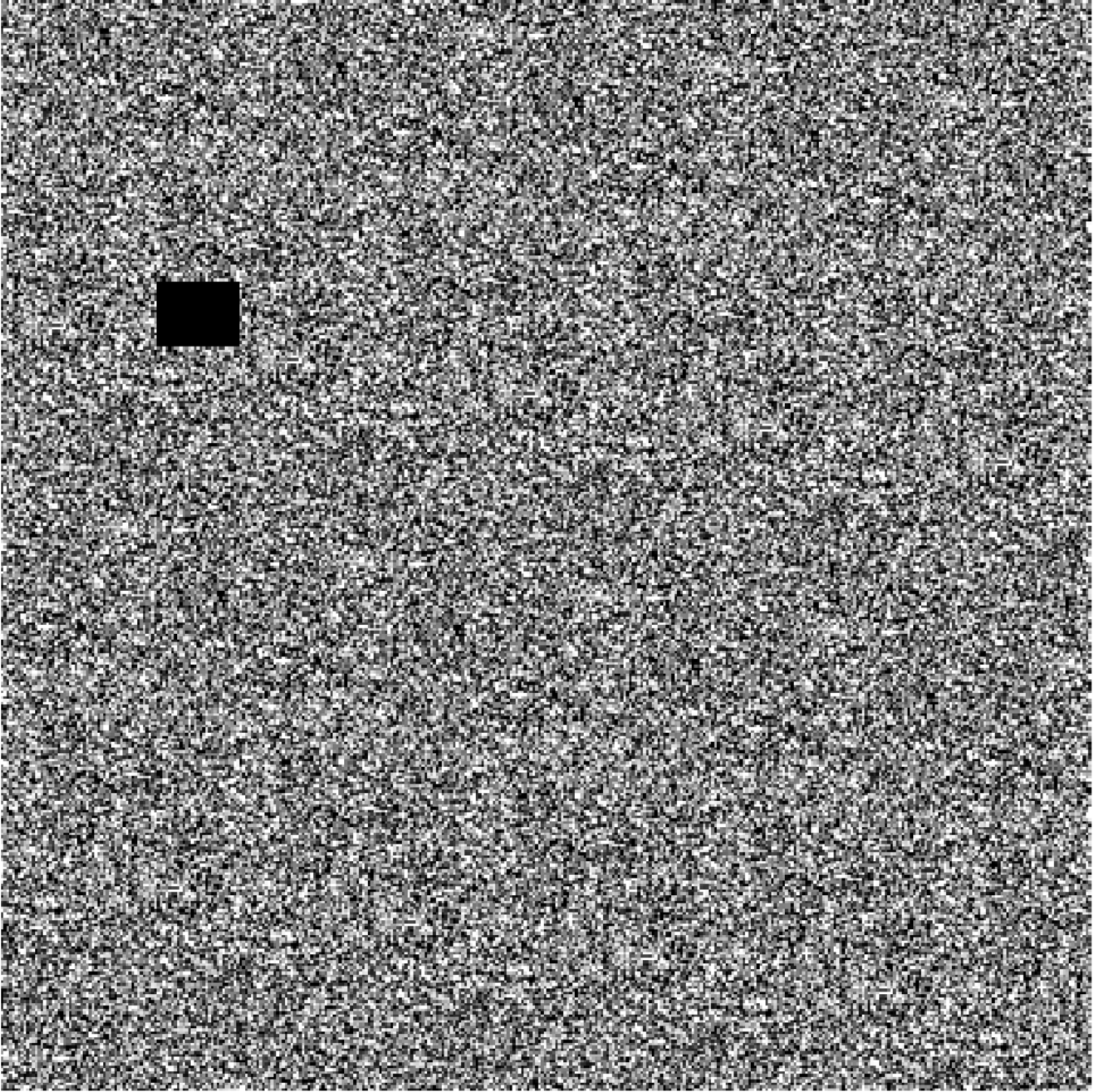}
		\caption{}
	\end{subfigure}
	\hfill
	\begin{subfigure}[c]{0.31\textwidth}
		\includegraphics[width=\textwidth]{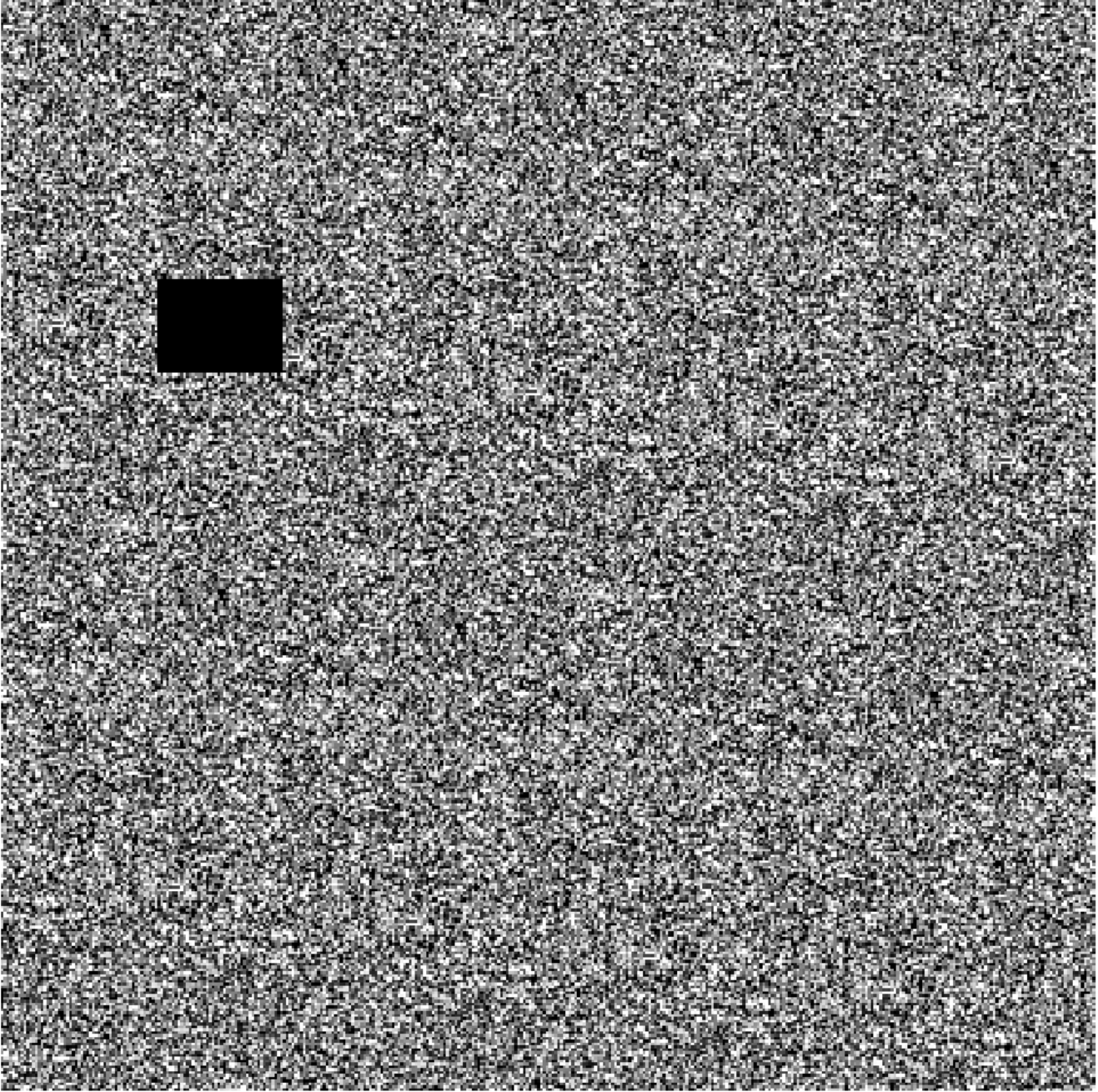}
		\caption{}
	\end{subfigure}
\end{figure}
\begin{figure}[H]\ContinuedFloat
	\begin{subfigure}[c]{0.31\textwidth}
		\includegraphics[width=\textwidth]{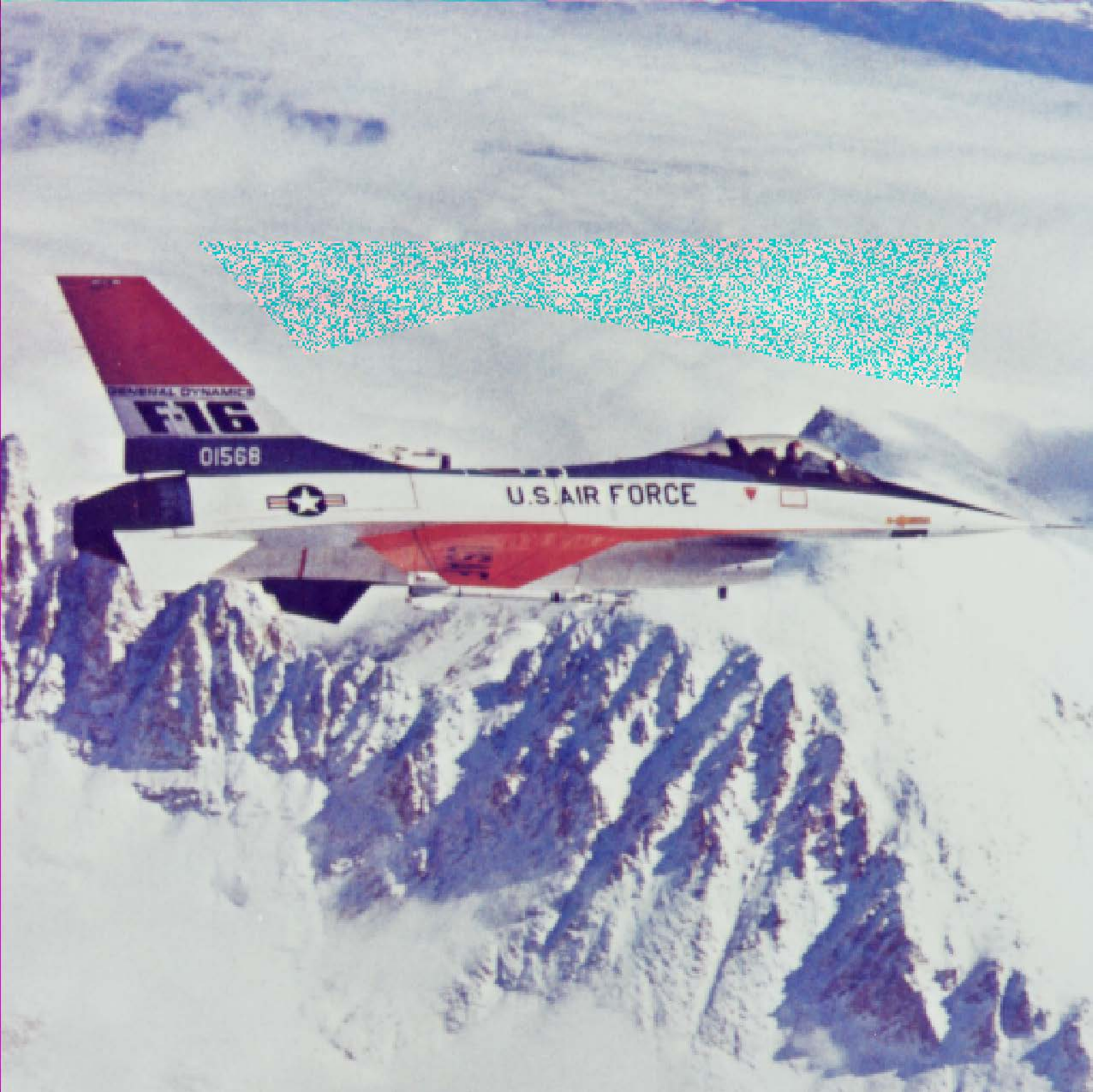}
		\caption{PSNR = 23.83 dB}
	\end{subfigure}
	\hfill
	\begin{subfigure}[c]{0.31\textwidth}
		\includegraphics[width=\textwidth]{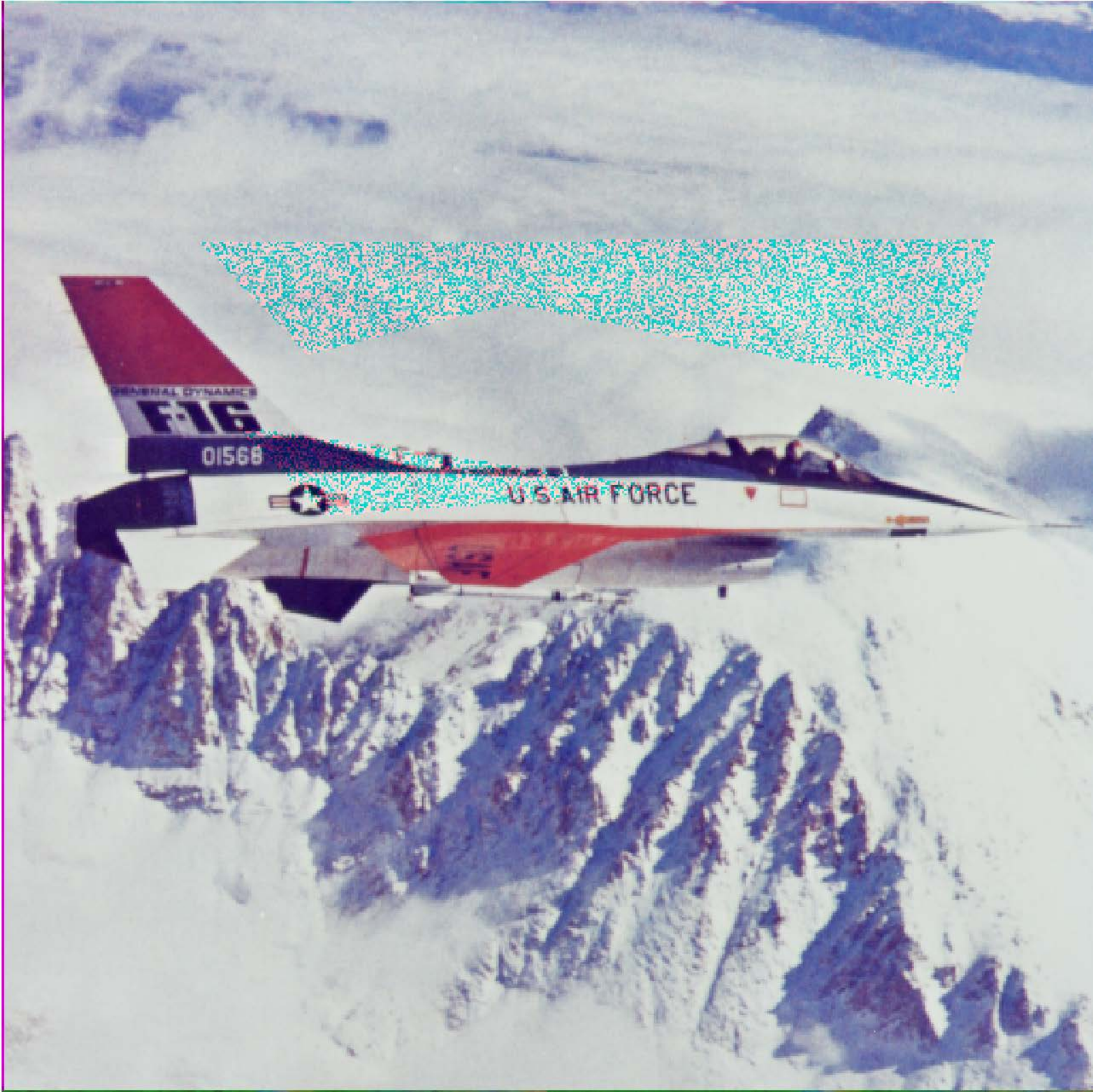}
		\caption{PSNR = 23.12 dB}
	\end{subfigure}
	\hfill
	\begin{subfigure}[c]{0.31\textwidth}
		\includegraphics[width=\textwidth]{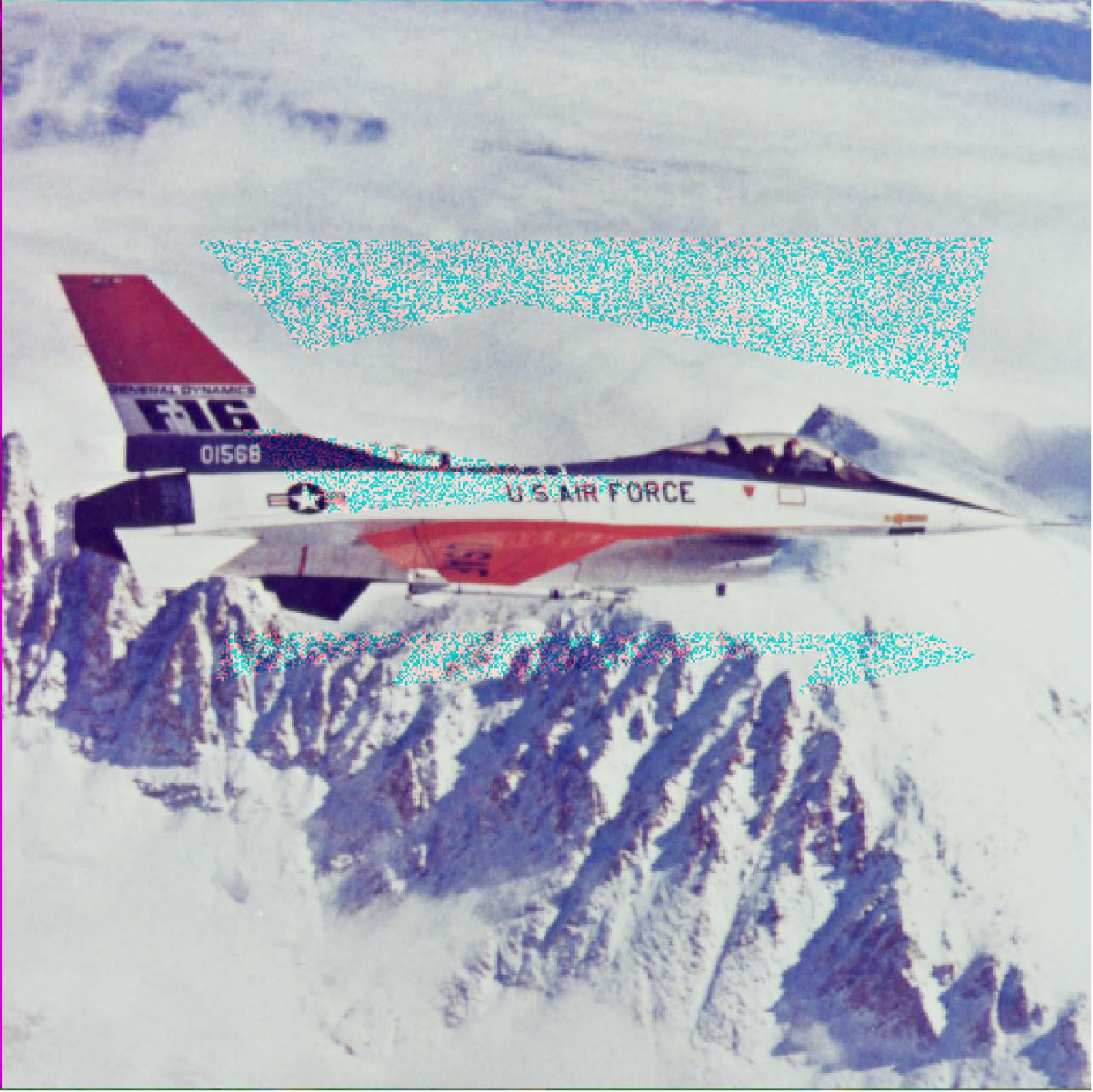}
		\caption{PSNR = 23.10 dB}
\end{subfigure}
	\caption{Occlusion attack results. (a)-(c) Stegano images with 1/36, 1/18, and 1/12 occlusion, (d)–(f) corresponding uncovered images, (g)-(i) results of decryption which their distortions are visible in cyan color.} \label{fig:stegocl}
\end{figure}

It is obvious from Figure \ref{fig:stegocl} that the proposed cryptology algorithm has certain robustness on resisting occlusion attack, and can recover original images in an approximately recognizable fashion. One could wonder why other local feature descriptors, e.g., speeded up robust features (SURF) \cite{bay2008speeded}, optimized SIFT \cite{abdollahi2015optimized}, or local binary pattern (LBP) \cite{guo2010completed}, \cite{ghorbani2015hog}, were not used to form the raw key. The idea of utilizing local feature descriptors can be expanded in a way that other local image descriptors were used instead of SIFT. However, the descriptor must be at least residence to the widely known occlusion attack.

\begin{figure}[H]
	\centering
	\begin{subfigure}[c]{0.31\textwidth}
		\includegraphics[width=\textwidth]{Fig6c}
		\caption{PSNR = 28.41 dB}
	\end{subfigure}
	\hfill
	\begin{subfigure}[c]{0.31\textwidth}
		\includegraphics[width=\textwidth]{Fig6f}
		\caption{}
	\end{subfigure}
	\hfill
	\begin{subfigure}[c]{0.31\textwidth}
		\includegraphics[width=\textwidth]{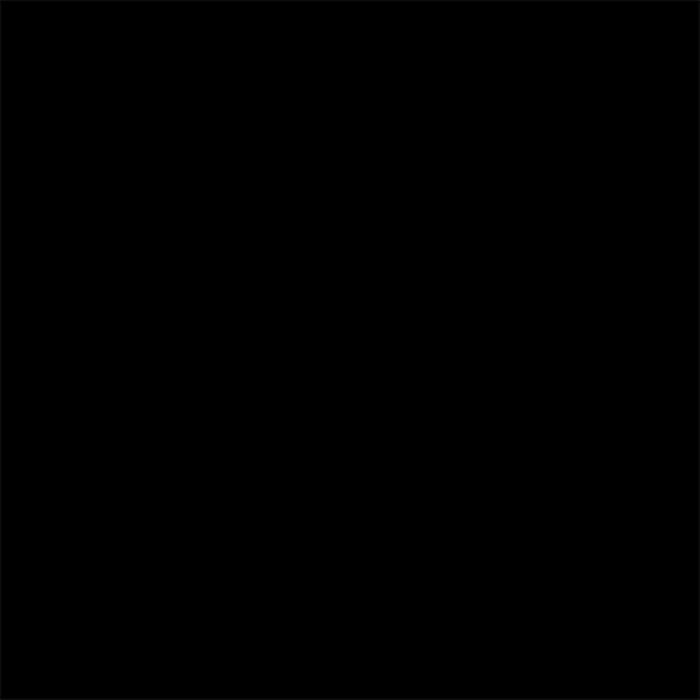}
		\caption{}
	\end{subfigure}
	\hfill
	\begin{subfigure}[c]{0.31\textwidth}
		\includegraphics[width=\textwidth]{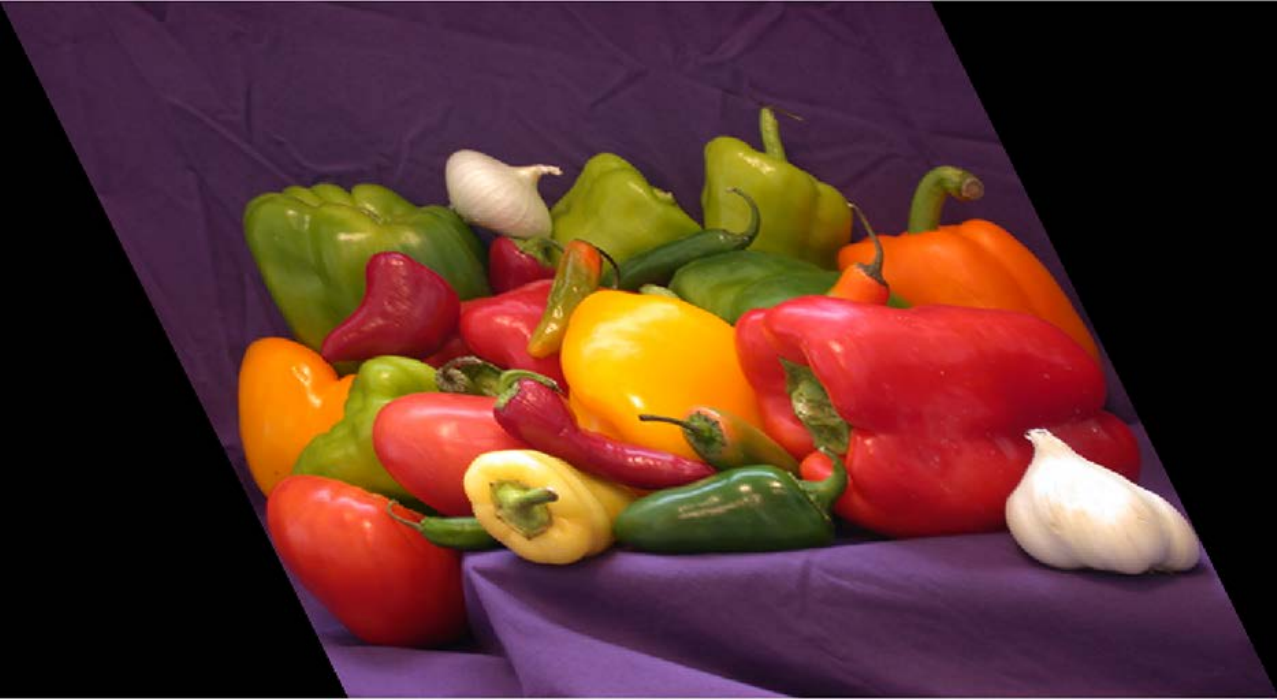}
		\caption{PSNR = 10.86 dB}
	\end{subfigure}
	\hfill
	\begin{subfigure}[c]{0.31\textwidth}
		\includegraphics[width=\textwidth]{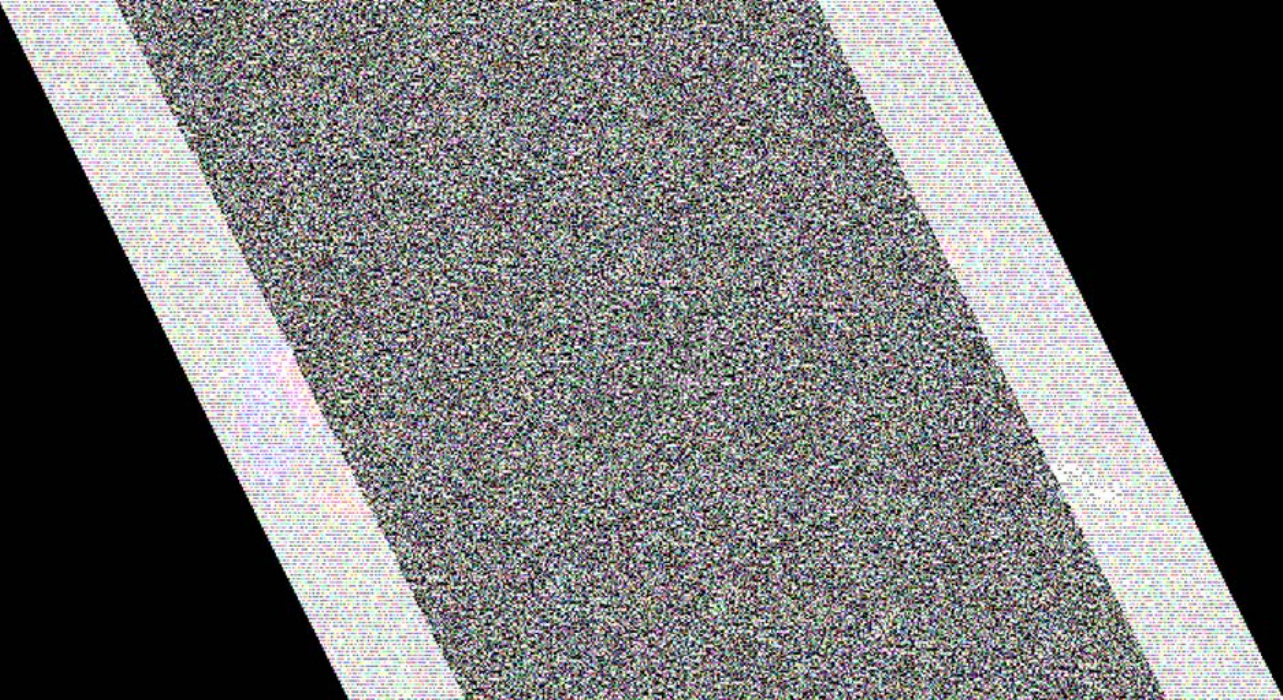}
		\caption{}
	\end{subfigure}
	\hfill
	\begin{subfigure}[c]{0.31\textwidth}
		\includegraphics[width=\textwidth]{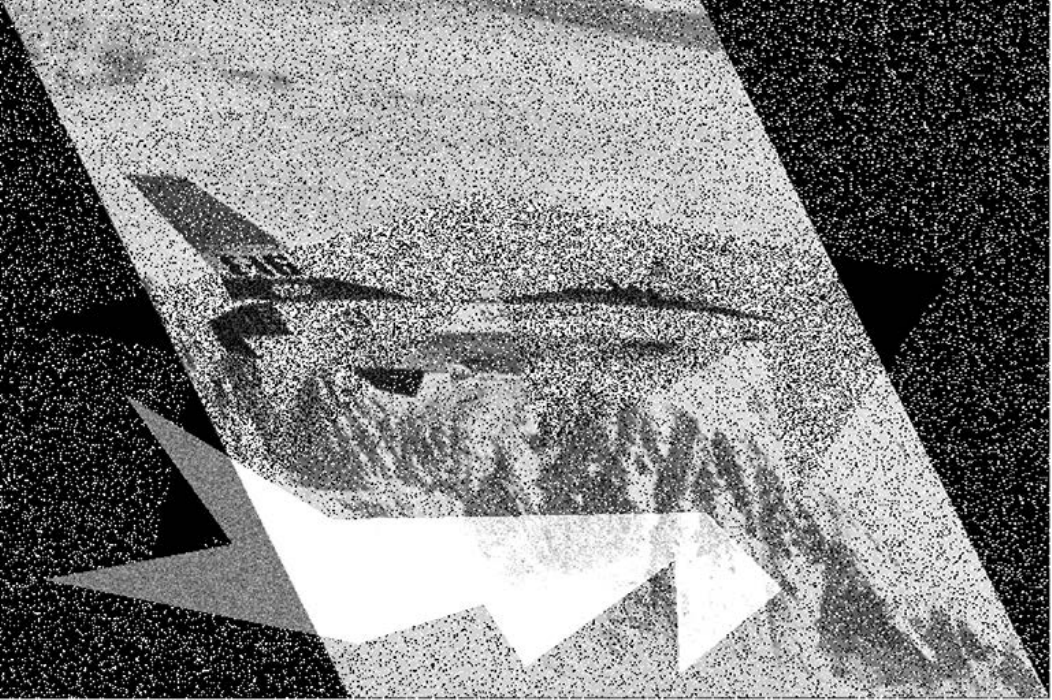}
		\caption{PSNR = 6.05 dB}
	\end{subfigure}
\caption{(a) Stegano images with 1/12 occlusion, (b) corresponding uncovered image, (c) result of decryption where LBP was used to form the raw key, (d) sheared Stegano image, (e) corresponding uncovered image, (f) result of the proposed de-cryptology method where in the presence of shearing attack.} \label{fig:lbp}
\end{figure}

Figure \ref{fig:lbp} demonstrates that if the the Stego image faces the occlusion attack or sever noise, revealing the original message is not possible. The main reason is in changing the image structure and consequently changing the feature vector, i.e., raw key. We can see in Figure \ref{fig:lbp}(d-f) that the proposed cryptology scheme is not robust enough when the image subject to shear attack. Although SIFT descriptor is robust enough against shear attack, decrypted message is not the same as the original message. It should be noted that previous to apply the de-cryptology procedure, obtained PSNR can inform the receiver that message has already interfered and the communication channel is no longer safe.

\subsection{Visual Test}
Applying cryptography and steganography algorithms generally could result in loss of image quality. Visual test is one of the criteria for evaluating the performance of cryptology algorithms which is divided into full-reference and no-reference categories. In the former category, the input image (crypto+stegano image) compares against a pristine reference image (cover image) which does not subject to any distortion. While in the latter category, assuming the reference image has changed, the comparison is made between statistical features of the input image and a set of features derived from a default model computed from images of natural scenes with similar distortions or a trained model on an image database. 

Full-reference algorithms \cite{sepas2012novel}, \cite{dehshibi2017hybrid} include (1) Mean-squared error (MSE), (2), Peak signal-to-noise ratio (PSNR) and (3) Structural Similarity (SSIM) Index. On the other hand, non-reference metrics include (1) Blind/Referenceless Image Spatial Quality Evaluator (BRISQUE) \cite{mittal2012no} and (2) Natural Image Quality Evaluator (NIQE) \cite{mittal2013making}. We compare the proposed cryptology algorithm with those are introduced in \cite{zhang2014symmetric} and \cite{hsiao2015fingerprint}. The mean and standard deviation of the evaluations are tabulated in Table \ref{tbl:visual}.

\begin{table}[H]
	\centering
	\caption{Comparison of image quality metrics.}
	\label{tbl:visual}
	\resizebox{\textwidth}{!}{%
		\begin{tabular}{lcccccccccc}
			\hline
			\multirow{2}{*}{} & \multicolumn{2}{c}{\textbf{MSE}} & \multicolumn{2}{c}{\textbf{PSNR}} & \multicolumn{2}{c}{\textbf{SSIM}} & \multicolumn{2}{c}{\textbf{BRISQUE}} & \multicolumn{2}{c}{\textbf{NIQE}} \\ \cline{2-11} 
			& Mean & STD & Mean & STD & Mean & STD & Mean & STD & Mean & STD \\ \hline
			Proposed Method & 0.0014 & 9.12e-07 & 54.4153 & 0.0027 & 0.9914 & 2.83e-05 & 29.5003 & 0.2546 & 3.0301 & 0.0302 \\
			Ref. \cite{zhang2014symmetric} + 3-LSB & 0.0015 & 8.76e-07 & 53.5898 & 0.0103 & 0.9871 & 1.46e-06 & 32.0002 & 0.521 & 3.9527 & 0.0045 \\
			Ref. \cite{hsiao2015fingerprint} + 3-LSB & 0.0015 & 9.51e-07 & 54.0087 & 0.0031 & 0.9899 & 3.71e-06 & 30.8596 & 0.8207 & 3.1802 & 0.0003 \\ \hline
			\multicolumn{11}{l}{$\mathrm{STD}^{\ast}$ is an abbreviation for Standard deviation.}
		\end{tabular}%
	}
\end{table}

\subsection{Complexity Analysis}

Complexity analysis was previously investigated in steganography systems \cite{hopper2009provably}. We examine the proposed cryptology algorithm with Kolmogorov complexity \cite{gholami2018}, and temporal complexity criteria \cite{adamatzky2010generative}, \cite{taghipour2016complexity} as complexity measurements. To represent the \enquote{temporal} complexity dynamics of the whole image, a series of numbers $C_{b_{1}}, C_{b_{2}}, \cdots, C_{b_{i}}$ is calculated using Eq. \ref{equ:12}. The temporal complexity calculates the dependence between two quantities regarding correlation coefficient between two images.

\begin{equation}
	\label{equ:12}
	\begin{split}
		C_{b_{i}}(Im^{i}_{tag}, Im_{ref}) = \frac{\mathrm{E}[(Im^{i}_{tag}-\mu_{Im^{i}_{tag}})(Im_{ref}-\mu_{Im_{ref}})]}{\sigma_{Im^{i}_{tag}}\sigma_{Im_{ref}}}, \\
		\mathrm{where} \quad 1 \leq i \leq 25
	\end{split}
\end{equation}
where $E$ is the expected value operator, $\mu$ is mean and $\sigma$ is the standard deviations. Here, we calculate temporal complexity for both encrypted and stegano images in which the $Im_{ref}$ is Cover image. Then, we applied locally weighted scatterplot smoothing (LOWESS regression) to find out the changing trend in each cryptology sub-module.

\begin{figure}[H]
	\centering
	\begin{subfigure}[b]{0.7\textwidth}
		\includegraphics[width=\textwidth]{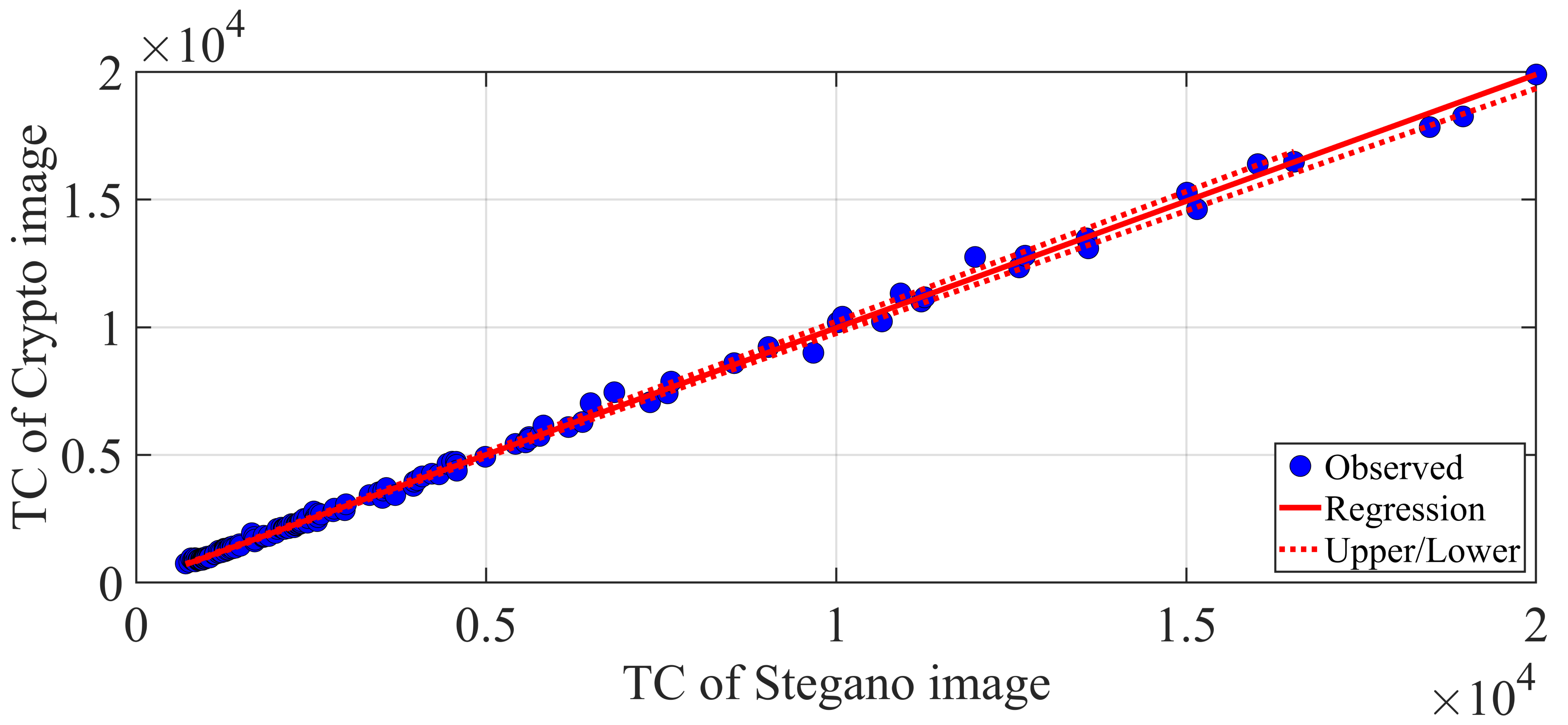}
		\caption{}
	\end{subfigure}
	\hfill
	\begin{subfigure}[b]{0.7\textwidth}
		\includegraphics[width=\textwidth]{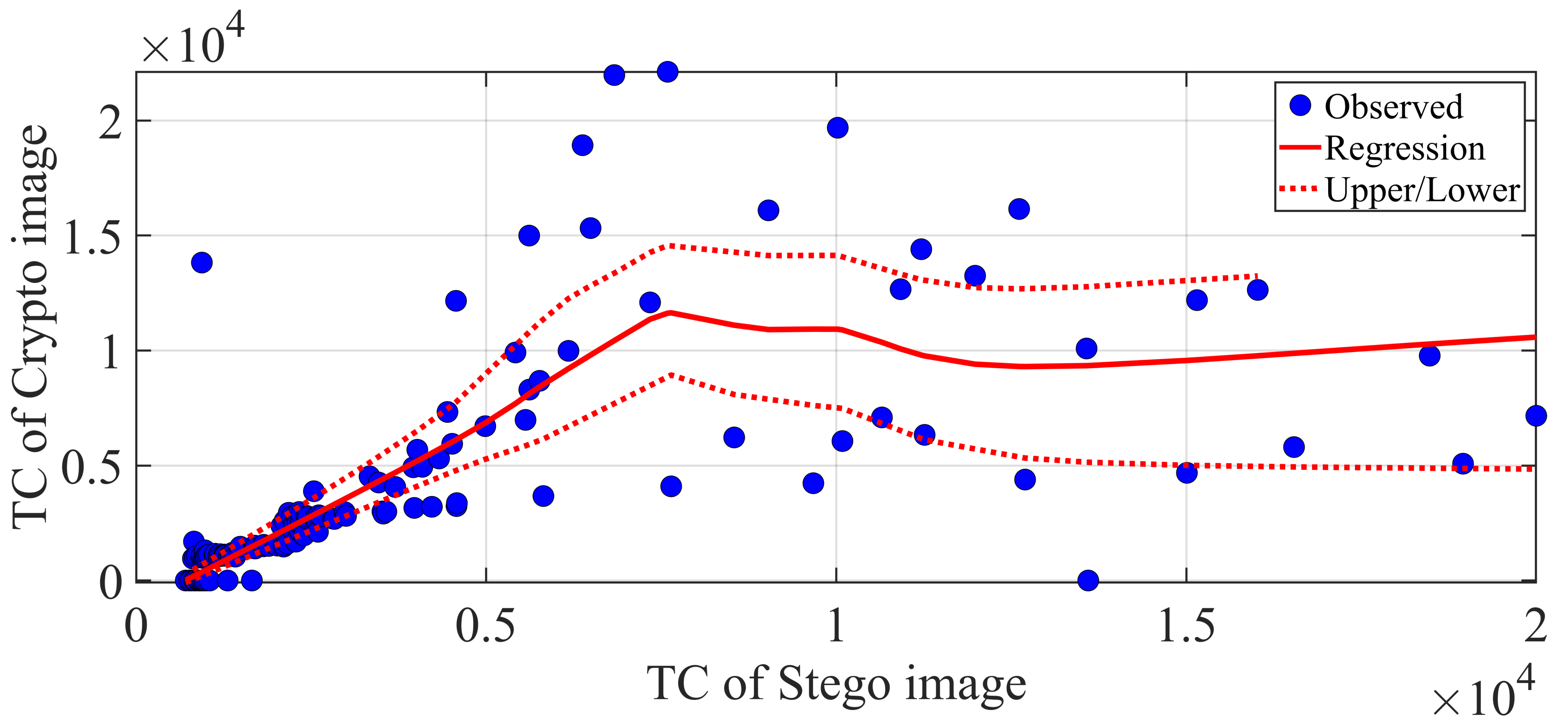}
		\caption{}
	\end{subfigure}
	\caption{Trends of temporal complexity (TC) in Crypto vs. Stegano images in (a) proposed method and (b) Ref. \cite{zhang2014symmetric} + 3-LSB. Monotone trend of LOWESS in the proposed method proves that CNN kernel could surpass high variance variations.} \label{fig:temporal}
\end{figure}

We set the smoothing factor to 0.9 to presented analysis with a smoother regression. Plot of Figure~\ref{fig:temporal} can quantify the amount of dispersion where the temporal complexity of encrypted images is drawn versus stegano images. Figure \ref{fig:temporal}(b) demonstrates that higher standard deviation causes data points spread out over a wider range. However, applying CNN kernel to cryptology process proves that it could surpass this diversity. Therefore, concerning previous experiments, one could conclude that higher complexity and dispersion are not always
analogous to each other (see Figure \ref{fig:temporal}(a)).

\begin{figure}[H]
	\centering
	\includegraphics[width=0.7\linewidth]{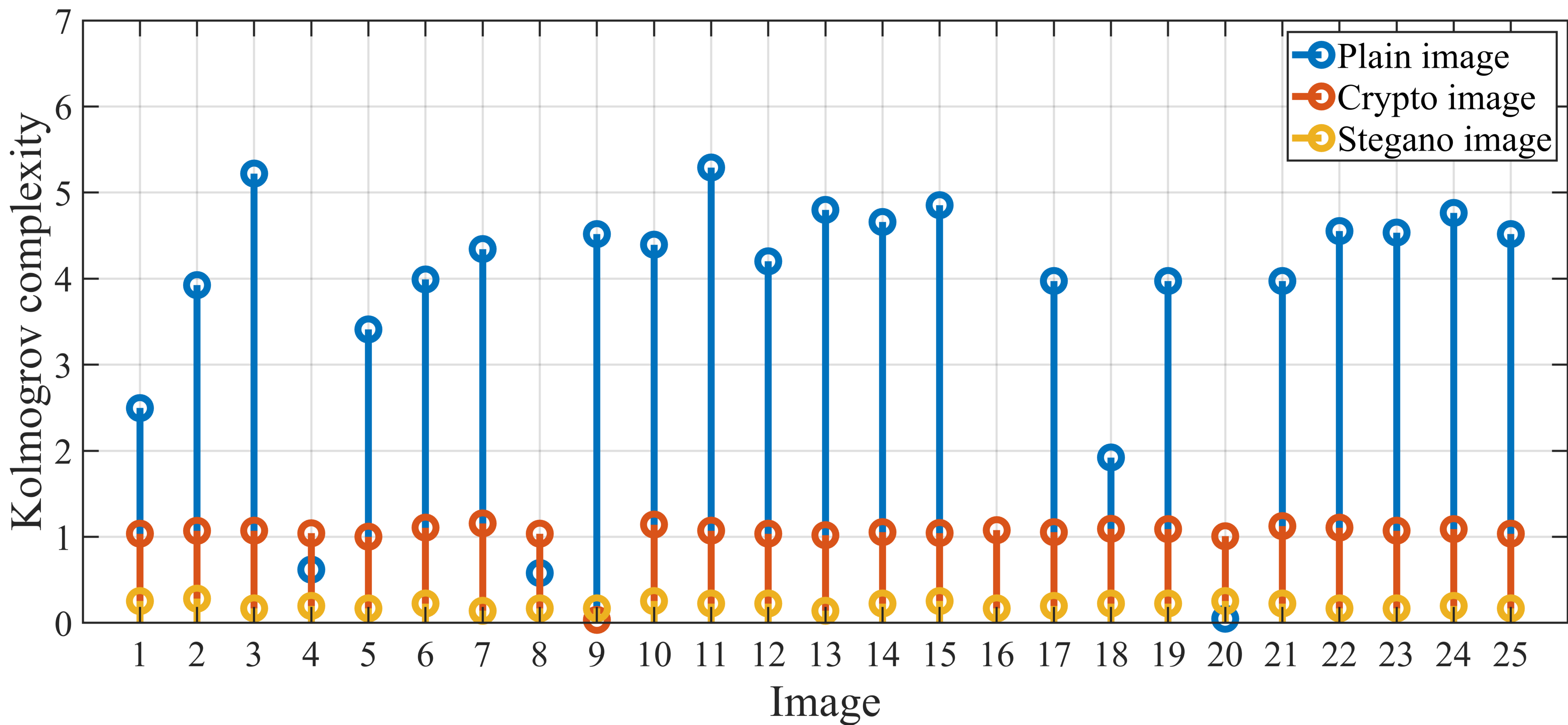}
	\caption{Kolmogorov complexity estimated by LBD method in plain, Crypto, and Stegano images.}
	\label{fig:kolmogrov}
\end{figure}

Dynamics of a system could help inferring if it has a stochastic nature. The Kolmogorov complexity (KC) $K(x)$ is used to quantify the randomness degree in encrypted and stegano images. $K(x)$ of an object $x$ is the length, in bits, of the smallest program that when run on a Universal Turing Machine produces object $x$. For calculating KC, we used the Layered Block Decomposition (LBD) method \cite{gholami2018} as a suitable approximation. Figure \ref{fig:kolmogrov} shows the KC of plain, Crypto, and Stegano images.

In LBD method, images are quantized and binarized in $q$ digital levels before aggregating. Therefore, the coarse graining of the KC estimation is defined by the number of digital levels in which an image is quantized. In Figure \ref{fig:kolmogrov}, one could infer two facts. (i) The range of changes in KC for plain images is larger than that of Crypto and Stegano images experience. Although limited variations in the KC values, due to the difference in the morphology of images, are acceptable, we appreciate a severe fall in Barbara (\#4), Clown (\#8), Pool (\#18), and Sailboat (\#20) images that we must inspect its cause by further experiments. (ii) The proposed method results in KC values with almost the least change. Therefore, it is reasonable to conclude that while the CNN kernel provides a system with a chaotic nature, its result in the proposed cryptology method indicates the least change in complexity, except for Couple (\#9).

Figure \ref{fig:temporal}(a) also confirms the validity of this finding. To make a better inference about KC, especially those images experiencing extreme changes in KC values, we fit all valid parametric probability distributions to the calculated KCs which is shown in Figure \ref{fig:parametric}.

\begin{figure}[H]
	\centering
	\begin{subfigure}[b]{0.7\textwidth}
		\includegraphics[width=\textwidth]{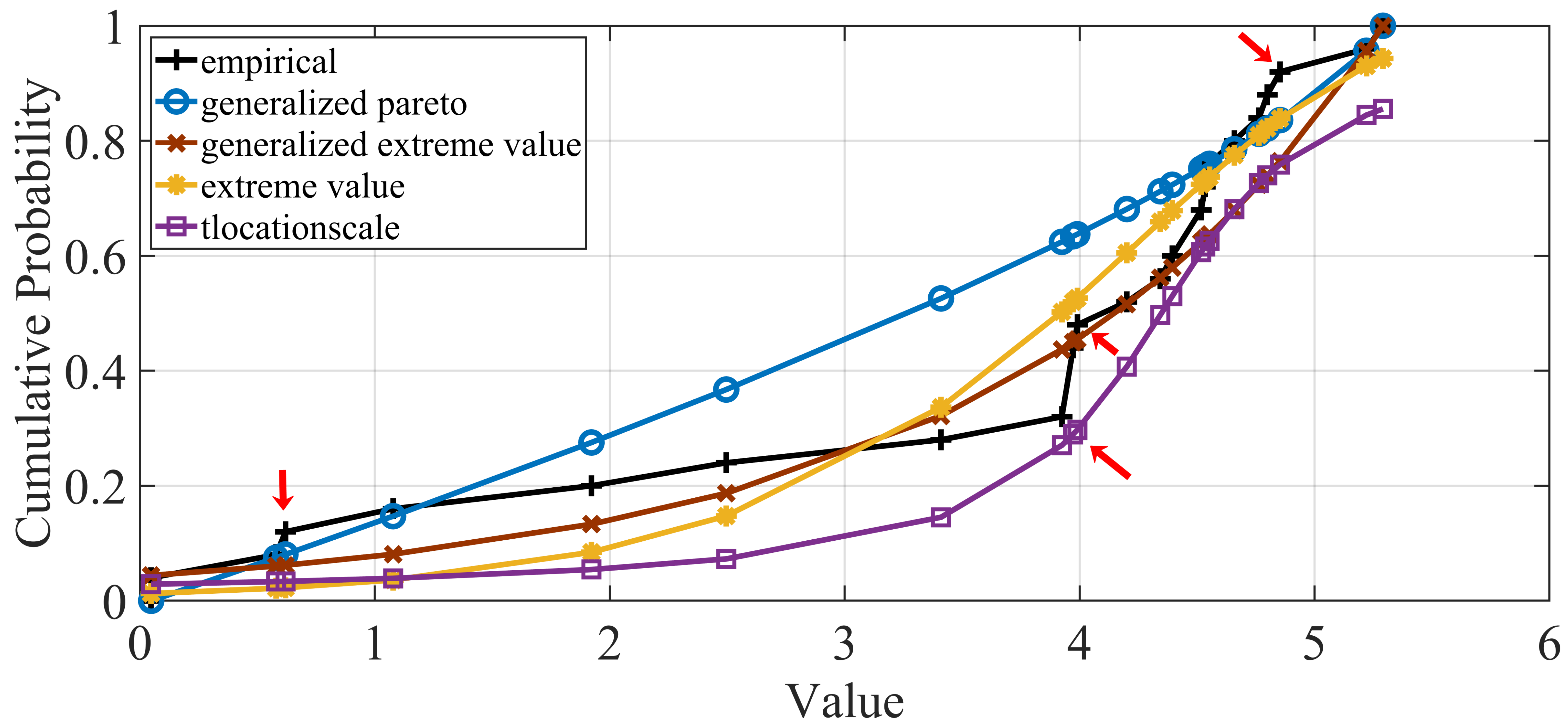}
		\caption{}
	\end{subfigure}
\end{figure}
\begin{figure}[H]\ContinuedFloat
\centering
	\begin{subfigure}[b]{0.7\textwidth}
		\includegraphics[width=\textwidth]{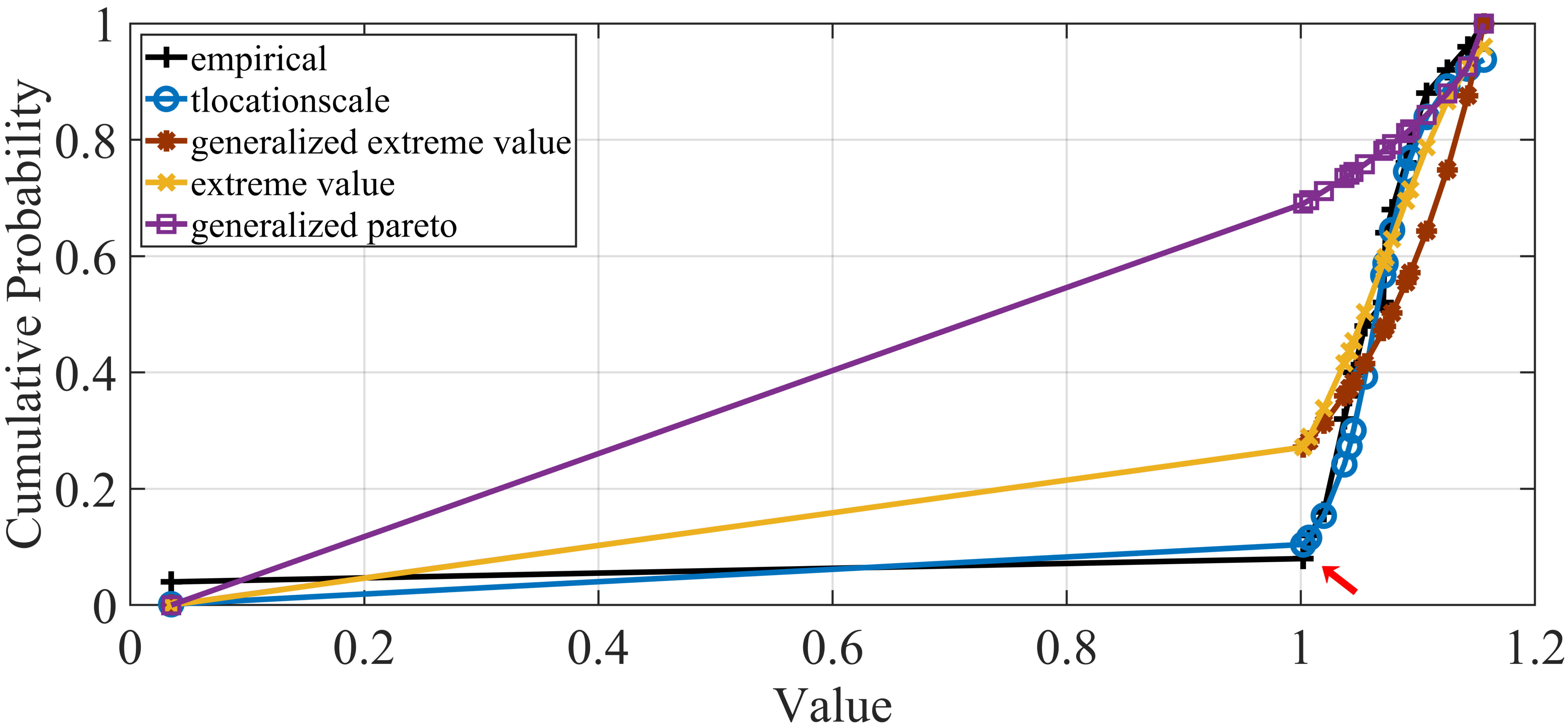}
		\caption{}
	\end{subfigure}
	\hfill
	\begin{subfigure}[b]{0.7\textwidth}
		\includegraphics[width=\textwidth]{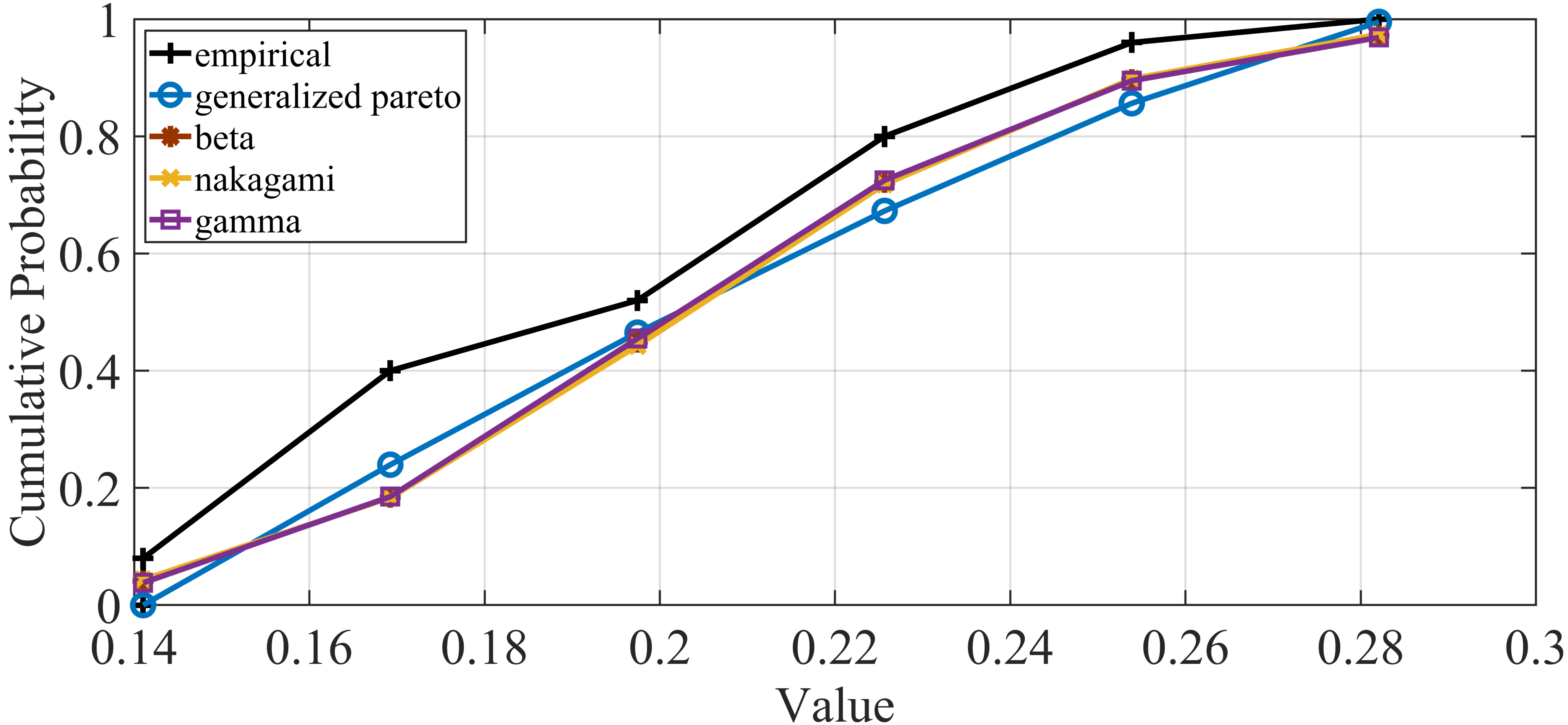}
		\caption{}
	\end{subfigure}
	\caption{A cumulative distribution function of KCs for (a) plain images, (b) Crypto images, and (c) Stegano images.}
	\label{fig:parametric}
\end{figure}

In Figure \ref{fig:parametric}(a), we can see that empirical cumulative distribution function (CDF) experiences four sharp changes in its slope while in Figure \ref{fig:parametric}(b) just one change is observable. Moreover, one could see that the most probable fit distribution to the KC values in plain and Stegano images remain the same, i.e., generalized Pareto \cite{coles2001introduction}, while in Crypto images it changed to ``T location-scale" \cite{epps2005tests} distribution which is useful for modeling prone to outliers data. Therefore, it is reasonable to conclude that the proposed method could reach to its chaotic nature in cryptography sub-module while it keep the nature of KC values distribution where the the steganography sub-module was applied.

\section{Conclusion}

In this paper, the dynamic property of a cellular non-linear network (CNN) is utilized to present a reliable and robust visual cryptology (cryptography + steganography) scheme. Common cryptographic algorithms use a pre-agreed key between the sender and receiver to encrypt the cipher. However, in the proposed method, both parties use the scale-invariant feature transform (SIFT) to extract the raw key from a cover image that is used in steganography module. Then, this raw key is fed to the proposed 3D CNN kernel to make the final key. While this type of key generation will increase the security level as an intruder cannot use well-known crypto-attack and needs knowledge in the field of image processing, it can increase the risk of key loss due to possible cover image distortion. To demonstrate the resilience of the proposed method against potential image distortion, e.g., occlusion, we conducted experiments that showed the proposed method can decrypt the cipher in the presence of such destruction in a visually acceptable manner. We also tested the effectiveness of the cryptology method using complexity criteria. The results of the experiments indicate that, contrary to the nature of a CNN which can potentially increase irregularities, crypto and stegano images show the least variation compared to the competitive methods. 

We believe that with the development of research and practice on cellular non-linear networks and memristor-based CNNs, speeding-up the computational routine of each cell is an urgent need that has to be answered. As each cell in the proposed CNN -- which belongs to uncoupled class of networks -- aims at solving the same differential equation, parallel processing would be the key for accelerating computational task in this lattice.

\bibliographystyle{unsrt}
\bibliography{manuscript}
\end{document}